\documentclass[10pt,onecolumn,preprintnumbers,amsmath,amssymb,nofootinbib,superscriptaddress,aps,prd,numberedappendix]{openjournal}

\usepackage{newtxtext,newtxmath}
\usepackage{lipsum}
\usepackage[T1]{fontenc}
\usepackage{ae,aecompl}
\usepackage[nodisplayskipstretch]{setspace}

\usepackage{graphicx}	% Including figure files
\usepackage{amsmath}	% Advanced maths commands
\usepackage{amssymb}	% Extra maths symbols

\DeclareUnicodeCharacter{2212}{-}

\begin{document}
\vspace{-0.5cm}
\title{Mapping Spatially Varying Additive Biases in Cosmic Shear Data}

\author{T. D. Kitching$^{1,\dagger}$ 
, A. C. Deshpande$^{1}$, P. L. Taylor$^{2}$}
\email{$^{\dagger}$t.kitching@ucl.ac.uk, © 2021. All rights reserved.}
\affiliation{
$^{1}$Mullard Space Science Laboratory, University College London, Holmbury St Mary, Dorking, Surrey RH5 6NT, UK\\
$^{2}$Jet Propulsion Laboratory, California Institute of Technology, 4800 Oak Grove Drive, Pasadena, CA 91109, USA}

\begin{abstract}
In this paper we address the challenge of extracting maps of spatially varying unknown additive biases from cosmic shear data. This is done by exploiting the isotropy of the cosmic shear field, and the anisotropy of a typical additive bias field, using an autocorrelation discrepancy map; which identifies significant non-Gaussian components of the map. We test this approach using simulations and find that the autocorrelation discrepancy map produces spatially varying features that are indicative of the additive bias field both in amplitude and spatial variation. We then apply this to the Dark Energy Survey Year 1 data, and find evidence for spatially varying additive biases of at most $2\times 10^{-3}$ on large-scales. The method can be used to empirically inform modelling of the spatially varying additive bias field in any cosmological parameter inference, and can act as a validation test for cosmic shear surveys. 
\end{abstract}

\maketitle
\vspace{-0.5cm}
\section{Introduction}
\label{S:Intro}
When measuring the weak lensing effect from data, biases can be introduced by several effects. These include inaccuracies in the algorithms used \citep{step1,step2,great08,great10,great3}, the size of the point spread function (PSF)  \citep{2017MNRAS.468.3295H,2019A&A...624A..92K, 2021MNRAS.504.4312G}, detector effects \citep{2014JInst...9C3048A}, or detection effects \citep{cccp, 2021arXiv210810057H}. The propagation of such biases into cosmic shear power spectra is shown in \cite{K19,K20}, that include multiplicative and additive biases and the impact of masked data sets. 

In \cite{K19,K20} it is shown that, as a good approximation, multiplicative biases only propagate into the cosmic shear power spectrum as an average (the mean taken over all observed angles on the celestial sphere), therefore they are readily accountable for via calibration of average properties with simulations, and subsequent division or multiplication of the statistics, or joint-marginalisation with cosmological parameters. However, the additive biases do not average over the celestial sphere and propagate into the observed masked cosmic shear power spectrum via an auto-correlation term and a cross-correlation term with the cosmic shear itself i.e. the variation of the additive bias on the celestial sphere needs to be fully accounted for; we will refer to angular variation on the celestial sphere as `spatial' variation for brevity.

%Treatments of additive biases to date have not fully accounted for such spatial variation. Instead biases are either assumed to be spatially constant \citep{DESY3_1,2021arXiv210513543A,2021A&A...646A.129J}, or propagated using an estimated auto-correlation term only \citep{2018PASJ...70S..25M}. A more complicated treatment was presented in \cite{2020A&A...633A..69H} that determined a field-of-view scale spatially varying additive term and used this to correct the auto-correlation contribution to the cosmic shear correlation function.

If one had a good model for the spatial variation of the additive biases then this could then be used to account, or remove, such biases. However, the difficulty with accounting for the spatial variation of the additive term is that with an unknown bias the model to apply is by nature unknown. We note here that in general cosmic shear power spectra cannot be sensitive to constant additive biases \cite[due to the field being spin-2; see][]{K19,K20}, and any random isotropic part will act as an additional shot noise term, therefore it is the anisotropic part of the additive field that is salient. In this paper we begin to address this by developing an empirical method that can extract a map of the additive bias from cosmic shear data. We compare our approach to alternatives such as smoothing the observed masked shear field, or determining excess expected variance.

In Section \ref{S:Method} we present the methodology and discuss requirements on additive biases, in Section \ref{S:Results} we present results of testing on simulations and an application to DES Year 1 data, and in Section \ref{Conclusions} we discuss conclusions.
\\
\section{Method}
\label{S:Method}
The observed masked shear field, including biases, noise, and mask effects, can be written 
\begin{eqnarray}
    \label{gamma}
    \widetilde\gamma(\mathbf{\Omega})&=&W(\mathbf{\Omega})\{[1+m_0(\mathbf{\Omega})][\gamma(\mathbf{\Omega})+n(\mathbf{\Omega})]+m_4(\mathbf{\Omega})[\gamma^*(\mathbf{\Omega})+n^*(\mathbf{\Omega})]+c(\mathbf{\Omega})\},
\end{eqnarray}
where each of the quantities are dependent on angular coordinates $\mathbf{\Omega}=(\theta,\phi)$, where $\theta$ and $\phi$ are latitude and longitude (or R.A. and dec). We include a spin-$0$ mask (in this nomenclature, spin-$s$ means spin positive $s$) $W(\mathbf{\Omega})$ ($W(\mathbf{\Omega})=1$ where data exists and $W(\mathbf{\Omega})=0$ where there is no data). The true spin-2 shear is $\gamma(\mathbf{\Omega})$, and the measured spin-2 shear is $\widetilde\gamma(\mathbf{\Omega})$. $m_0(\mathbf{\Omega})$ and $m_4(\mathbf{\Omega})$ are spin-0 and spin-4 position-dependent multiplicative bias terms respectively; in general each term needs to be spin-2, shear is spin-2 so at linear order only a spin-0 multiplicative bias or a spin-4 multiplied by a spin-minus 2 field can contribute. $c(\mathbf{\Omega})$ is a spin-2 position-dependent additive bias. $n(\mathbf{\Omega})$ is the un-lensed uncorrelated galaxy ellipticity, or the zero-lag intrinsic ellipticity field \citep{2001ApJ...559..552C,2016MNRAS.461.4343L,2015JCAP...08..015B}, which for a finite number of galaxies is a shot noise term \citep{2019PhRvD.100j3506B}. Any intrinsic alignment terms are captured in the $\gamma(\mathbf{\Omega})$ term. $^*$ is a complex conjugate. In general we use $\widetilde{x}$ to mean an observed quantity.

One can take a spherical harmonic transform of the observed masked shear field and separate the E-mode component (that contains the cosmological information),
\begin{eqnarray}
    \label{gtoe}
    \widetilde\gamma^E_{\ell m}&=&\frac{1}{2}\int {\rm d}\mathbf{\Omega}\, [
    \widetilde\gamma(\mathbf{\Omega})\,{}_2Y^*_{\ell m}(\mathbf{\Omega})+
    \widetilde\gamma^*(\mathbf{\Omega})\,{}_{-2}Y^*_{\ell m}(\mathbf{\Omega})]
\end{eqnarray}
where $\ell$ and $m$ are angular wavennumbers, and ${}_sY_{\ell m}(\mathbf{\Omega})$ are spin-weighted spherical harmonics for spin-$s$. The power spectra of the observed masked shear can now be determined
\begin{eqnarray}
\label{eqsize}
\widetilde C^{EE}_{\ell} := \frac{1}{2\ell+1}\sum_m 
\widetilde\gamma^E_{\ell m}\widetilde\gamma^{E,*}_{\ell m}
\end{eqnarray}
where the sum runs over $m=[-\ell,\ell]$, and care must be taken disambiguate $m$ (wavenumber) and $m(\mathbf{\Omega})$ (position-dependent multiplicative bias). 

As shown in \cite{K20} if we assume no strong coupling between the mask and the multiplicative bias fields, that terms $\mathcal{O}(m(\mathbf{\Omega})c(\mathbf{\Omega}),m^2(\mathbf{\Omega}))=0$, and that the B-modes are negligible \citep[see][for justification, and relaxations of, these assumptions]{K19,K20} then the E-mode auto-correlation power spectrum is 
\begin{eqnarray}
\label{lindec}
\widetilde C^{EE}_{\ell}&\approx&{\sum^L_{\ell'}}
{\mathcal M}^{++}_{\ell\ell'} [(1+2\langle m^R(\mathbf{\Omega})\rangle)(C^{EE}_{\ell'}+N_{\ell})+2C^{E c_E}_{\ell'}+C^{c_E c_E}_{\ell'}],
\end{eqnarray}
where $m^R(\mathbf{\Omega})=\mathbb{R}(m_0(\mathbf{\Omega})+m_4(\mathbf{\Omega}))$, and the angular brackets denote an angular average. Superscript $c_E$ denotes a correlation of the E-mode part of the $c(\mathbf{\Omega})$ field, superscript $E$ denotes a correlation of the E-mode part of the $\gamma(\mathbf{\Omega})$ field. $\smash{N_{\ell}=\langle n_{\ell' m'}n^*_{\ell m}\rangle=\delta^K_{\ell\ell'}\delta^K_{mm'}\sigma_e^2/N_{\rm gal}}$ is the shot noise term where $\sigma_e^2$ is variance of the unlensed ellipticities, and $N_{\rm gal}$ is the effective number of galaxies \citep[for a discussion of the effective number density see][]{2019PhRvD.100j3506B, 2013MNRAS.434.2121C}. Throughout we denote maximum $\ell$-mode as $L$, where sums over $\ell$ run from $2$ to $L$, and sums over $m$ run from $-\ell$ to $\ell$. The mixing matrix ${\mathcal M}^{++}_{\ell\ell'}$ incorporates the correlation of angular modes $\ell$ due to the mask and is defined in \cite{K20} (equation 12). This general methodology for propagating the affect of masks into power spectra is known as the `pseudo-$C_{\ell}$' approach \citep[see e.g.][that is particularly relevant for spin-2 fields]{BCT}. 

If the $m_4(\mathbf{\Omega})$ term is small then an unbiased estimate for the shear, neglecting higher order terms $\mathcal{O}(m(\mathbf{\Omega})c(\mathbf{\Omega}))$, is $\widetilde\gamma(\mathbf{\Omega})/[1+m_0(\mathbf{\Omega})]$ and this is the approach commonly used \citep[e.g.][]{DESY3_1,2018PASJ...70S..25M,2020A&A...633A..69H}. However the treatment of the additive term is more complex. The power spectrum is cannot be dependent on spatially constant terms, because $_2Y_{\ell m}(\mathbf{\Omega})=0$ for $\ell<2$. Furthermore, whilst on average over an ensemble of realisations of the shear field $\smash{\langle C^{Ec_E}_{\ell}\rangle=0}$ for a single realisation (as we have with data) this is non-zero, and both the terms in equation (\ref{lindec}) must be accounted for. The change in the power spectrum caused by an additive bias is
\begin{eqnarray}
\label{deltaCl}
\delta\widetilde C^{EE}_{\ell}&=&{\sum^L_{\ell'}}
{\mathcal M}^{++}_{\ell\ell'} [2C^{E c_E}_{\ell'}+C^{c_E c_E}_{\ell'}]=2\widetilde C^{E c_E}_{\ell'}+\widetilde C^{c_E c_E}_{\ell'}.
\end{eqnarray}
Furthermore without a model for the $c(\mathbf{\Omega})$ field (and hence a model for the $c_E$ correlations) measurements of the power spectra alone cannot be used to disentangle the shear contribution from the additive bias contribution; this is fundamentally because, neglecting the multiplicative bias term, at any given $\ell$-mode there is one observable $\widetilde\gamma^E_{\ell m}$ and two unknowns $\gamma^E_{\ell m}+c^E_{\ell m}$ that contribute. The same argument is true for EB, BE and BB contributions; however if $\gamma^B(\mathbf{\Omega})=0$ then the BB power spectra can be used to determine the B-mode part of $c(\mathbf{\Omega})$ since after correction for the noise term in such a case $\widetilde\gamma^B(\mathbf{\Omega})=c^B(\mathbf{\Omega})$. Therefore we need a way to determine the unknown E-mode component of $c(\mathbf{\Omega})$ such that it can be removed or modelled, and this is focus of this paper.
\begin{figure*}
\centering
\includegraphics[width=0.33\columnwidth]{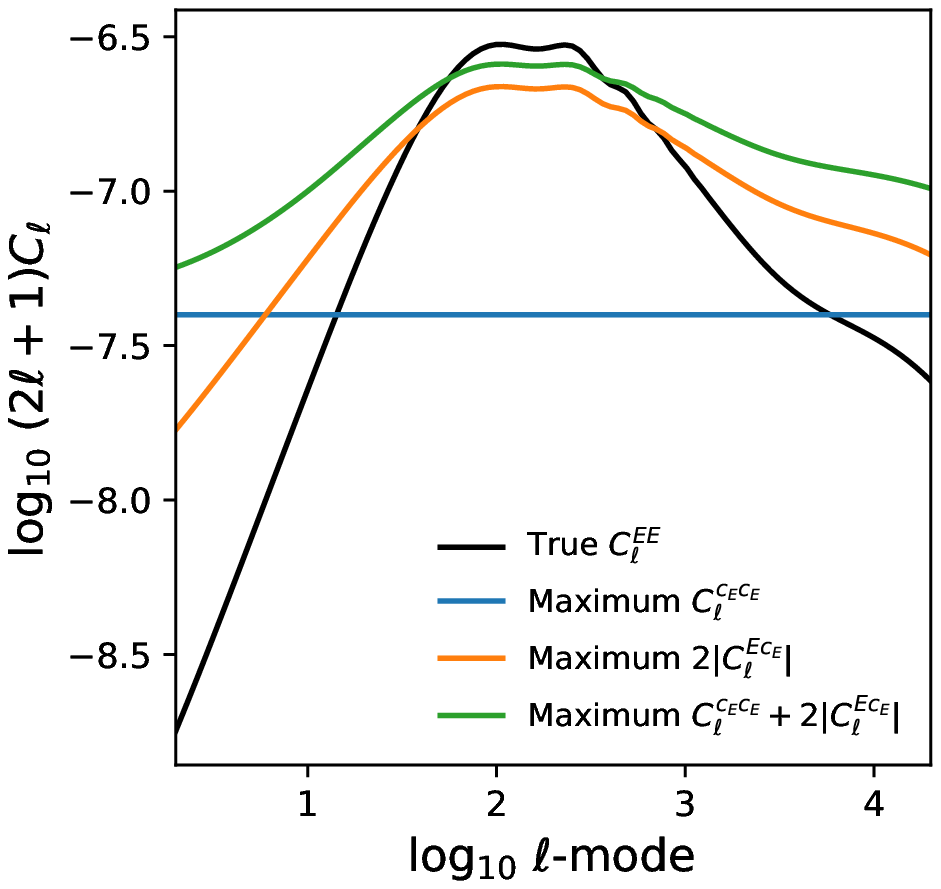}
\includegraphics[width=0.33\columnwidth]{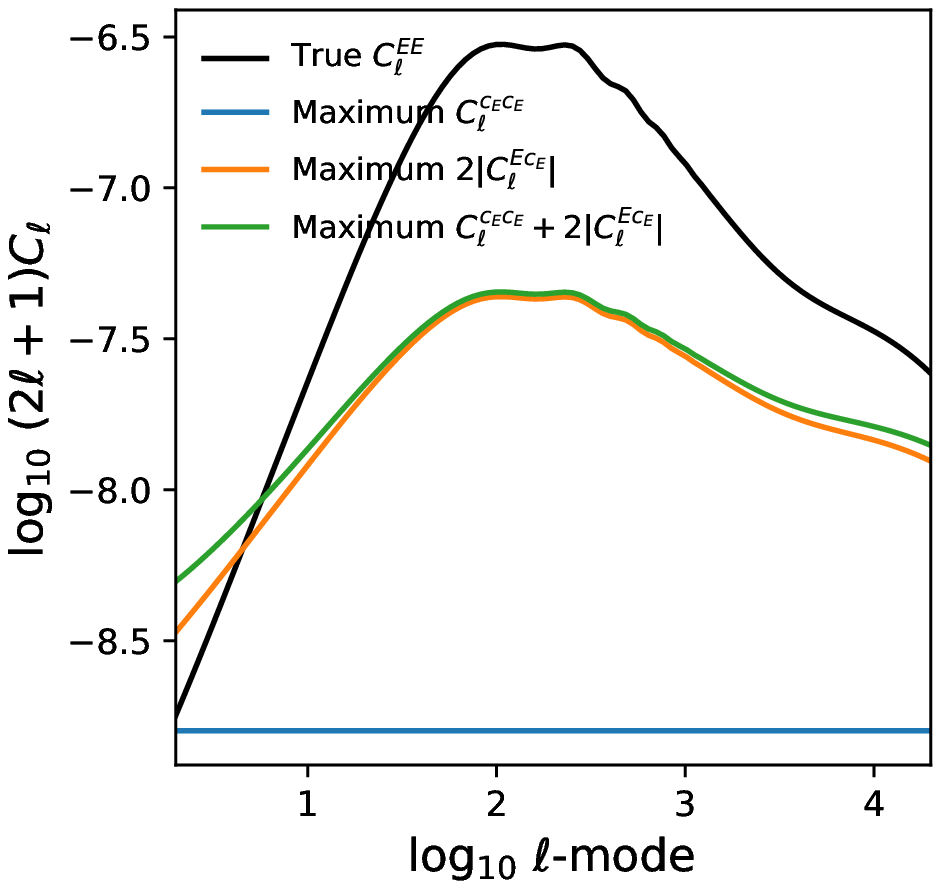}
\includegraphics[width=0.33\columnwidth]{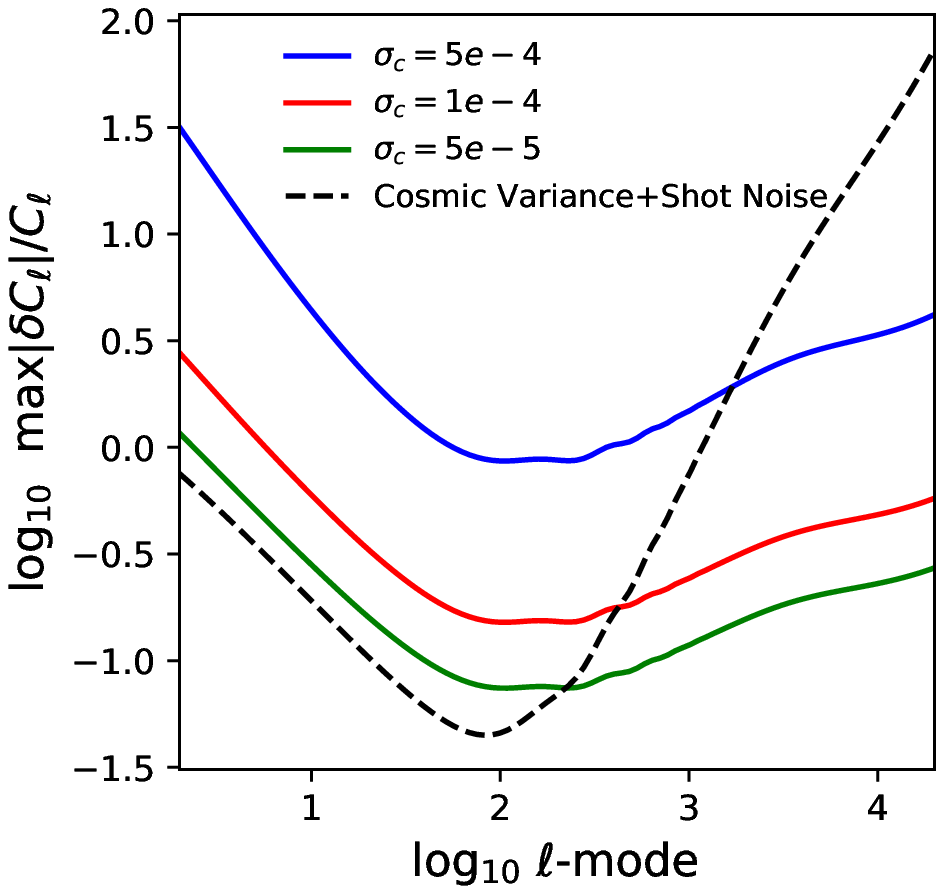}
\caption{Left and Middle: the cosmic shear power spectrum (see Section \ref{S:Method}), compared to the maximum imposed bounds on $\smash{C^{cc}_{\ell}}$ and $\smash{C^{Ec_E}_{\ell}}$ by  requirements of $\sigma_c<5\times 10^{-4}$ (left) and  $\sigma_c<10^{-4}$ (middle). Right: the maximum fractional change on the cosmic shear power spectrum for $\sigma_c<5\times 10^{-4}$ (blue) and $\sigma_c<10^{-4}$ (red), compared to the fractional noise contribution (cosmic variance plus shot noise) to the observed power spectrum (dashed line).} 
\label{Reqs}
\end{figure*}

\subsection{Additive Requirements}
\label{S:Req}
Before we describe how to identify and remove additive biases it is useful to understand how requirements on additive biases propagate into the cosmic shear power spectrum. Typically requirements on biases are set either on the absolute value of the mean additive bias field components, $|\langle c_i(\mathbf{\Omega})\rangle|\leq R_{c_i}$, where $R_x$ is a requirement on $x$; or on the standard deviation of the additive bias field per component $\sigma_{c_i}\leq R_{\sigma_{c_i}}$. For example $\smash{R_{c_i}\simeq 10^{-4}}$ and $\smash{R_{\sigma_{c_i}}\simeq 5\times 10^{-4}}$ for a dark energy Stage-IV like experiment \citep{2006astro.ph..9591A, cropper,massey}. As we have commented any requirement on the mean of the additive field is nugatory because of the impossibility of the power spectrum to be dependent on constant terms. The impact of setting a requirement on $\sigma_{c_i}$ is more complicated, but can be understood via Parseval's theorem applied to the spin-weighted spherical harmonic setting \citep{seibert2018spin}, 
\begin{eqnarray}
\frac{1}{4\pi}\int {\rm d}{\mathbf{\Omega}} |f^2({\mathbf{\Omega}})|=\sum_m\sum_{\ell}f_{\ell m}f^*_{\ell m}\,\,\,\,\,\,\,{\rm and}\,\,\,\,\,\,\,\frac{1}{4\pi}\int {\rm d}{\Omega} |c_E^2({\mathbf{\Omega}})|=\frac{1}{2\pi}\sigma^2_{c_i}=\sum_{\ell}(2\ell+1)C^{c_Ec_E}_{\ell},
\end{eqnarray}
which we show for an arbitrary function on the sphere $f({\Omega})$, and in our case of an additive bias field, assuming that the error on each of the component of the additive field $c_i({\mathbf{\Omega}})$ are equal, where $c^E({\mathbf{\Omega}})=c_1({\mathbf{\Omega}})+c_2({\mathbf{\Omega}})$ so that $i=1$, $2$. Since by definition $C^{c_Ec_E}_{\ell}>0$ this leads to a maximum condition on the amplitude of the $c_Ec_E$ power spectrum at any given $\ell$-mode (corresponding to the case of a single spike at that mode) so that a requirement $R_{\sigma_{c_i}}$ strictly leads to
\begin{eqnarray}
{\rm max}[C^{c_Ec_E}_{\ell}]\leq \frac{R^2_{\sigma_{c_i}}}{[2\pi(2\ell+1)]}
\end{eqnarray}
for all $\ell$. In practice the power spectrum will not be a spike so this is very much a worst case. 

Whilst the $\smash{C^{c_Ec_E}_{\ell}>0}$ the cross-correlation term $\smash{C^{Ec_E}_{\ell}}$ can be either positive of negative. To understand the impact of a requirement on this term we can use the Cauchy-Schwarz inequality applied to complex value fields such that $\smash{2|C^{Ec_E}_{\ell}|\leq 2[C^{c_Ec_E}_{\ell}C^{EE}_{\ell}]^{1/2}}$ for all $\ell$. Given a requirement this leads to a maximum possible value of this term to be 
\begin{eqnarray}
{\rm max}[2|C^{Ec_E}_{\ell}|]\leq 2[C^{EE}_{\ell}]^{1/2}\frac{R_{\sigma_{c_i}}}{[2\pi(2\ell+1)]^{1/2}}.
\end{eqnarray}
Again in practice the auto-correlation power spectrum will not be a spike so this is very much a worst case. Therefore we find that by setting a requirement on $R_{\sigma_{c_i}}$ the amplitude of the change in the power spectrum per mode is constrained to be in a worst case
\begin{eqnarray}
{\rm max}[\delta C^{EE}_{\ell}]\leq {\rm max}\left\{\frac{R^2_{\sigma_{c_i}}}{[2\pi(2\ell+1)]}\pm
2[C^{EE}_{\ell}]^{1/2}\frac{R_{\sigma_{c_i}}}{[2\pi(2\ell+1)]^{1/2}}\right\}
\end{eqnarray}
for all $\ell$. Therefore over scales where ${\rm max}[|\delta C^{EE}_{\ell}|]/C^{EE}_{\ell}\ll 1$ any residual spatially varying additive field will have a minimal impact, if the overall variance of the bias field is constrained by $R_{\sigma_{c_i}}$. This has an interesting effect of a requirement $R_{\sigma_{c_i}}$ being much more constraining than one may expect by also constraining the $\ell$-mode ranges over which one needs to be concerned.

In Figure \ref{Reqs} we show the maximum ${\rm max}[C^{cc}_{\ell}]$,  ${\rm max}[2C^{Ec_E}_{\ell}]$, and ${\rm max}[\delta C^{EE}_{\ell}]$ for $R_{\sigma_{c_i}}=5\times 10^{-5}$, $1\times 10^{-4}$ and $5\times 10^{-4}$. It can be seen that above $\ell\approx 1000$ the maximum changes imposed by these requirements are subdominant to the noise. The requirements suppress the maximum allowable changes at intermediate scales $\ell\approx 100$, but at large-scales do not. Since already modern shape measurement methods have performances with $\sigma_{c_i}\leq 5\times 10^{-4}$ \citep[e.g.][]{2021A&A...646A.124H, 2021arXiv210810057H} this means we only need to be predominately concerned with the impact of additive biases for low-$\ell$ modes. Furthermore it suggests that one can set a requirement on $\sigma_{c_i}$, without modelling a full spatially varying field and that if the requirement is small enough the  impact of the spatial variation is well-constrained. These results confirm the findings of \cite{2020A&A...635A.139E} and \cite{2016MNRAS.455.3319K} who found that for typical systematic effects constrained with a requirement on $\sigma_{c_i}$ that the power spectrum changes were predominately on large-scales. Furthermore that small-scale systematic effects were sub-dominant to the cosmic shear power spectrum and had a minimal impact on cosmological parameter inference.

\subsection{Mapping the additive bias} 
\label{Mapping the additive bias} 
To detect and remove additive biases the property that we exploit is that the shear field and the additive bias field are expected to have different generic spatial variations. The shear field is expected to be isotropic, and on large-scales close to a Gaussian random field\footnote{In the pedagogical derivation in this subsection 2.2 the fields we consider are strongly isotropic, using the definition of \cite{2008arXiv0807.0687M}, which implies that the spherical harmonic coefficients $\gamma_{\ell m}$ are independent, Gaussian distributed and the power spectrum is only dependent on $\ell$. In the masked case, section 2.3, the field is only weakly isotropic, because the coefficients are no longer independent (they are correlated) but the power spectrum is still only dependent on $\ell$. See \cite{2008arXiv0807.0687M} and \cite{2010JMP....51d3301M} for further discussions.}. The additive bias field is expected to be anisotropic, and generally not a Gaussian random field; since such biases are caused by underlying processes that are not isotropic or Gaussian.

For an isotropic all-sky Gaussian field the spherical harmonic coefficients $\gamma^E_{\ell m}$ are uncorrelated independent complex-valued variables that satisfy $\smash{\langle\gamma^E_{\ell m}\rangle=0}$ and $\smash{\langle\gamma^E_{\ell m}\gamma^{E,*}_{\ell' m'}\rangle=C_{\ell}\delta^K_{\ell\ell'}\delta^K_{mm'}}$. In time-series analysis departures from \emph{iid} (independent and identically distributed random variables) distributed data can be determined via the autocorrelation (or autocovariance) function that finds significant non-\emph{iid} behaviour by looking for excess covariance between different time steps separated by a lag (essentially the off-diagonal part of the observed time-dependent covariance matrix). In \cite{2019arXiv191111442H} this concept was generalised to spin-0 isotropic Gaussian distributed data on the sphere, and applied to Cosmic Microwave Background (CMB) data. We will review and slightly generalise the \cite{2019arXiv191111442H} result for the case of masked cosmic shear data.

In the following we consider a bias field $c(\mathbf{\Omega})=c_1(\mathbf{\Omega})+{\rm i}c_2(\mathbf{\Omega})$, and use the subscript $i=1$, $2$ to label the real and imaginary components of any quantity respectively. The spherical harmonic transform of the shear field is
\begin{eqnarray}
\label{eeq5}
\widetilde\gamma_{\ell m}=\int {\rm d}\mathbf{\Omega}\,\widetilde\gamma(\mathbf{\Omega})\,{}_2Y^*_{\ell m}(\mathbf{\Omega}),
\end{eqnarray}
we define this for $\ell\geq 2$ to reflect the spin-2 nature of the shear field. We can now define the autocorrelation probe coefficients
\begin{eqnarray}
\label{tprobe}
\widetilde{T}_{\ell}(\mathbf{\Omega})=\sum_m \widetilde\gamma_{\ell m}\,_2Y_{\ell m}(\mathbf{\Omega}).
\end{eqnarray}
These $\widetilde{T}_{\ell}(\mathbf{\Omega})$ are the shear field filtered on scale $\ell$ (modulo normalisation). If the field in question is an isotropic all-sky (unmasked) Gaussian random field then \citep{2019arXiv191111442H}
\begin{eqnarray}
\label{mu}
\langle T_{\ell}(\mathbf{\Omega})\rangle=0\,\,\,\,\,\,\,{\rm and}\,\,\,\,\,\,\,
\langle T_{\ell}(\mathbf{\Omega})T^*_{\ell'}(\mathbf{\Omega})\rangle=\frac{(2\ell+1)}{4\pi}\widetilde{C}_{\ell}\delta^K_{\ell\ell'}
\end{eqnarray}
where angular brackets are ensemble averages\footnote{The ensemble average refers to an average over a hypothetical set of random realisations of the field in question.} and $\smash{\langle\widetilde\gamma_{\ell m}\widetilde\gamma^{*}_{\ell' m'}\rangle=\widetilde{C}_{\ell}\delta^K_{\ell\ell'}\delta^K_{mm'}}$ is the observed power spectrum. It can also be shown for an isotropic all-sky Gaussian random field that the variance of $T_{\ell}(\mathbf{\Omega})T^*_{\ell'}(\mathbf{\Omega})$ is 
\begin{eqnarray}
\label{var}
{\rm var}(T_{\ell}(\mathbf{\Omega})T^*_{\ell'}(\mathbf{\Omega}))=\frac{(2\ell+1)(2\ell'+1)}{16\pi^2}[\widetilde{C}^2_{\ell}\delta^K_{\ell\ell'}+\widetilde{C}_{\ell}\widetilde{C}_{\ell'}]
\end{eqnarray}
through use of Wick's theorem and the additive properties of the spherical harmonic functions; this agrees with \cite{10.2307/2985145} who present the variance of autocorrelation estimators in the time-series case.

The central methodology is to compute the autocorrelation function $\widetilde{T}_{\ell}(\mathbf{\Omega})\widetilde{T}^*_{\ell'}(\mathbf{\Omega})$ and for each $\mathbf{\Omega}$ to identify values of this observable that exhibit an excess value with respect to the isotropic Gaussian expected values. This is done by making an \emph{autocorrelation discrepancy map} as follows 
\begin{eqnarray}
\label{Dalpha}
\widetilde{D}_i(\mathbf{\Omega}):=\sum^L_{\ell'}{\rm max}\{|\widetilde\alpha_{i,\ell'}(\mathbf{\Omega})|-t_{i,\ell'}(\mathbf{\Omega}),0\}{\rm sgn}[\widetilde\alpha_{i,\ell'}(\mathbf{\Omega})]
\end{eqnarray}
where
\begin{eqnarray}
\widetilde\alpha_{i,\ell'}(\mathbf{\Omega})=\sum^L_{\ell} |\widetilde{T}_{i,\ell}(\mathbf{\Omega})\widetilde{T}^*_{i,\ell'}(\mathbf{\Omega})|w_{\ell\ell'}{\rm sgn}[\widetilde{T}_{i,\ell}(\mathbf{\Omega})], 
\end{eqnarray}
that can alternatively be written like $\smash{\widetilde\alpha_{i,\ell'}(\mathbf{\Omega})=\widetilde{T}^*_{i,\ell'}(\mathbf{\Omega})\sum^L_{\ell} |\widetilde{T}_{i,\ell}(\mathbf{\Omega})|w_{\ell\ell'}}$. 
$w_{\ell\ell'}=1$ if $0\leq \ell'-\ell\leq k_{\rm max}$, and zero otherwise, where $k=\ell'-\ell$ is referred to as the `lag', and a maximum $k_{\rm max}$ is chosen to optimise speed and accuracy (noting that for large lag correlations tend to be very small for most plausible additive bias fields), but in this paper we set $k_{\rm max}=L$ which if computationally feasible should be done. ${\rm sgn[x]}$ is defined to be $(-1,0,1)$ for $(x<0,x=0,x>1)$ respectively and is introduced to maintain the sign information of the anisotropic field. $t_{i,\ell'}(\mathbf{\Omega})$ is a position-dependent threshold value that we define generically as the sum of the expected mean plus uncertainty
\begin{eqnarray}
\label{tL}
t_{i,\ell'}(\mathbf{\Omega})=\sum^L_{\ell}\langle T_{i,\ell}(\mathbf{\Omega})T^*_{i,\ell'}(\mathbf{\Omega})\rangle w_{\ell\ell'}+N_{\sigma}\left[\sum^L_{\ell}{\rm var}( T_{i,\ell}(\mathbf{\Omega})T^*_{i,\ell'}(\mathbf{\Omega}) w_{\ell\ell'})\right]^{1/2}
\end{eqnarray}
that can be computed analytically in the all-sky case (equations 
\ref{mu} and \ref{var}), where $N_{\sigma}$ is the number of standard deviations that a detection threshold is set. We note that the $\widetilde\gamma_{\ell m}$ are Gaussian distributed, and therefore so are the $\widetilde{T}_{\ell}(\mathbf{\Omega})$, the $\widetilde\alpha_{\ell'}(\mathbf{\Omega})$ follow a normal product distribution with zero means \citep[which can be well approximated by a Gaussian distribution for independent variables see e.g.][]{2018arXiv180703981G}. For an isotropic Gaussian random field the probability of $\widetilde{D}(\mathbf{\Omega})\not=0$ is given by $p(\widetilde{D}(\mathbf{\Omega})\not=0)\simeq k_{\rm max}{\rm erf}[N_{\sigma}/2^{1/2}]$ i.e. for large $N_{\sigma}$ it should be dominantly zero in the isotropic case. Any deviation away from zero is therefore a detection of a departure from the isotropic Gaussian case. For a pedagogical explanation of autocorrelation discrepancy maps we refer to \cite{2019arXiv191111442H}. 

The structure of the function $\widetilde{D}(\mathbf{\Omega})$ is as follows. The $\smash{\widetilde\alpha_{\ell'}(\mathbf{\Omega})}$ function encodes the observed local correlations between scales $\ell$ and $\ell'$. In principle, for computational reasons, one may only want to consider neighbouring scales and apply a maximum lag, that we encode via $w_{\ell\ell'}$ \citep[][instead set a limit on the sum, see their equations 2.4-2.7]{2019arXiv191111442H}. $t_{\ell'}(\mathbf{\Omega})$, is similar to $\smash{\widetilde\alpha_{\ell'}(\mathbf{\Omega})}$ but encodes the expected local correlation between scales for a model isotropic field. The ${\rm max}\{x,0\}$ part of the function finds $\ell'$ modes where the observed local correlation is larger than expected, otherwise is set to zero. In finding excess correlation the modulus of the observed local correlation is used, therefore in order to retain the sign information ${\rm sgn}[x]$ is introduced.

One can consider the observed field to be the sum of an isotropic and an anisotropic field $\widetilde\gamma_{\ell m}=\widetilde\gamma^{\rm iso}_{\ell m}+\widetilde\gamma^{\rm ani}_{\ell m}$, and this can be substituted into equation (\ref{tprobe}) and then equation (\ref{Dalpha}). The thresholding acts to eliminate the isotropic contribution for large $N_{\sigma}$ -- which guarantees that the probability of $\sum_{\ell}[\sum_m \widetilde\gamma^{\rm iso}_{\ell m}(\mathbf{\Omega}){}_0Y_{\ell m}(\mathbf{\Omega})][\sum_m' \widetilde\gamma^{\rm iso}_{\ell' m'}(\mathbf{\Omega}){}_0Y_{\ell'm'}(\mathbf{\Omega})]^*$ being non-zero in the contribution to $\widetilde D(\mathbf{\Omega})$ is negligible, hence that either $\widetilde\gamma^{\rm iso}(\mathbf{\Omega})$ and/or $[\sum_m' \widetilde\gamma^{\rm iso}_{\ell' m'}(\mathbf{\Omega}){}_0Y_{\ell'm'}(\mathbf{\Omega})]^*$ are negligible in the sum (this removes the $\widetilde\gamma^{\rm iso}_{\ell m}$ autocorrelation and the cross correlation with $\widetilde\gamma^{\rm ani}_{\ell m}$). Therefore in this case  
\begin{eqnarray}
\widetilde{D}(\mathbf{\Omega})\simeq \sum^L_{\ell'}\sum^L_{\ell} \left(\sum_m \widetilde\gamma^{\rm ani}_{\ell m}\,_2Y_{\ell m}(\mathbf{\Omega})\right)\left(\sum_{m'} \widetilde\gamma^{\rm ani}_{\ell'm'}\,_2Y_{\ell'm'}(\mathbf{\Omega})\right)^* w_{\ell\ell'}+\,{\rm constant}\simeq f[\widetilde\gamma^{\rm ani}(\mathbf{\Omega})]^2+\,{\rm constant},
\end{eqnarray}
via inverse spherical harmonic transforms. The constant is equal to a sum over the variance term in equation (\ref{tL}), and the factor $f=(1/[4L^2])k_{\rm max}(4L-2k_{\rm max})$ accounts for a maximum lag being $0<k_{\rm max}\leq L$ (and therefore not covering the entire $(\ell,\ell')$ plane); which for $k_{\rm max}=L$ is $f=1/2$. Therefore in the case of additive biases in cosmic shear data we find that
\begin{eqnarray}
\label{ADM}
\frac{{\rm sgn}[\widetilde{D}(\mathbf{\Omega})]}{f^{1/2}}\left[|\widetilde{D}(\mathbf{\Omega})|-|{\rm min}[\widetilde{D}(\mathbf{\Omega})]|\right]^{1/2}\simeq c(\mathbf{\Omega})
\end{eqnarray}
is an estimator for the additive bias field.
\begin{figure*}
\centering
\includegraphics[width=0.33\columnwidth]{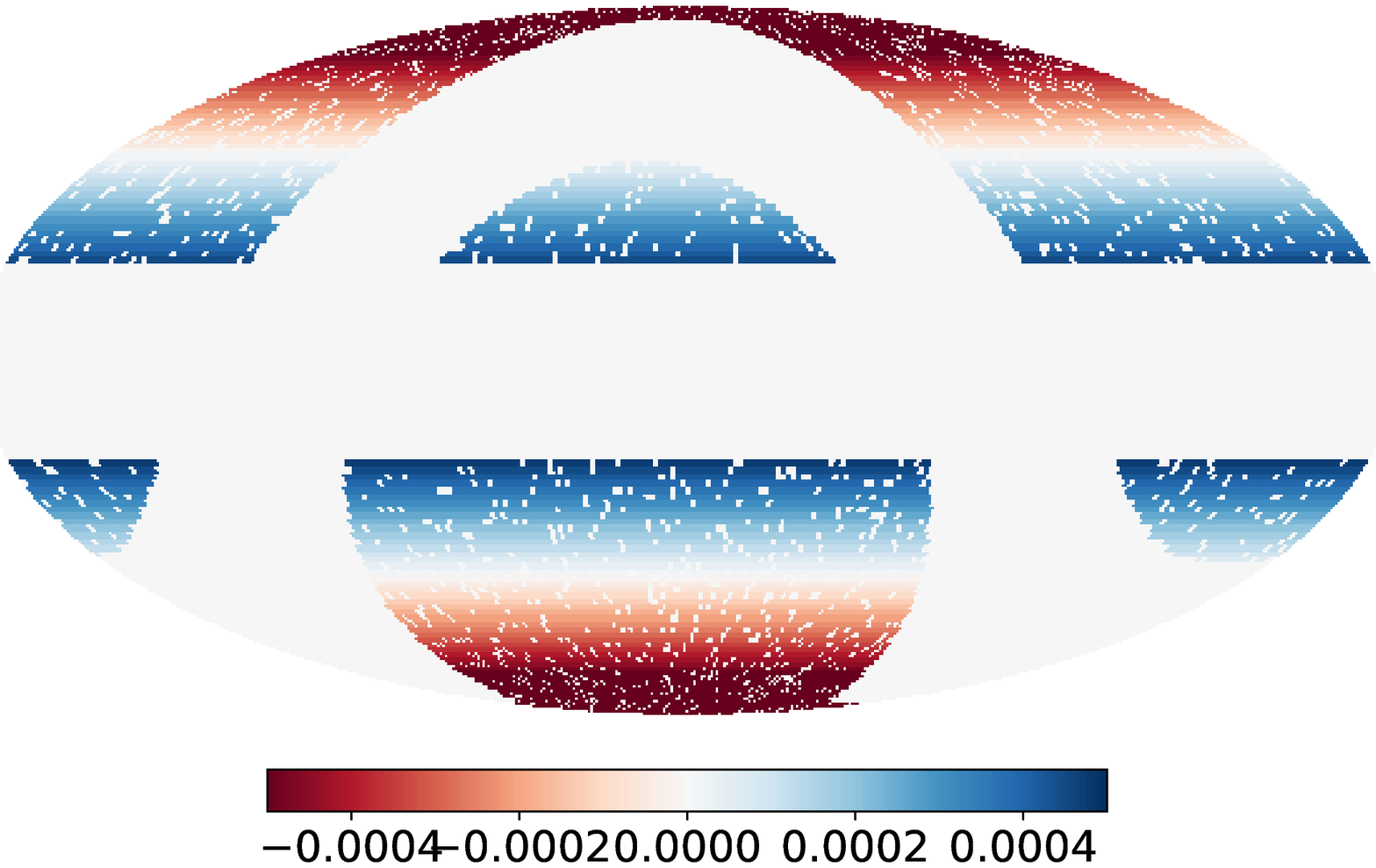}
\includegraphics[width=0.33\columnwidth]{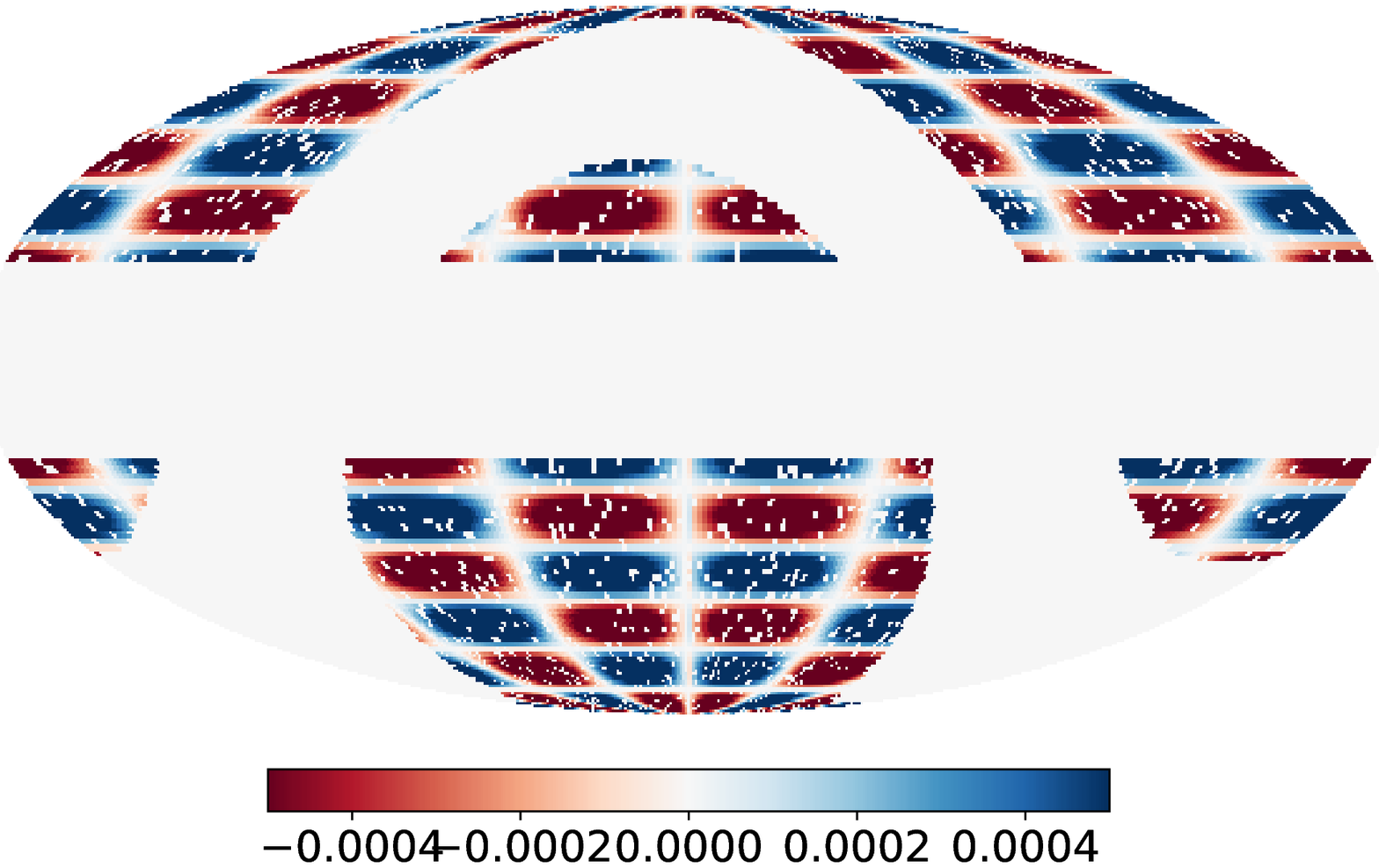}
\includegraphics[width=0.33\columnwidth]{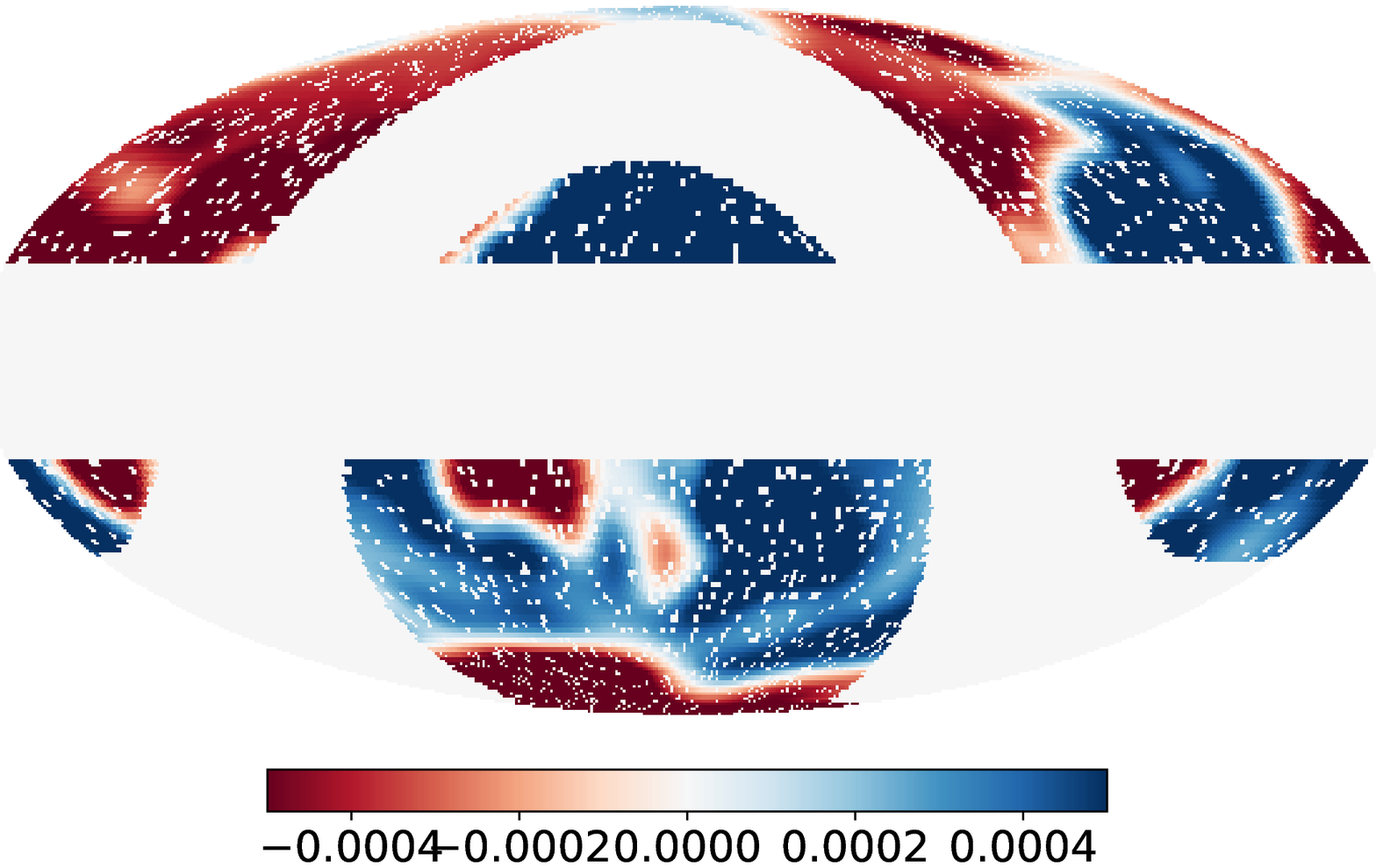}\\
\includegraphics[width=0.33\columnwidth]{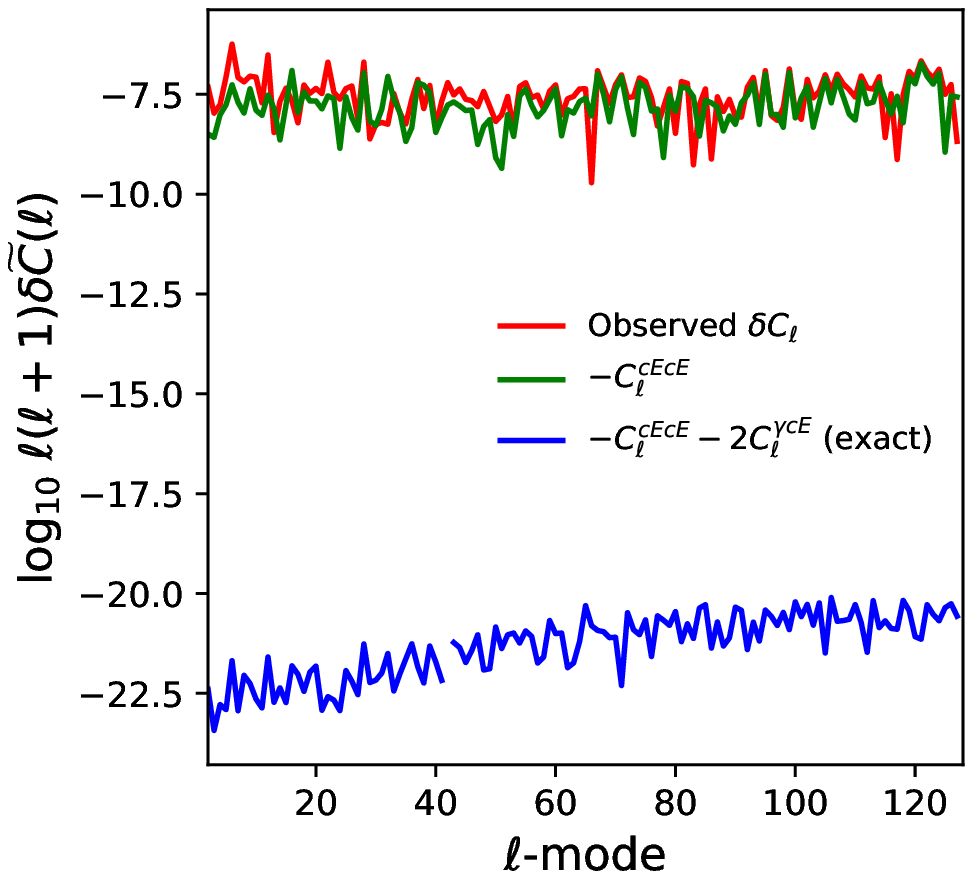}
\includegraphics[width=0.33\columnwidth]{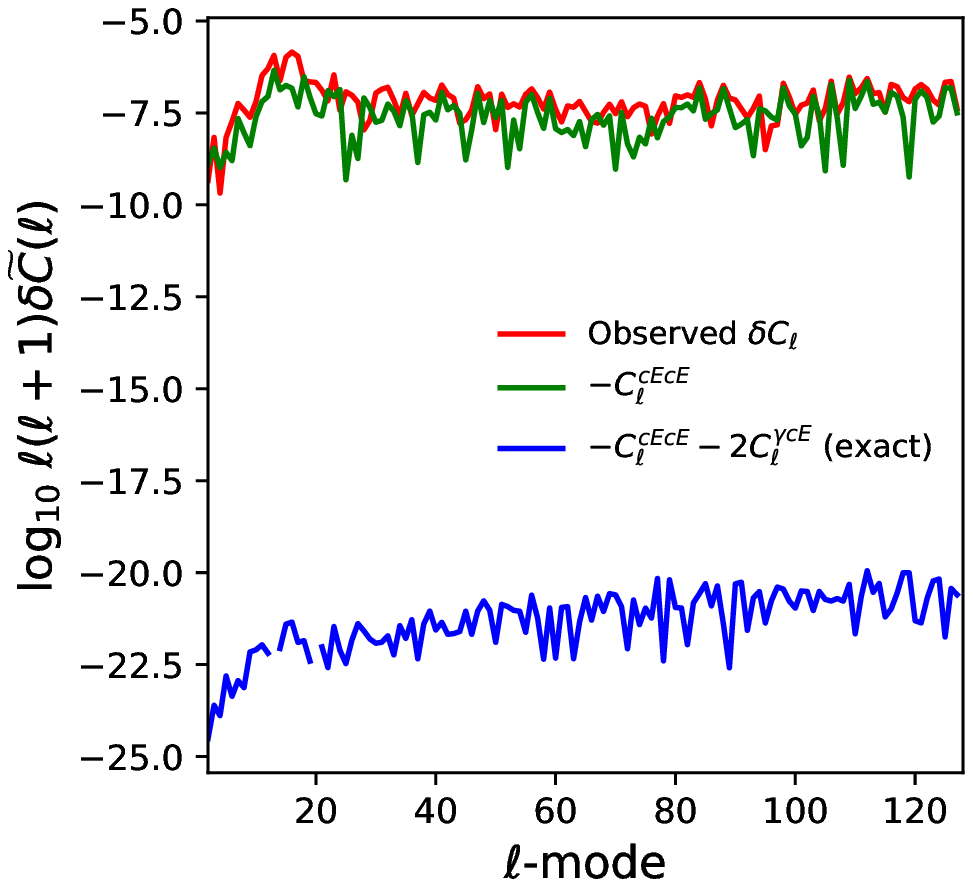}
\includegraphics[width=0.33\columnwidth]{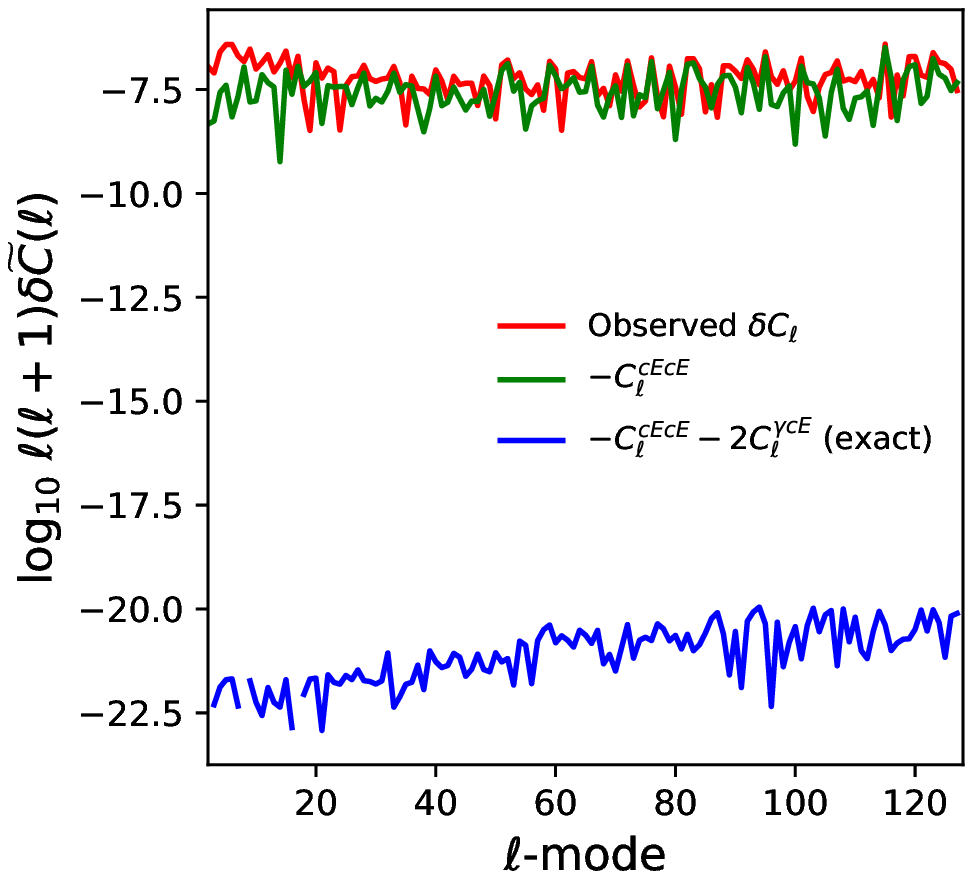}
\caption{Top row: the $c(\mathbf{\Omega})$ of the three masked additive bias fields used in the simulation tests, described in Section \ref{S:Results}. Shown using a Mollweide projection. Left to right are i) a simple galactic plane; ii) a simple patch pattern; and iii) a smoothed Earth topography map. Bottom row: the  change the observed EE power spectrum caused by the additive bias field. Red is the uncorrected power spectrum change; green accounts for the autocorrelation of the additive bias field, the $C^{c_Ec_E}_{\ell}$ term in equation (\ref{deltaCl}); blue also subtracts the shear-additive field cross-correlation accounts for both terms in equation (\ref{deltaCl}).}
\label{cases}
\end{figure*}

\subsection{Masked Data}
When constructing the threshold values $t_{\ell'}(\mathbf{\Omega})$, in the all-sky case the mean and variance are only dependent on the observed $\widetilde{C}_{\ell}$ (equations \ref{mu} and \ref{var}). However all cosmic shear data is always masked. In the case that a mask is present in the data the observed $\widetilde\gamma_{\ell m}$ are no longer independent and uncorrelated -- only the coefficients derived from the underlying, true, shear are $\gamma_{\ell m}$. For a mask $W(\mathbf{\Omega})$ the observed transform coefficients are related to the underlying isotropic Gaussian distributed variables via $\smash{\widetilde\gamma_{\ell m}=\sum^L_{\ell'}\sum_{m'} \gamma_{\ell m}W_{\ell\ell' m m'}}$, where $W_{\ell\ell'mm'}=\int{\rm d}\mathbf{\Omega}\,_0Y_{\ell m}(\mathbf{\Omega})\,_0Y^*_{\ell' m'}(\mathbf{\Omega})W(\mathbf{\Omega})$. This leads to an expression for $\langle T_{\ell}(\mathbf{\Omega})T^*_{\ell'}(\mathbf{\Omega})\rangle$ that is dependent on the true $C_{\ell}$ with a complex mode-mixing relationship 
\begin{eqnarray}
\langle T_{\ell}(\mathbf{\Omega})T^*_{\ell'}(\mathbf{\Omega})\rangle=\sum^L_{\ell''}C_{\ell''}\sum_{m}\sum_{m'}\,_0Y_{\ell m}(\mathbf{\Omega})\,_0Y^*_{\ell' m'}(\mathbf{\Omega})\sum_{m''}W_{\ell\ell''mm''}W_{\ell'\ell''m'm''}
\end{eqnarray}
the variance is even more complex, involving combinations of four multiplied spherical harmonic functions. In this case pursuing an analytic approach is not sensible because i) the computations are very complex/intractable in a reasonable time, ii) the computations depend on the true $C_{\ell}$ not the observed $\widetilde{C}_{\ell}$. 

Therefore a better approach is to empirically determine the expected mean and variance of the autocorrelation functions, in order to construct the threshold values, from an ensemble of masked Gaussian random field simulations (in Appendix A we also test log-normal simulations). Therefore we employ the following approach: 
for a given cosmology we compute the cosmic shear power spectrum $C_{\ell}$; we generate a set of Gaussian random field simulations $r=\{1,\dots,N_{\rm sim}\}$, and for each simulation we apply the observed mask $W(\mathbf{\Omega})$; the threshold values are  computed by taking the mean and variance over the ensemble 
\begin{eqnarray}
t_{i,\ell'}(\mathbf{\Omega})={\rm mean}[\beta_{i,\ell'}(\mathbf{\Omega}; r)]+N_{\sigma}{\rm var}^{1/2}[\beta_{i,\ell'}(\mathbf{\Omega}; r)]\,\,\,{\rm where}\,\,\,\beta_{i,\ell'}(\mathbf{\Omega}; r)=\sum^L_{\ell} |\widetilde{T}_{i,\ell}(\mathbf{\Omega};r)\widetilde{T}^*_{i,\ell'}(\mathbf{\Omega};r)|w_{\ell\ell'}{\rm sgn}[\widetilde{T}_{i,\ell}(\mathbf{\Omega};r)],
\end{eqnarray}
where $T_{i,\ell}(\mathbf{\Omega};r)$ is for a given realisation, and the mean and variance are taken over the ensemble. This procedure leads to a set of threshold values that take into account the same mask, and associated correlations caused in the autocorrelation functions, as observed. 

The final autocorrelation discrepancy map is therefore a detection of deviation from isotropy \emph{for a fixed cosmology} in the masked case. In practice this should not affect the utility of the statistic since the assumed cosmology could be chosen to be that derived from the data without correcting for additive biases or a fiducial cosmology, furthermore sensitivity to the statistic to small changes in cosmology should be minimal for a sufficiently stringent threshold ($N_{\sigma}\geq 3$). 

We note that the uncertainty on the autocorrelation discrepancy map is not the $N_{\sigma}$ value, which only serves as a threshold to isolate the anisotropic signal. In the masked case the uncertainty on the autocorrelation discrepancy map is determined by the number of simulations that one can produce i.e. the threshold value itself is noisy. One can then either estimate the uncertainty via boostrap re-sampling given a sufficiently larger number of simulations, or by creating multiple independent sets of simulations; we demonstrate this in Appendix A. An additional choice that must be made is the orientation of the local coordinate frame with respect to which $c_1$ and $c_2$ are defined; our fiducial frame is to orientate negative $c_1$ to the galactic North pole (top of the Figures in this paper), and positive $c_2$ at $45$ degrees rotated in a clockwise direction. We investigate the dependency of local coordinate frame choice in Appendix A. 
For very small masked fields with strong mode mixing over many scales. We also test this in Appendix A, using the Dark Energy Survey Year 1 mask (see Section \ref{Application to Data}). An alternative for small fields, where one is sure that curved-sky effects are negligible, would be to recast the formalism into a flat-sky approach using Fourier transforms instead of spherical harmonic transforms.

Finally we note that the approach can be iterative in that the estimate from equation (\ref{ADM}) can be used to create a new corrected observed masked shear field $\widetilde\gamma_{i+1}(\mathbf{\Omega})=\widetilde\gamma(\mathbf{\Omega})-\sum_i c_{i}(\mathbf{\Omega})$ where $i$ is an iteration and the total estimate of the additive bias field over all iterations is $\sum_i c_{i}(\mathbf{\Omega})$. As $i$ increases the residuals should tend to zero $c_{i}(\mathbf{\Omega})\rightarrow 0$. We find in practice that this is the case even after one iteration, but we don't recommend this approach because iteration can amplify noisy parts of the estimate. Further work could extend this approach by introducing regularisation and an iterative scheme.

\subsection{Alternatives}
Instead of the autocorrelation discrepancy map there are a few alternatives that one can try. A straightforward approach is to apply a smoothing function to the observed masked shear field (i.e. a convolution with a filter) 
\begin{eqnarray}
\label{gsmooth}
\widetilde\gamma_{\rm smooth}(\mathbf{\Omega})=F[\widetilde\gamma(\mathbf{\Omega}); \sigma_F]
\end{eqnarray}
where $F$ is a filter function, that we take to be an isotropic Gaussian, and $\sigma_F$ is the width of the Gaussian filter in radians. By applying a filter one may hope that on the scale of the filter the shear field may average to zero, leaving parts of the observed field that do not average to zero, including the additive bias field. A second alternative is to create an ensemble of realisations and create a threshold in real (angular space),
$
s(\mathbf{\Omega})={\rm mean}[\widetilde\gamma(\mathbf{\Omega}; r)]+N_{\sigma}{\rm var}^{1/2}[\widetilde\gamma(\mathbf{\Omega}; r)]
$
and determine excess values of the shear field  $\widetilde{E}(\mathbf{\Omega}):={\rm max}\{\widetilde\gamma(\mathbf{\Omega})-s(\mathbf{\Omega}),0\}$.
However, this estimator is not expected to perform well since for a given position the shear field may randomly fluctuate low, or high, which may counteract the local additive field. In comparison with the autocorrelation functions at particular angular position the shear field $\widetilde\gamma(\mathbf{\Omega})$ does not have any unique statistical properties for an isotropic Gaussian random field, whereas  $\widetilde\alpha_{\ell}(\mathbf{\Omega})$ does.

\section{Results}
\label{S:Results}
To test the ability of the autocorrelation discrepancy maps to identify additive biases we use masked-sky simulations that include a Gaussian random field shear $\gamma(\mathbf{\Omega})$, an isotropic shot noise term $n(\mathbf{\Omega})$, and an anisotropic additive bias field $c(\mathbf{\Omega})$. 

\subsection{Models}
We model $\gamma(\mathbf{\Omega})$ as a Gaussian random field using the {\tt massmappy} code \citep{wallis}, assuming a DES Year 1 cosmology \citep{des1,des2,des3} to compute the EE cosmic shear power spectrum. The EE cosmic shear power spectrum, assuming the the Limber \citep{1953ApJ...117..134L,2017MNRAS.469.2737K,2017JCAP...05..014L}, flat-Universe \citep{2018PhRvD..98b3522T}, reduced shear \citep{ad2}, flat-sky \citep{10.1046/j.1365-8711.1998.02054.x} and prefactor-unity \citep{2017MNRAS.469.2737K} approximations is given by:
\begin{equation}
    \label{eq:cltheory}
    C^{EE}_{\ell} = \int_0^{\chi_{\rm H}} {\rm d}\chi \frac{q^2(\chi)}{\chi^2} P_{\delta} \left(\frac{\ell + 1/2}{\chi}, \chi\right),\,\,\,\,\,\,\,{\rm where}\,\,\,\,\,\,\,
    q(\chi) = \frac{3}{2}\Omega_{\rm M} \frac{H^2_0}{c^2} \frac{\chi}{a(\chi)} \int^{\chi_{\rm H}}_\chi {\rm d}\chi'\, n(\chi')\, \frac{\chi'-\chi}{\chi};
\end{equation}
where $P_{\delta}$ is the power spectrum of matter overdensities that we calculate using \texttt{CAMB} \citep{cambcite} \citep[we include the corrections from][for the non-linear corrections]{2015MNRAS.454.1958M}. $\chi$ and $\chi_{\rm H}$ are the comoving distance and comoving distance to the horizon respectively calculated using the \texttt{astropy} package \citep{astropy1, astropy2}, $H_0$ is the Hubble constant, $a$ is the scale factor of the Universe, $\Omega_{\rm M}$ is the dimensionless total matter density of the Universe,  and $c$ is the speed of light in a vacuum.  $n(\chi)$ is the galaxy distribution function of the survey, where we use the photometric DES Year 1 galaxy distribution (available at \url{http://desdr-server.ncsa.illinois.edu/despublic/y1a1\_files/redshift\_bins/}) \citep{des2}. 

We make a mask that removes data within $20^{\circ}$ in both the galactic and ecliptic planes; and also $20\%$ of pixels at random, to represent an all-sky mask with random patches removed (each pixel is  $1.98$ square degrees, as determined by the maximum $\ell$-mode) -- this gives a total observed sky fraction of $f_{\rm sky}=0.4$; the mask can be seen the following Mollweide-projected Figures of the celestial sphere (e.g. Figure \ref{cases}) where the galactic and ecliptic planes can be clearly seen as a horizontal and inverted-U feature, and the random pixel-removed are scattered over the sphere.
The shot noise term is modelled using $\sigma_e=0.3$, and $N_{\rm gal}=30f_{\rm sky}3600(4\pi[180/\pi]^2)$, which may be expected for a Stage IV experiment \citep{2006astro.ph..9591A} with $30$ galaxies per square arcminute.

For the additive bias field we investigate three different cases. The models we use are similar to those in \cite{K19,K20}: i) a simple galactic plane,  $c(\mathbf{\Omega})=\alpha[\pi-|\phi-\pi|]$; ii) a simple patch pattern, $c(\mathbf{\Omega})=\alpha\sin(7|\phi-\pi|)\sin(7|\theta-\pi|)$; and iii) a Gaussian-smoothed Earth topography map \citep{2013A&A...558A.128L} where $c(\mathbf{\Omega})=\alpha F({\rm Earth}(\mathbf{\Omega}); 0.05)$ projected onto the celestial sphere to represent a non-analytic function. We express these in terms of an arbitrary amplitude $\alpha$ since these are all normalised to have a mean of zero and a variance of $\sigma^2_c$ -- this is an optimistic scenario where it is assumed there is no mean additive bias after calibration, but only a residual variance about zero. As a fiducial value we use $\sigma_c=5\times 10^{-4}$, which represents the current performance of the best shape measurement methods \citep[e.g.][]{2021A&A...646A.124H, 2021arXiv210810057H}. Throughout we use {\tt SSHT} \cite{ssht} to sample the sphere \citep{mw}, and use a maximum $\ell$-mode of $L=128$ (resulting in pixels that are $1.98$ square degrees), and we use $N_{\rm sim}=100$ (see Section \ref{S:Method}), for the masked data case.

In Figure \ref{cases} we show the three models using $\sigma_c=5\times 10^{-4}$. We also show the change in the observed EE power spectrum caused by the additive bias, and subtract from this the autocorrelation of the additive bias field and cross-correlation of the additive bias field with the shear field. It is clear that only accounting for the autocorrelation of the additive bias field is insufficient to model the change in the power spectrum, and that the cross-correlation with the shear field must also be included, i.e. both terms in equation (\ref{deltaCl}) are important.
\begin{figure*}
\centering
\textsf{\scriptsize Observed Shear Map}\\
\includegraphics[width=0.33\columnwidth]{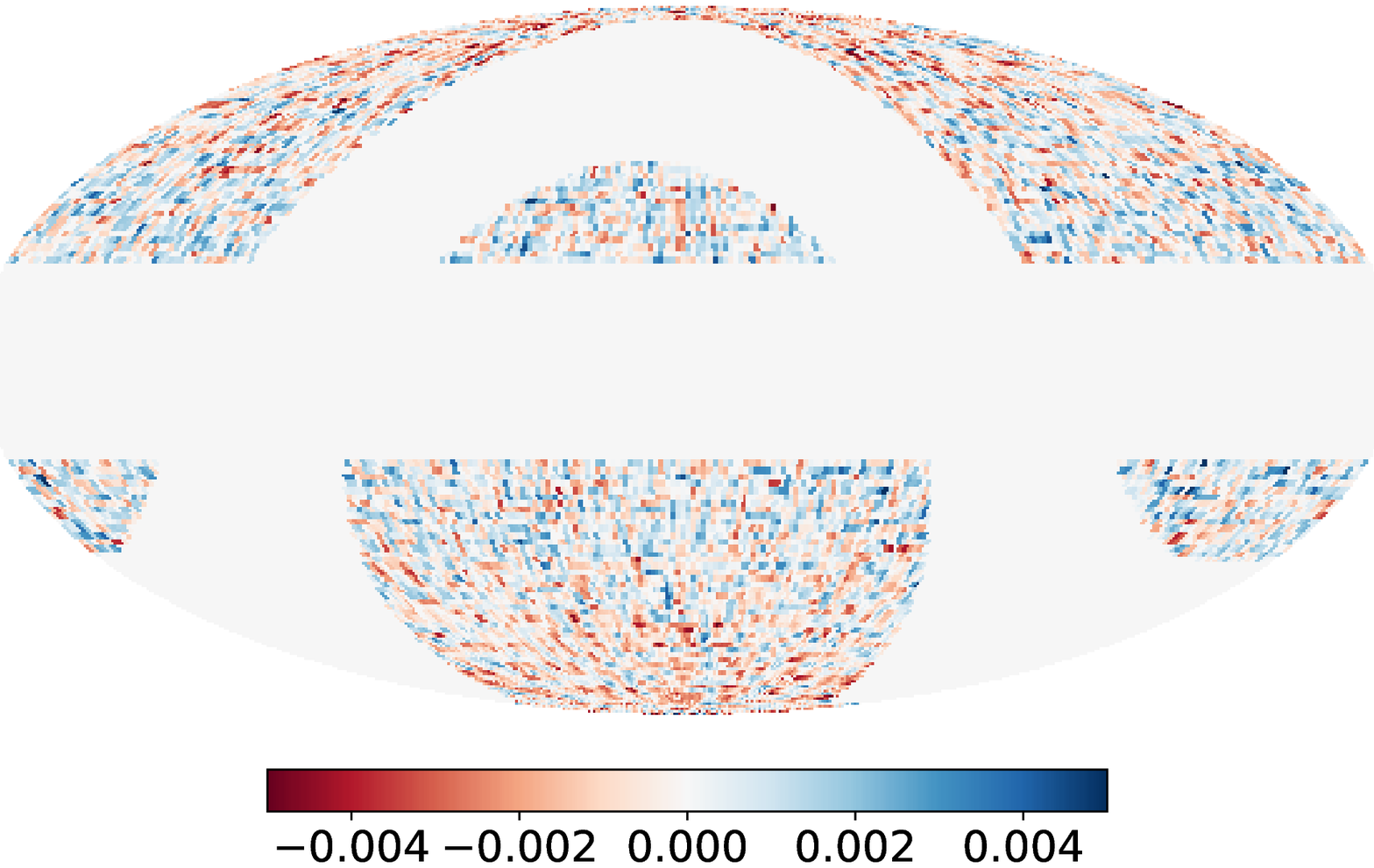}
\includegraphics[width=0.33\columnwidth]{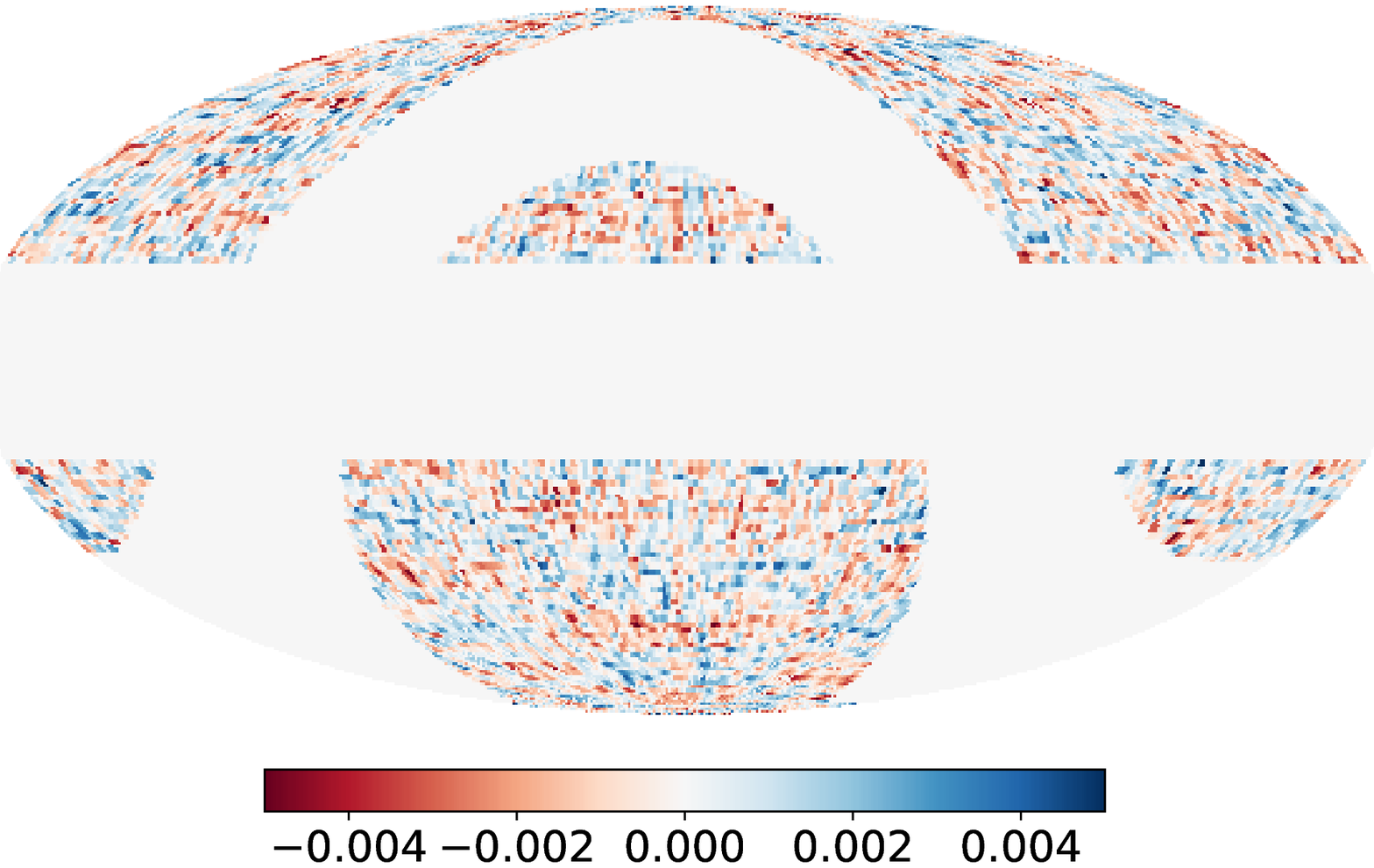}
\includegraphics[width=0.33\columnwidth]{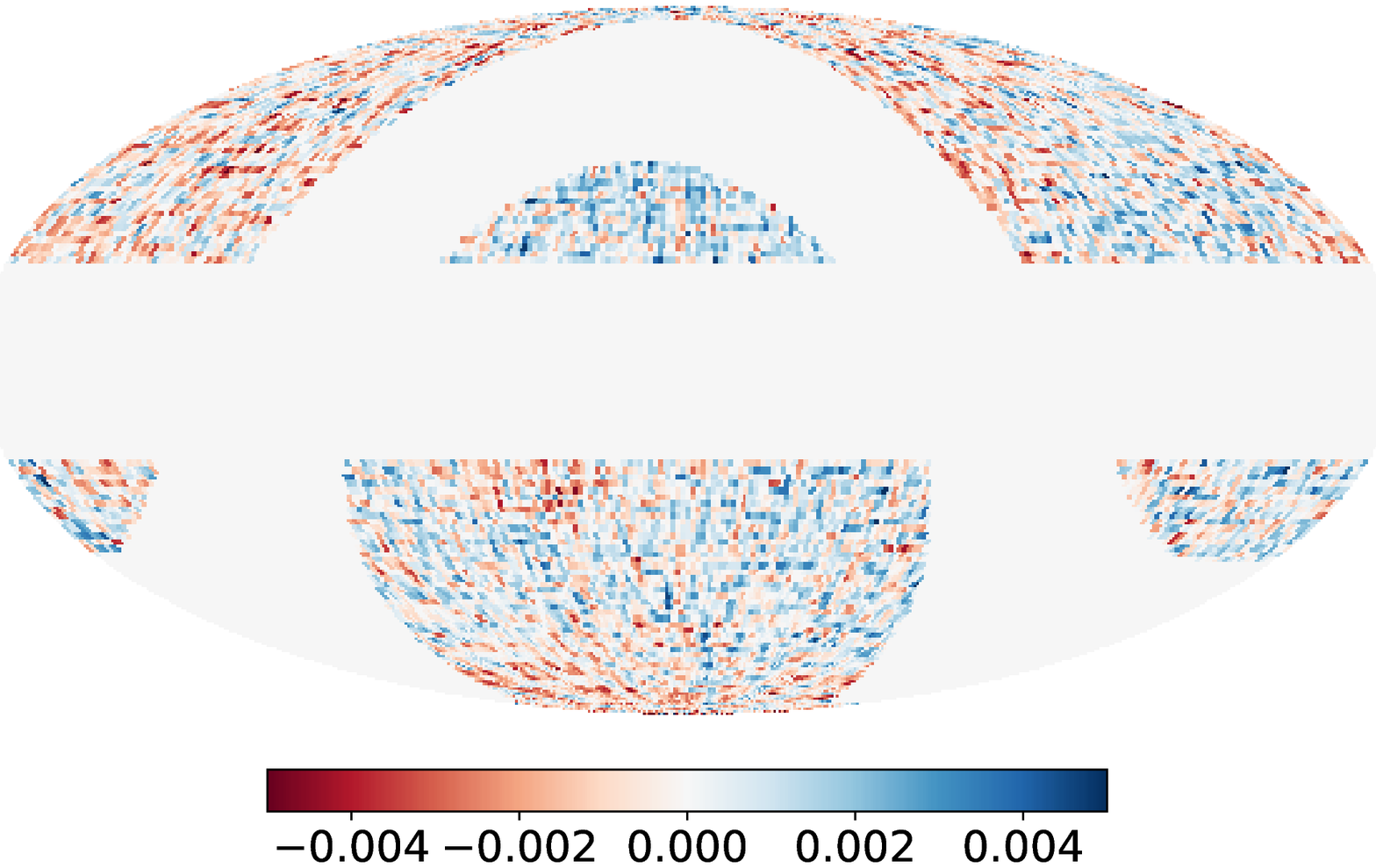}\\
\textsf{\scriptsize Smoothed Shear Map}\\
\includegraphics[width=0.33\columnwidth]{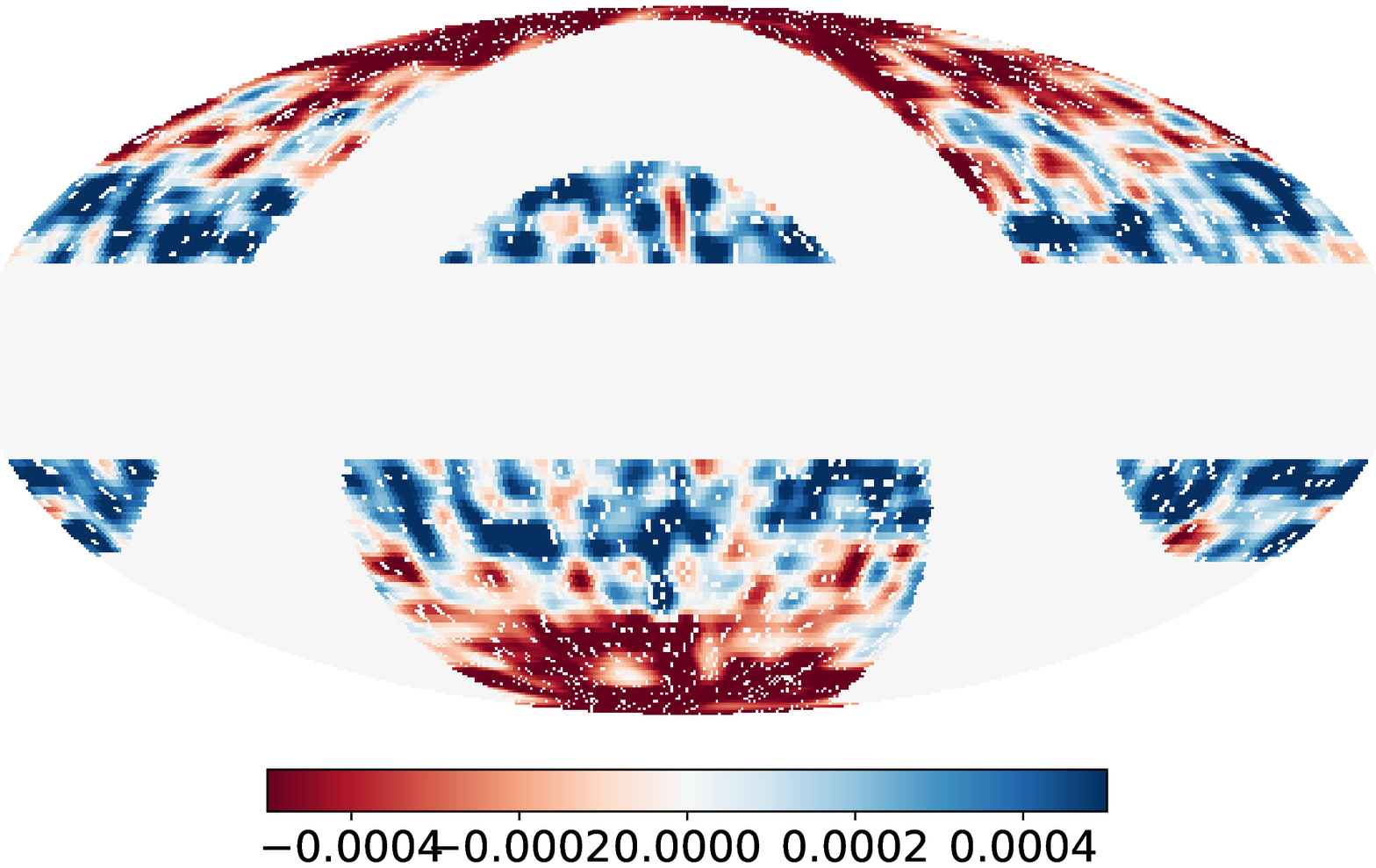}
\includegraphics[width=0.33\columnwidth]{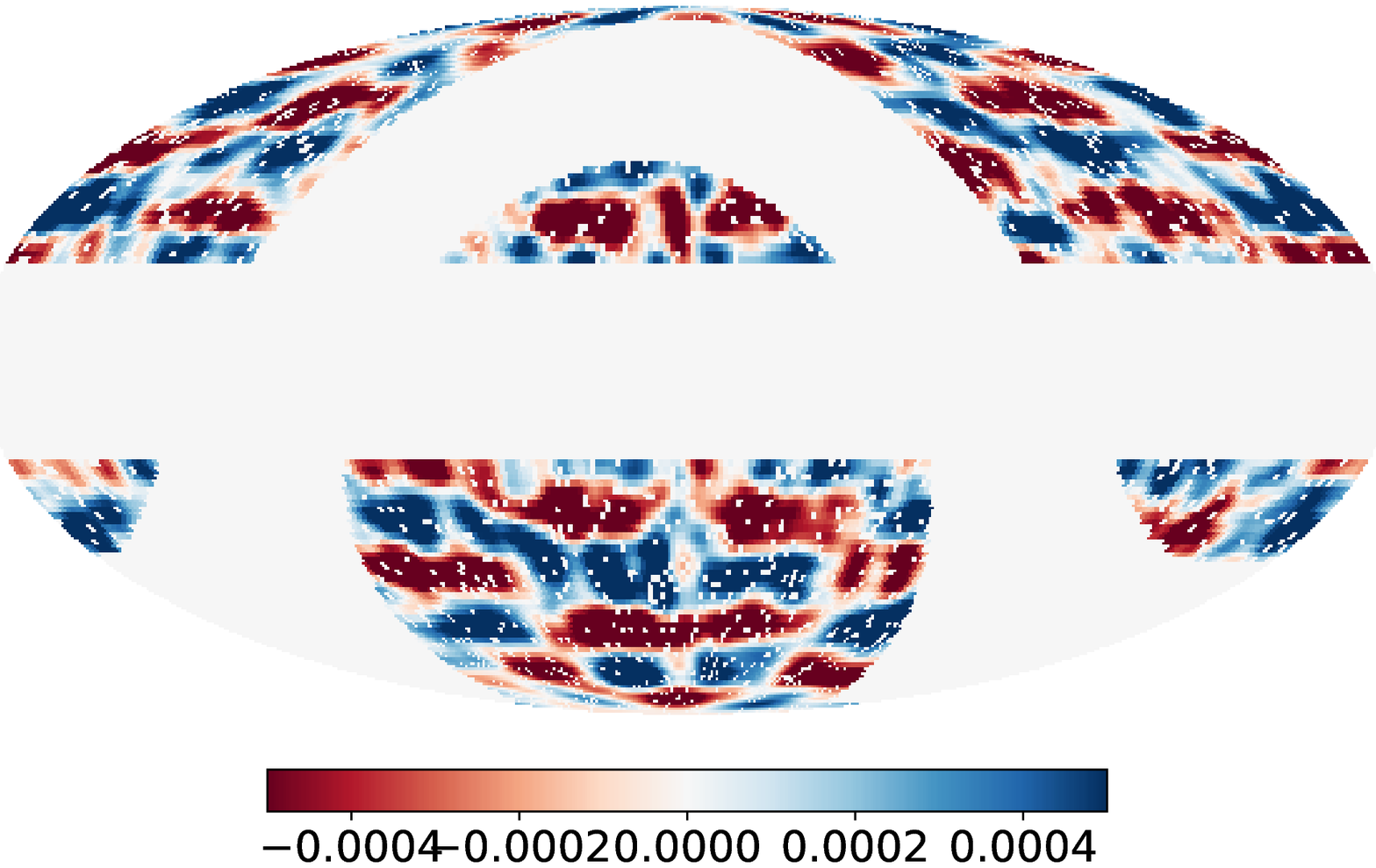}
\includegraphics[width=0.33\columnwidth]{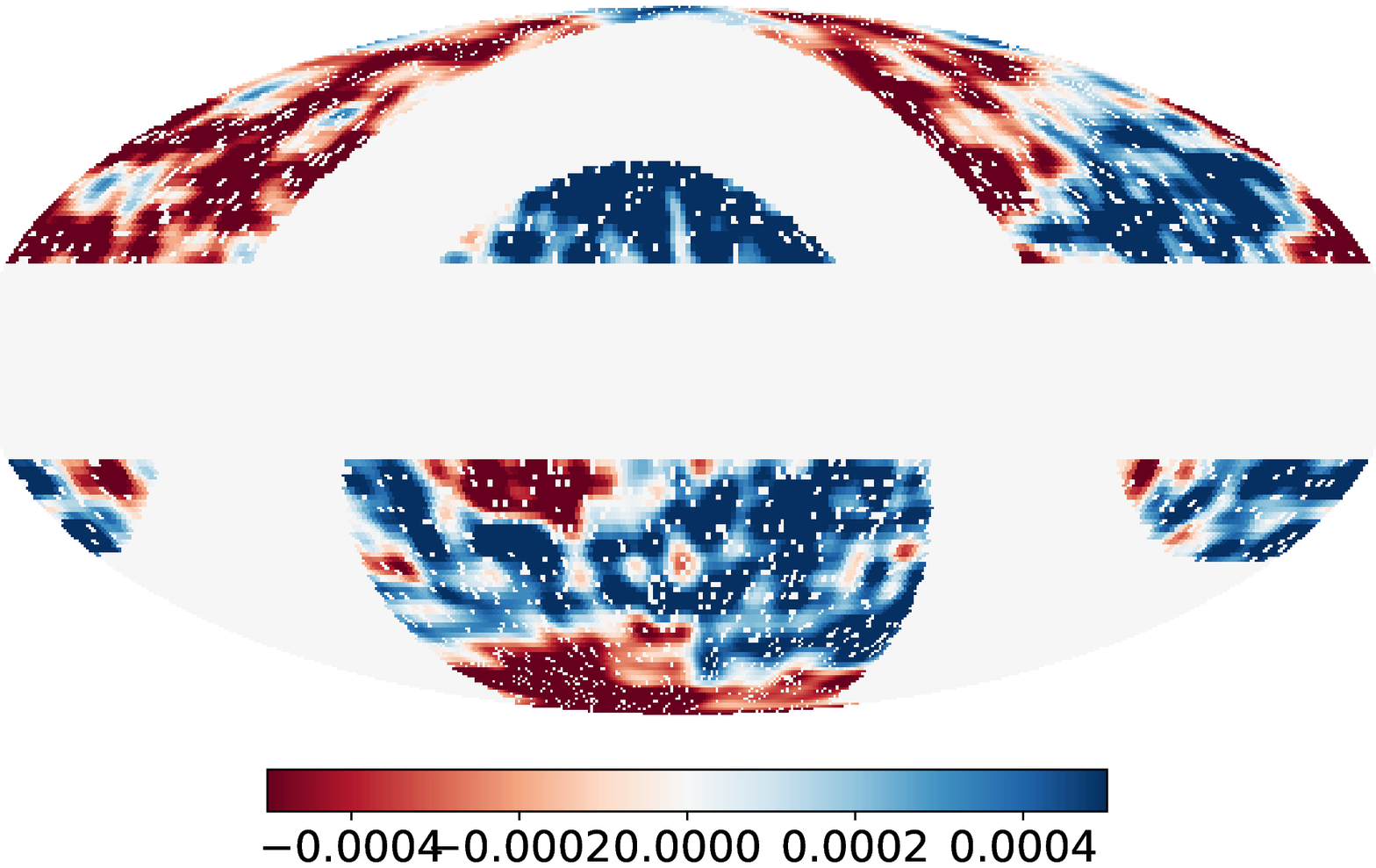}\\
\textsf{\scriptsize Autocorrelation Discrepancy Maps}\\
\includegraphics[width=0.33\columnwidth]{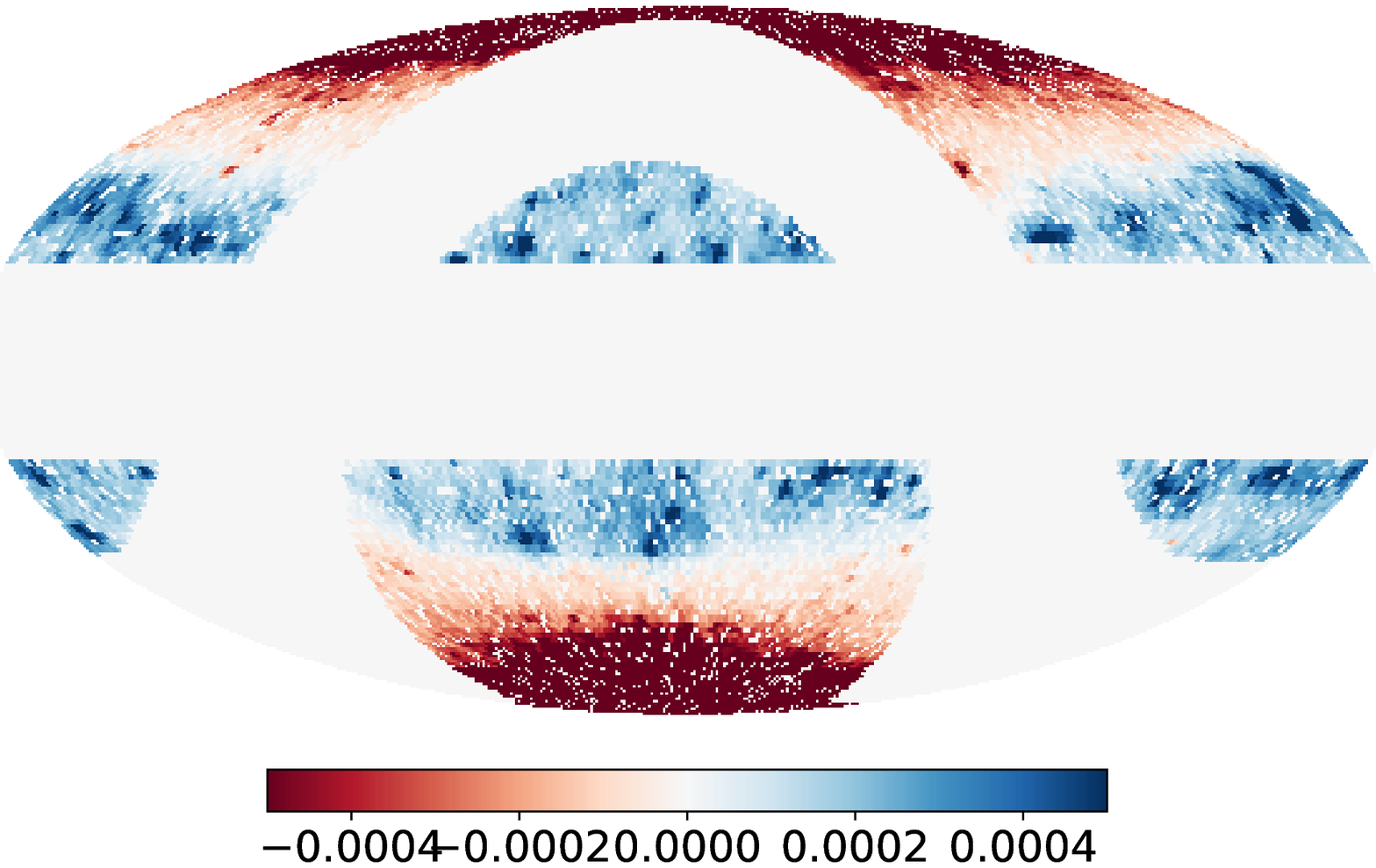}
\includegraphics[width=0.33\columnwidth]{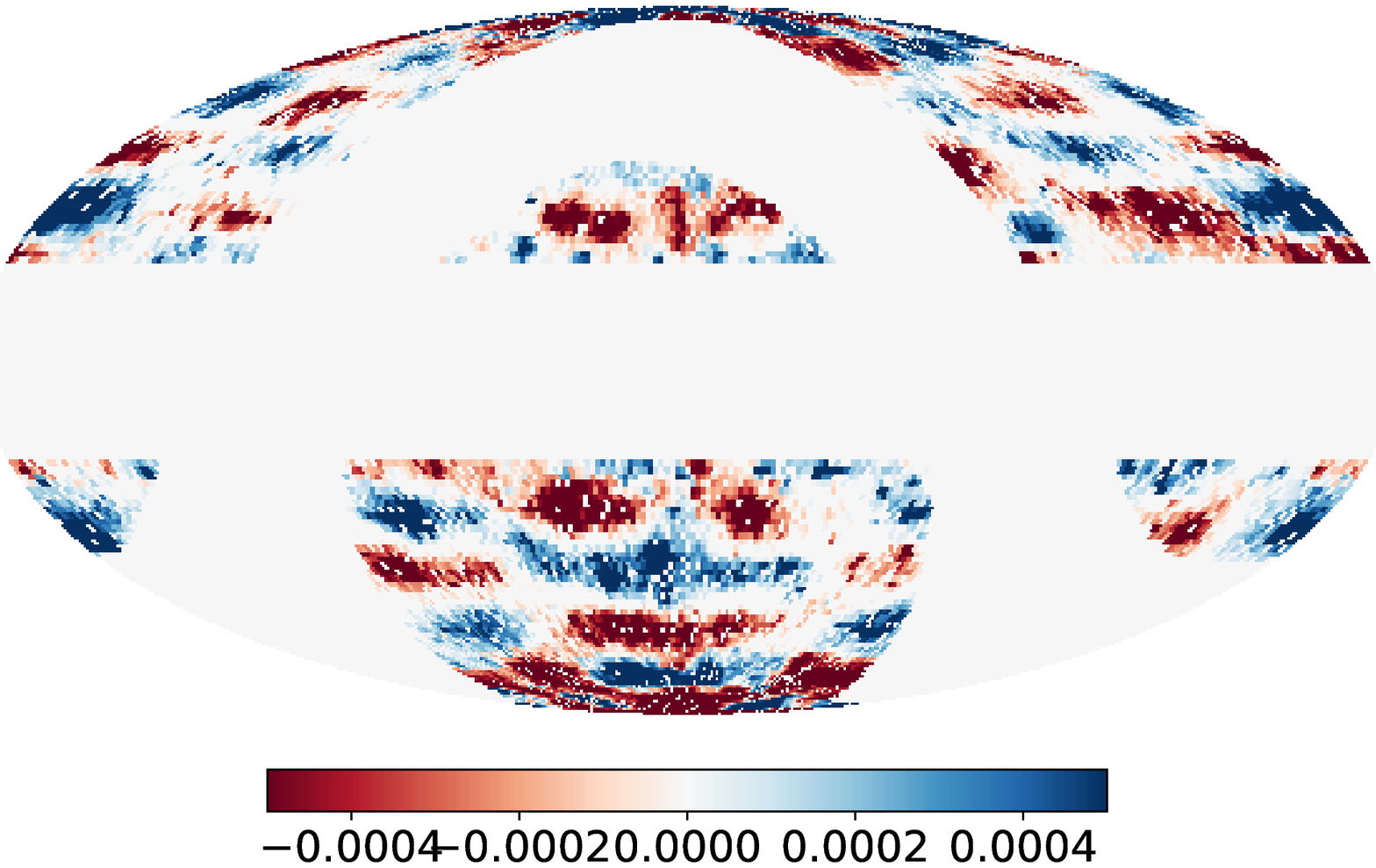}
\includegraphics[width=0.33\columnwidth]{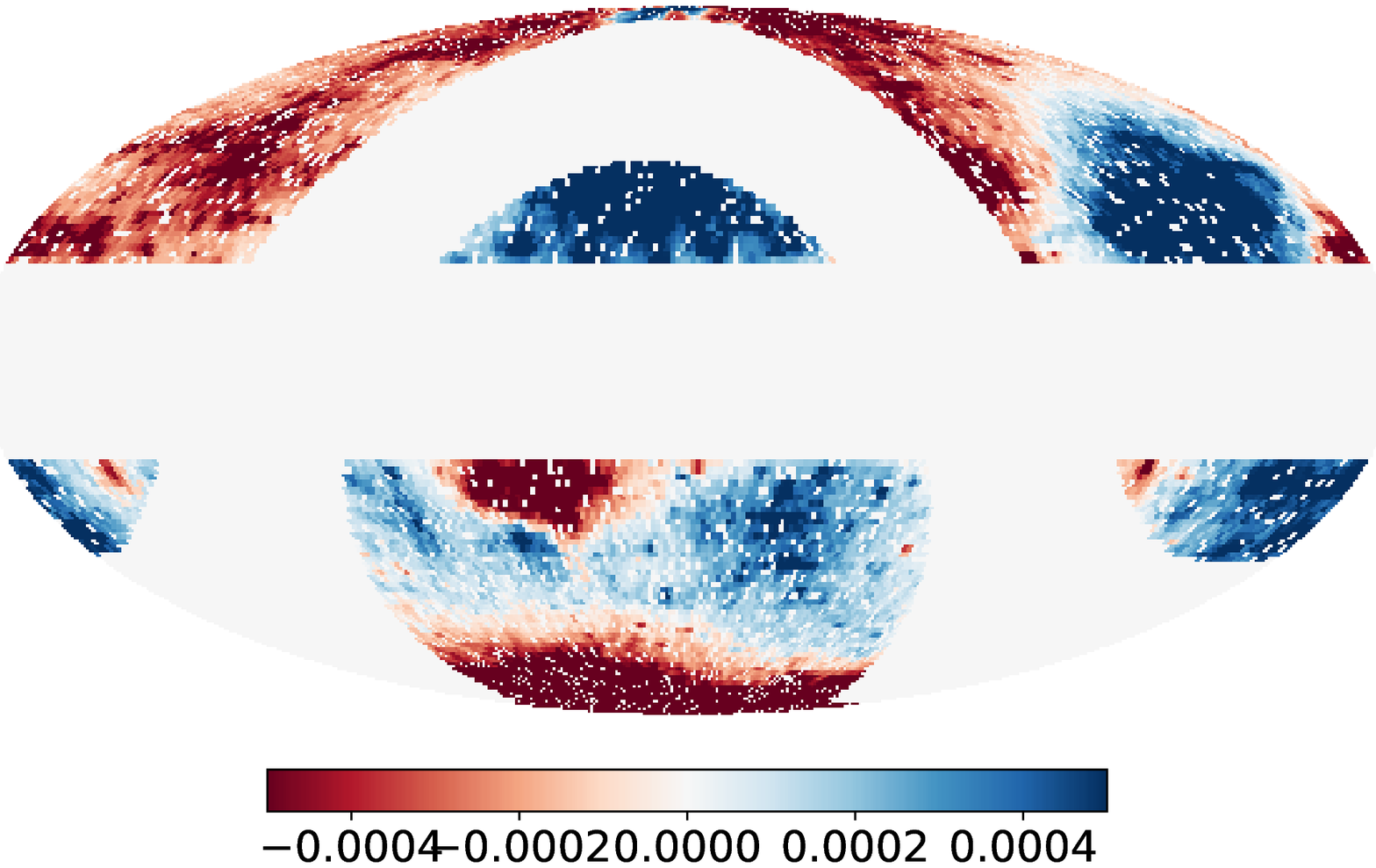}\\
\includegraphics[width=0.33\columnwidth]{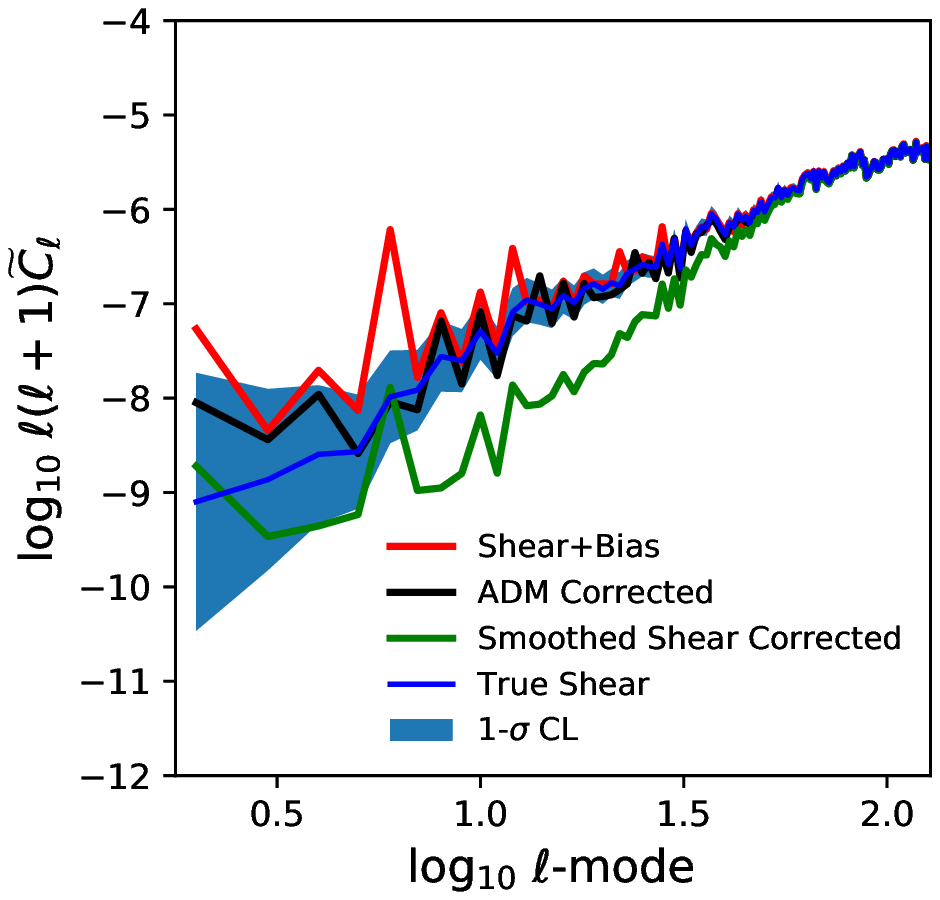}
\includegraphics[width=0.33\columnwidth]{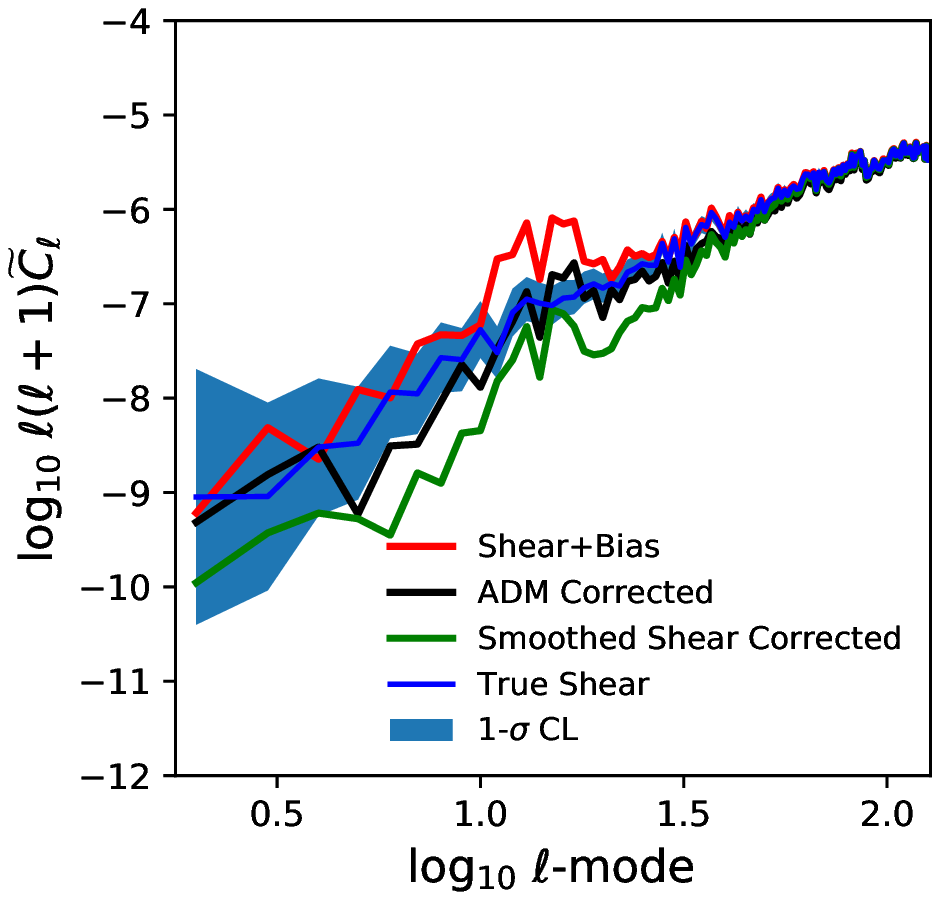}
\includegraphics[width=0.33\columnwidth]{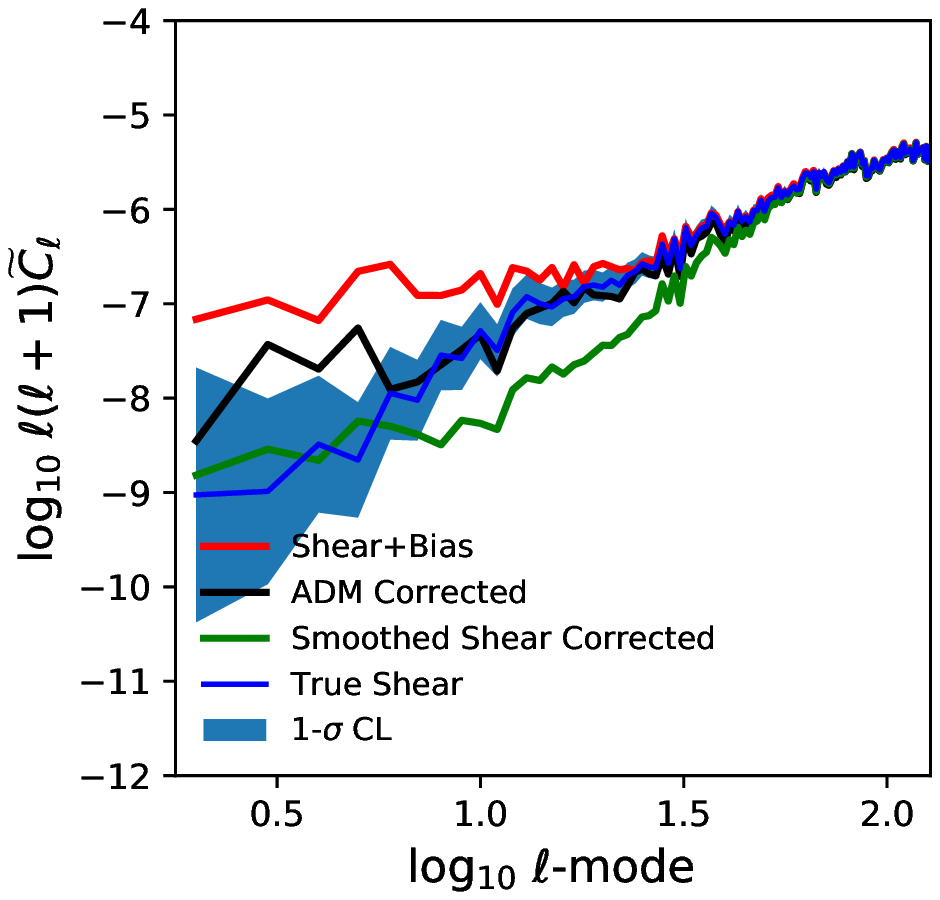}
\caption{Top row: a realisation of an observed  shear field, that  contains the cosmic shear, shot noise and an additive bias field. Second row: the smoothed observed shear field, smoothed with a Gaussian filter of width $0.03$ radians. Third row: the extracted autocorrelation discrepancy map with $N_{\sigma}=3$. Columns left to right are for i) a simple galactic plane; ii) a simple patch pattern; and iii) a smoothed Earth topography map; which correspond to Figure \ref{cases}. Bottom row: the observed shear power spectra; red includes the additive bias; black is corrected using the autocorrelation discrepancy map; green is corrected using the smoothed shear field; blue shows the true shear (without any bias) and the blue band is the $1\sigma$ cosmic variance confidence limit.}
\label{tests}
\end{figure*}

\subsection{Tests}
In Figure \ref{tests} we show the result of testing the autocorrelation discrepancy map to extract the additive bias field from an observed shear field. For each case we show a single realisation of an observed shear field, the smoothed shear field, and the extracted autocorrelation discrepancy map. For these tests we use $N_{\sigma}=3$. We find that in all cases the autocorrelation discrepancy map extracts a field that is indicative of the additive bias field both in amplitude and spatial variation. 

We also show in the change in the observed EE power spectrum caused by the additive bias field, and the corrected power spectrum as a result of subtracting the derived additive shear field from the autocorrelation discrepancy map (equation \ref{ADM}) or the smoothed shear field (equation \ref{gsmooth}). We compare this with the cosmic (sample) variance about the true input shear power spectrum $\pm 2C^{EE}_{\ell}/[f_{\rm sky}(2\ell+1)]$. We find that the autocorrelation discrepancy map results in a corrected power spectrum that is consistent with the input power spectrum. The smoothed shear correction always over-corrects the power spectrum because it removes the shear as well as the additive bias field i.e. it does not remove large-scale shear at scales larger than the smoothing radius of the filter.

\begin{figure*}
\centering
\raisebox{.3\height}{\includegraphics[width=0.33\columnwidth]{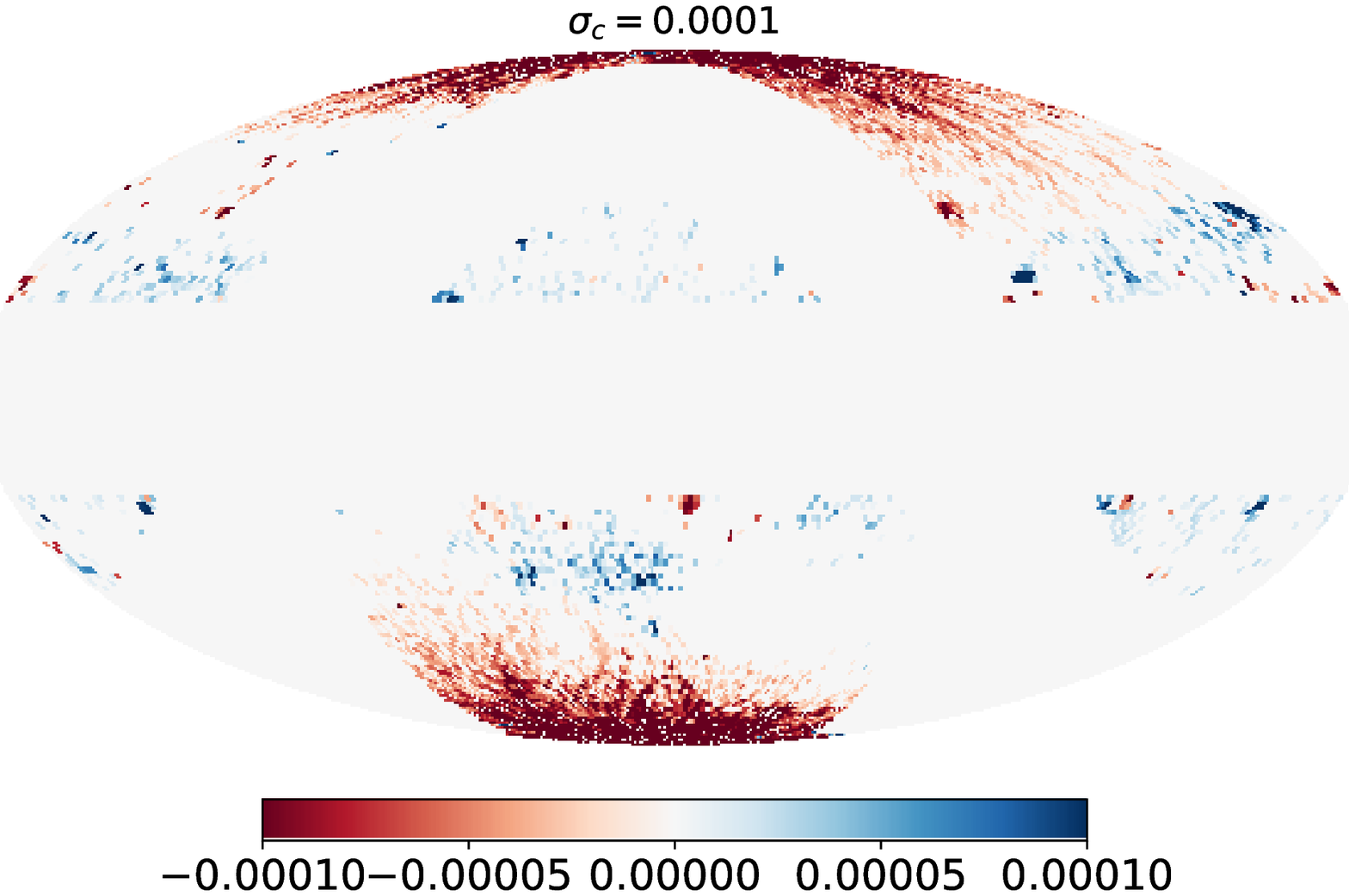}}
\raisebox{.3\height}{\includegraphics[width=0.33\columnwidth]{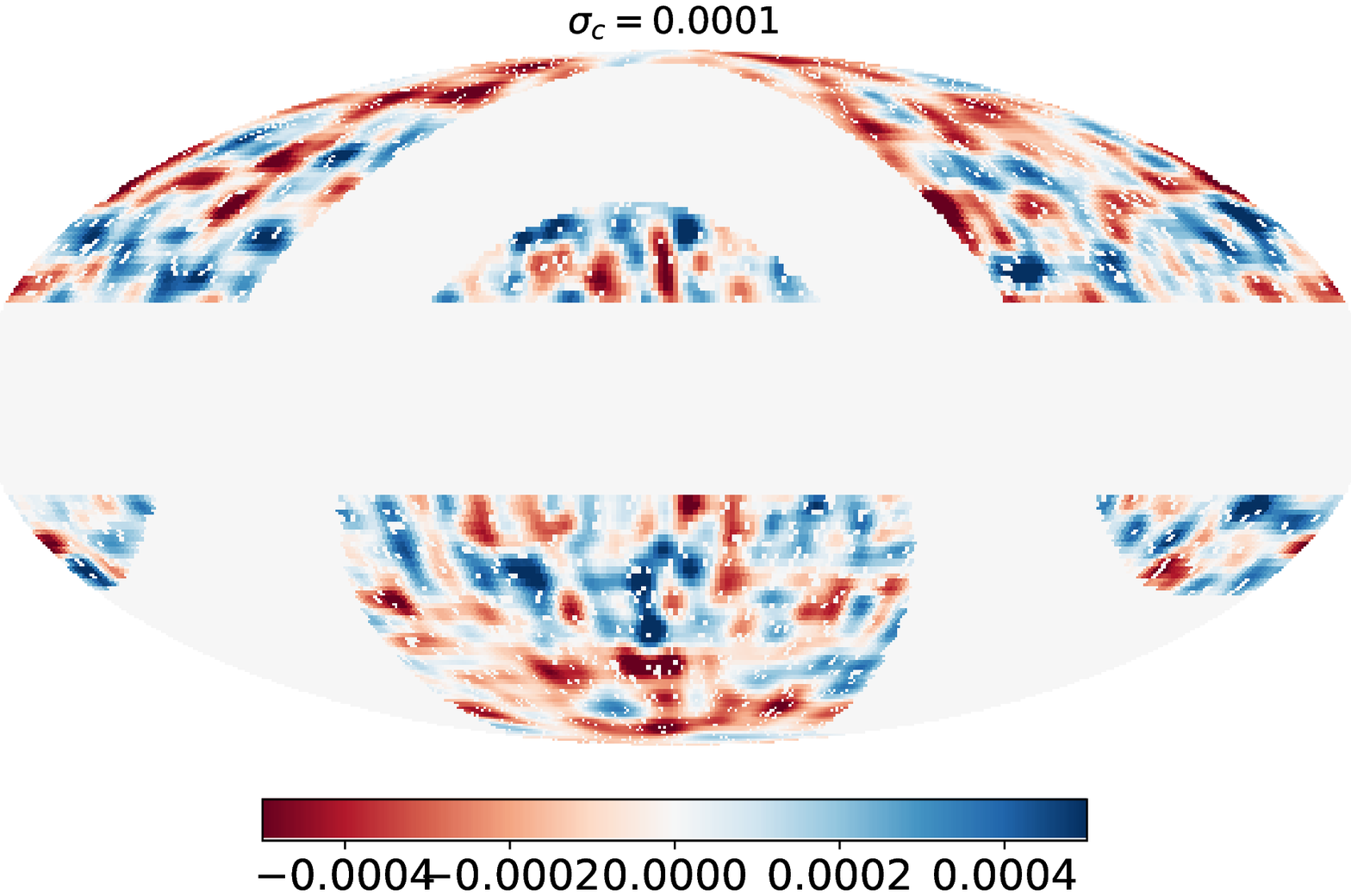}}
\includegraphics[width=0.33\columnwidth]{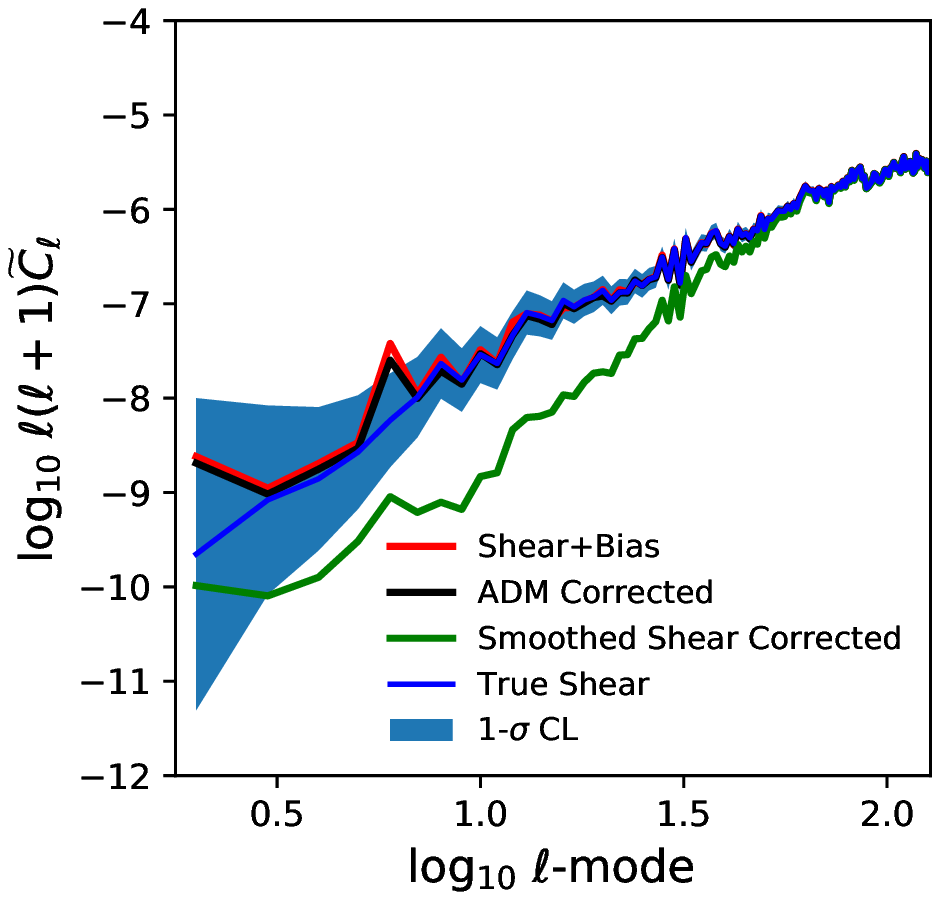}
\caption{For the simple galactic 
plane pattern we show the autocorrelation discrepancy map for varying $\sigma_c=1\times 10^{-4}$, the case for $\sigma_c= 5 \times 10^{−4}$ 
is the fiducial value shown in Figure 3. Note that the colour-scale varies in these plots relative to 
Figures 2, 3 and 8. 
The middle panel shows the equivalent 
smoothed shear map, and 
the right panel shows the power spectrum change.}
\label{sigs}
\end{figure*}

As $\sigma_c$ varies we will expect the performance of the autocorrelation discrepancy map to also change. But we also expect the impact of the additive bias field on the power spectrum to vary as described in Section \ref{S:Req}. In Figure \ref{sigs} we find for $\sigma_c=10^{-4}$ that the method produces an indicative map, and a power spectrum consistent with the input, but that the change in the power spectrum for this level of residual is already small, as expected.

\begin{figure*}
\centering
\includegraphics[width=0.33\columnwidth]{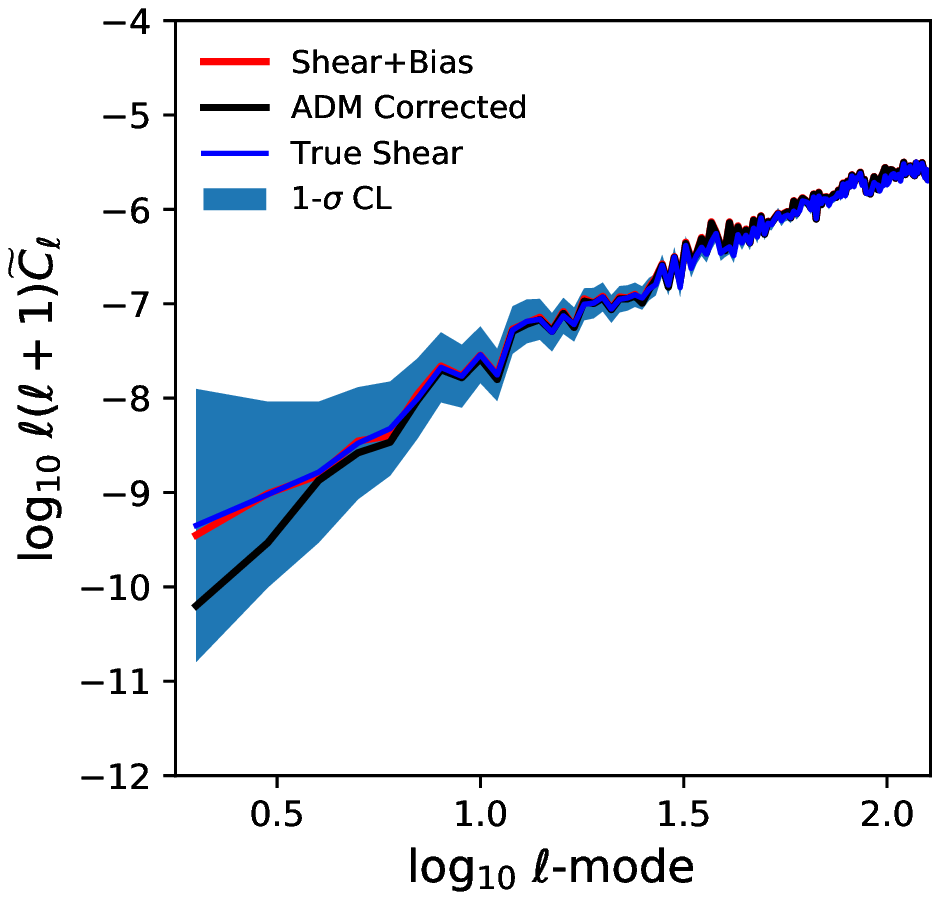}
\includegraphics[width=0.33\columnwidth]{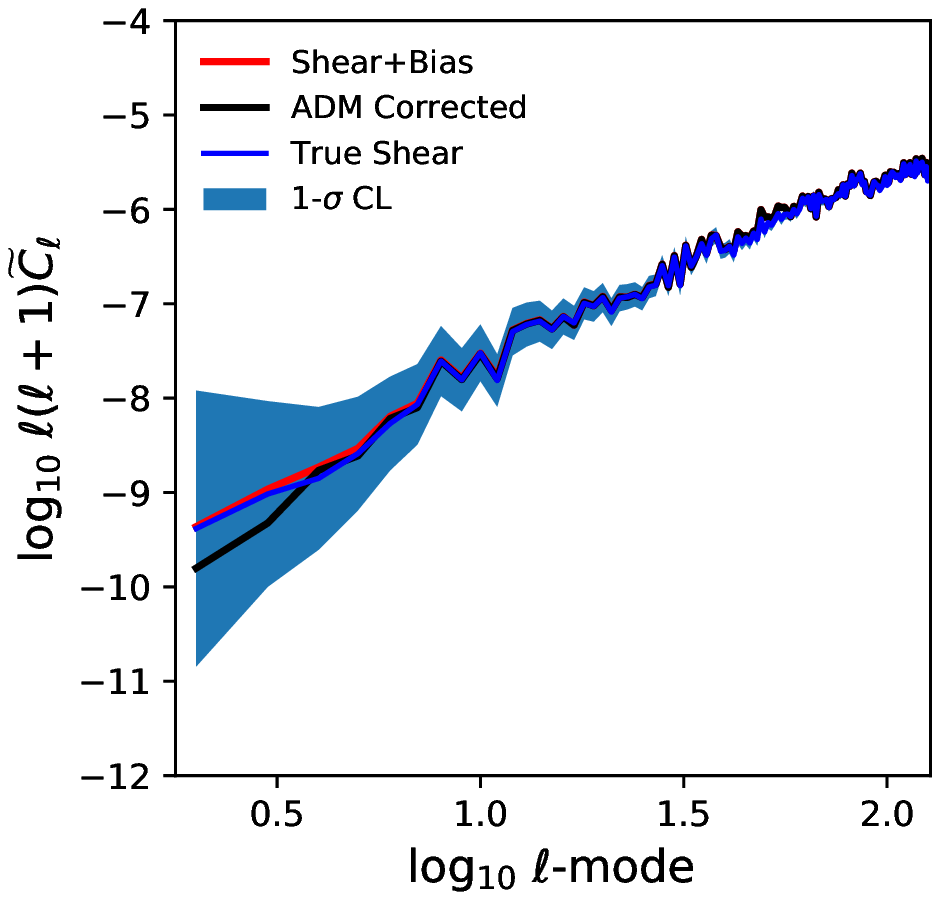}
\includegraphics[width=0.33\columnwidth]{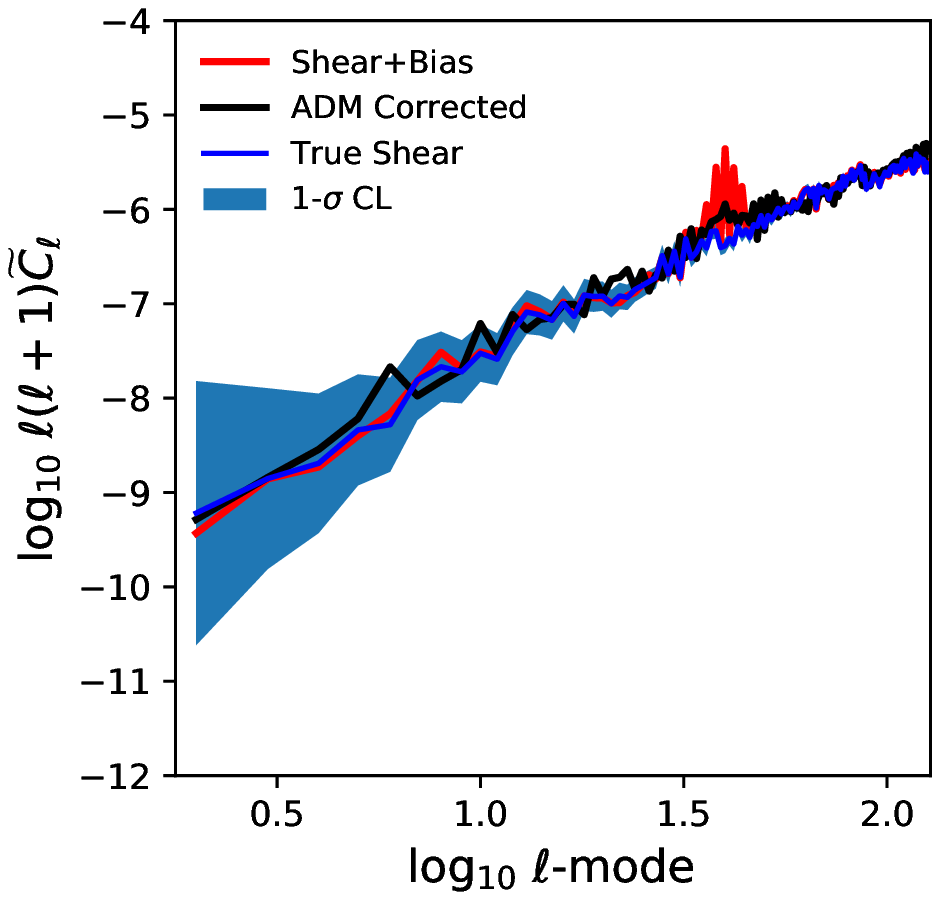}
\caption{A test of the scale-dependency of the reconstructed power spectrum, for the simple pattern case. In the left and middle plots we increase the scale at which the power spectrum peaks, to $\ell\simeq 40$ and $\ell\simeq 60$, by changing the additive bias fild to  
$c(\mathbf{\Omega})=\alpha\sin(10|\phi-\pi|)\sin(10|\theta-\pi|)$ and $c(\mathbf{\Omega})=\alpha\sin(20|\phi-\pi|)\sin(20|\theta-\pi|)$ respectively for a fixed $\sigma_c=5\times 10^{-4}$. In the right plot we use a different pattern that is $c(\mathbf{\Omega})=\sum_m {}_2Y_{\ell_c m}(\mathbf{\Omega})$ with $\ell_c=40$. In all cases the autocorrelation discrepancy map (ADM) recovers the input power spectrum.}
\label{scales}
\end{figure*}

In Figure \ref{scales} we vary the scale of the simple patch pattern such that the peak in the power occurs at smaller scales, for a fixed $\sigma_c$. As expected from Section \ref{S:Req} the impact of the power decreases as the scale is decreased (i.e. as the peak in the power moves to higher $\ell$), and whilst the autocorrelation discrepancy map can reconstruct such features the impact on the power spectrum is negligible. We also use a different pattern that is $c(\mathbf{\Omega})=\sum_m {}_2Y_{\ell_c m}(\mathbf{\Omega})$ with $\ell_c=40$, this produces a more pronounced feature, which we find is reconstructed by the autocorrelation discrepancy map.

In Appendix A we also present some further tests. The ability of the autocorrelation discrepancy map to extract the additive bias field will depend on the choice of the free parameter in the method, $N_{\sigma}$, for a given maximum multipole $L$ and a given $\sigma_c$; in Appendix A we show a test of varying this parameter. In Appendix A we also investigate the noise on the reconstructed autocorrelation discrepancy map. Finally we investigate relaxing the assumption of using Gaussian random fields, by investigating log-normally distributed shear fields.

\subsection{Application to Data}
\label{Application to Data}
We now test the autocorrelation discrepancy map on DES Year 1 data \citep{des1,des2,des3}. All data products used in this Section are available here \url{https://doi.org/10. 5281/zenodo.3980652} \cite[first compiled in][]{2021MNRAS.500.5436P}. We select galaxies with non-zero catalog weight $w_{i,{\rm DES}} > 0$ and correct for the catalogue-provided multiplicative and additive biases using
\begin{eqnarray}
\label{qndees}
\widetilde\gamma_{j,{\rm DES}}=\frac{1}{N_{\rm DES}}\sum_i^{N_{\rm gal}}w_{i,{\rm DES}}[e_{j,i,{\rm DES}}-c_{j,{\rm DES}}]\,\,\,\,\,\,\,{\rm where}\,\,\,\,\,\,\,
N_{\rm DES}=\sum_i^{N_{\rm gal}}w_{i,{\rm DES}}[1+{\rm mcorr}_{j,{\rm DES}}]
\end{eqnarray}
where $e_{1,i,{\rm DES}}$ is the DES Year 1 observed ellipticity for galaxy $i$ for component $j={1,2}$, $c_{j,{\rm DES}}$ is the additive bias provided, ${\rm mcorr}_{j,{\rm DES}}$ is the multiplicative bias correction. We use the same theoretical $C^{EE}_{\ell}$ as in Section \ref{S:Method}, which is based on the DES Year 1 cosmology, and the same $L=128$. We use $N_{\sigma}=3$ and $k_{\rm max}=L$.

\begin{figure*}
\centering
\raisebox{.0\height}{\includegraphics[width=0.33\columnwidth]{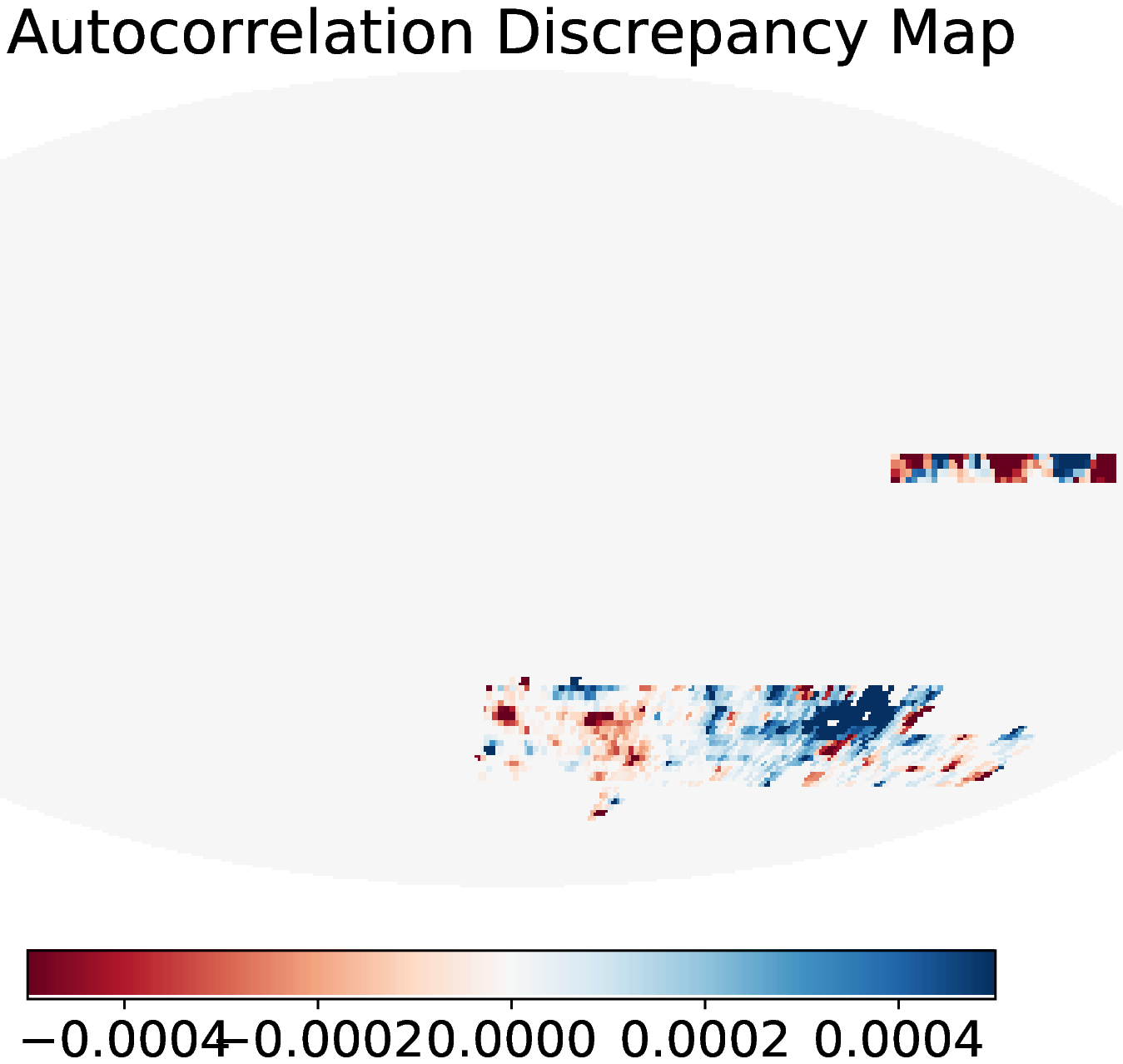}}
\includegraphics[width=0.2\columnwidth]{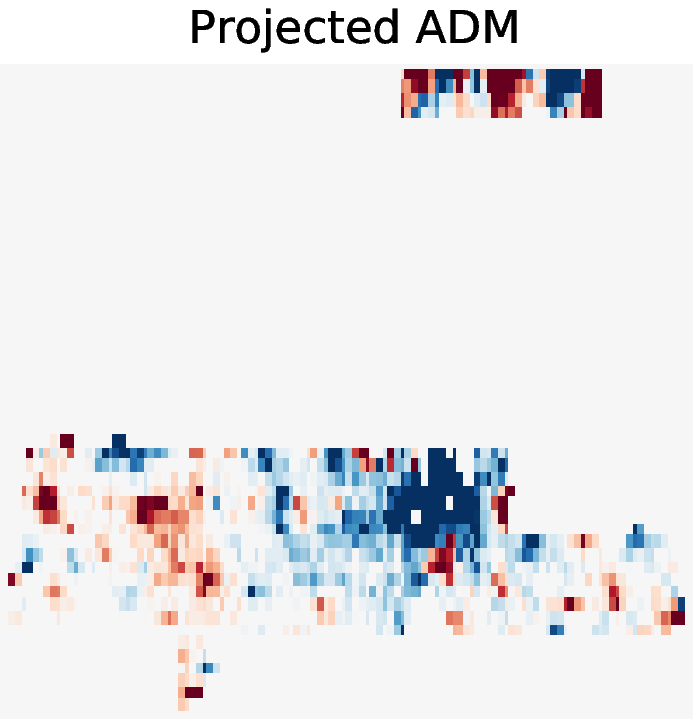}
\includegraphics[width=0.2\columnwidth]{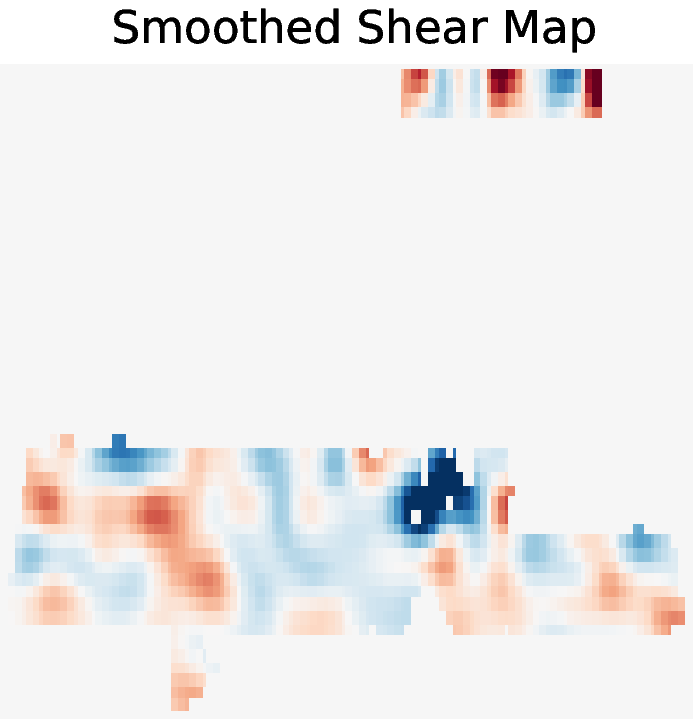}
\includegraphics[width=0.2\columnwidth]{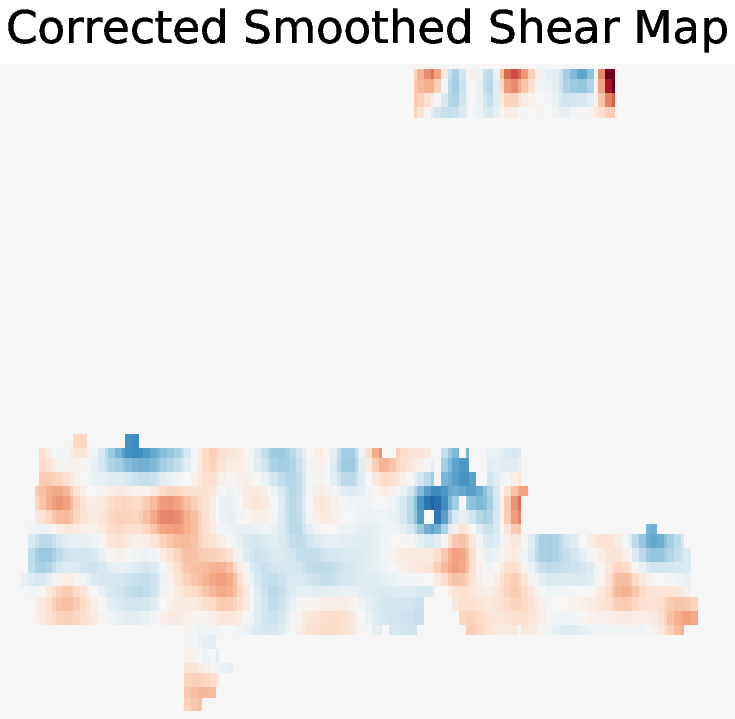}\\
\raisebox{.0\height}{\includegraphics[width=0.33\columnwidth]{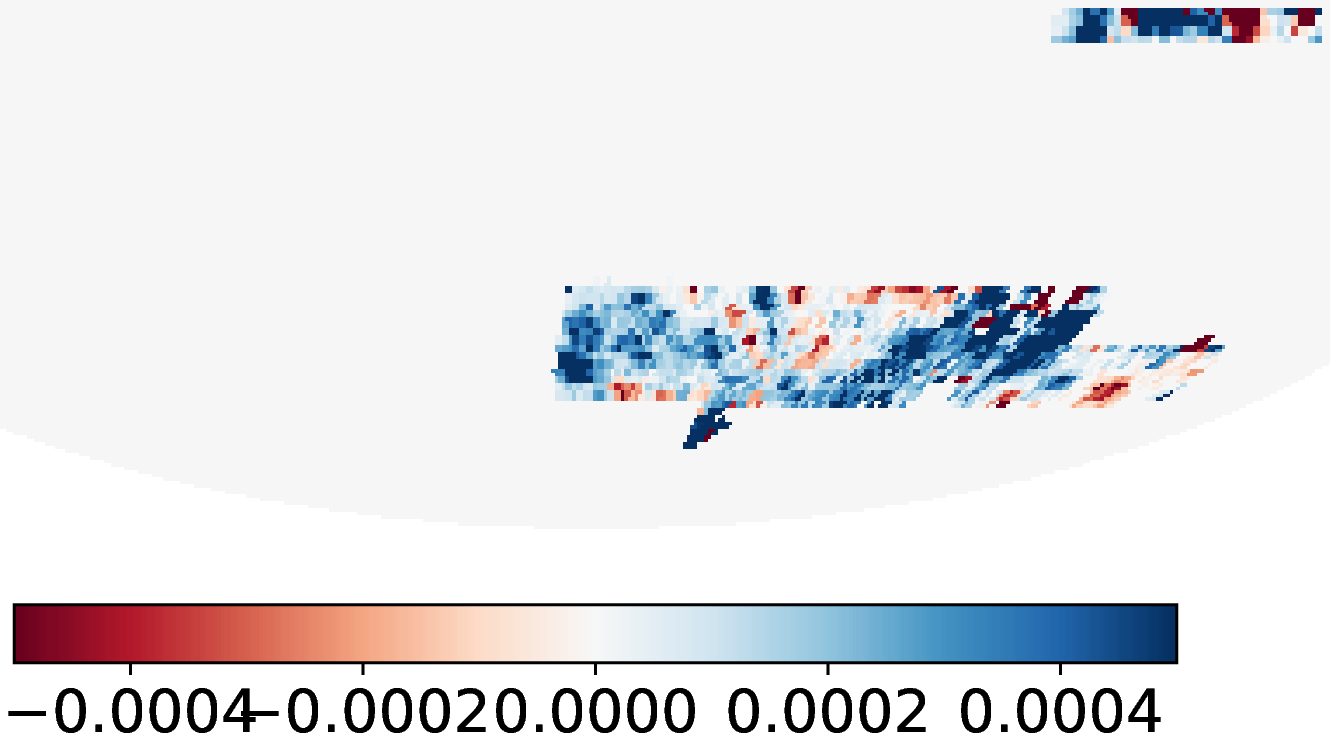}}
\includegraphics[width=0.2\columnwidth]{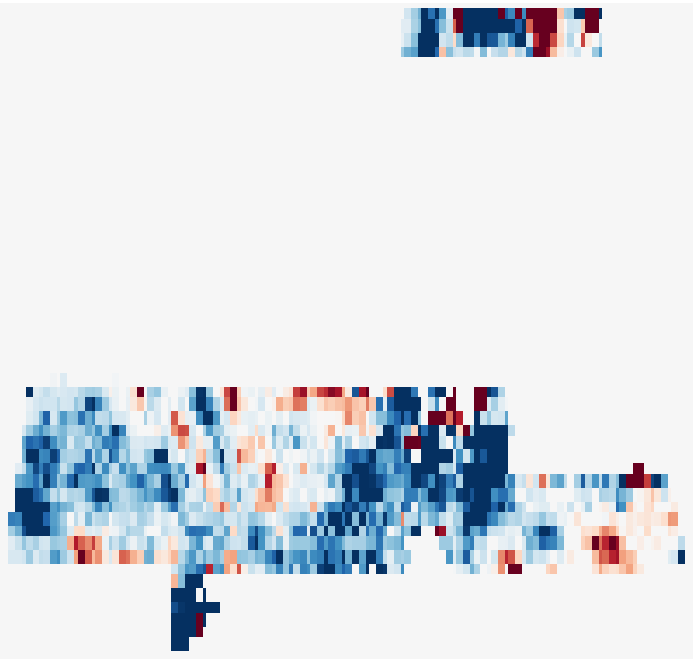}
\includegraphics[width=0.2\columnwidth]{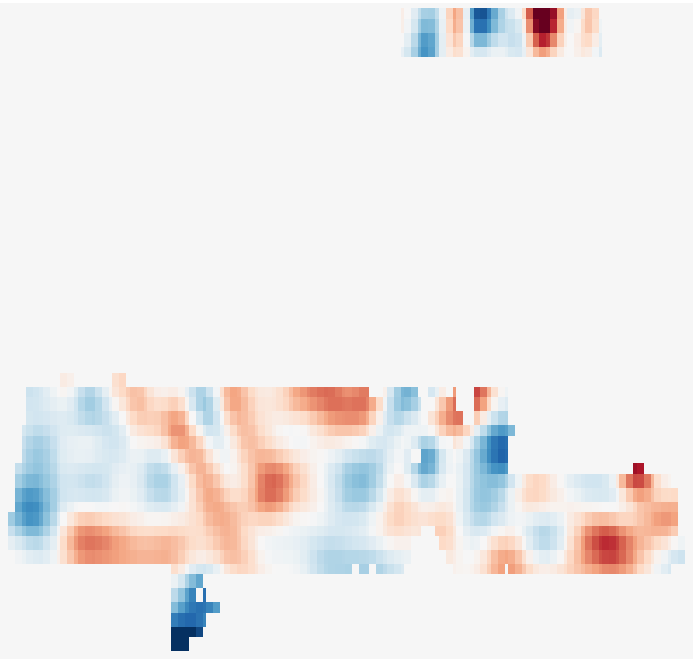}
\includegraphics[width=0.2\columnwidth]{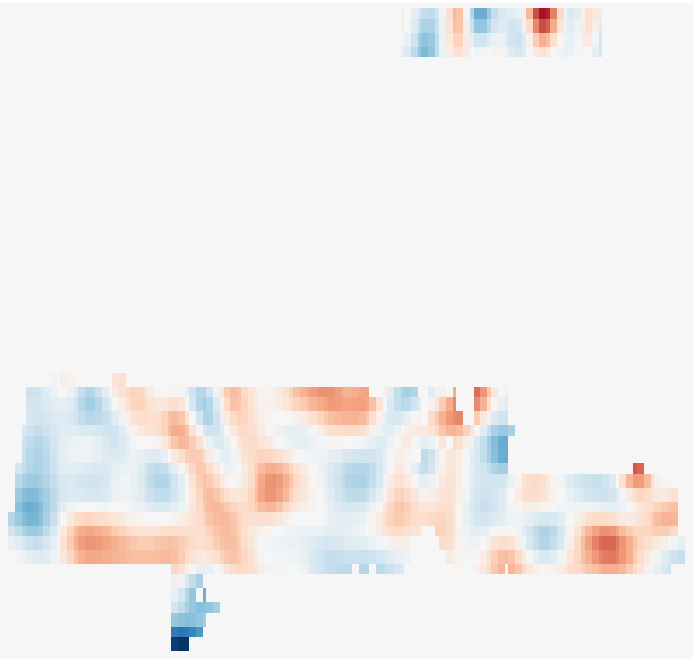}
\caption{An application to DES Year 1 data. The top panels show the autocorrelation discrepancy map for the real part of the observed shear field. The left panel shows the projection on the celestial sphere with ecliptic pole at the north, the second left panel shows a planar projection of the zoomed in region where there is data, using the same colour scale as the left panel. The third left panels show the smoothed shear field, using a Gaussian filter with width $0.03$ radians, using the same colour scale as the right panel. The bottom row shows the same as the top row except for the imaginary part of the observed shear field. The right panels show the smoothed shear field with the estimated additive bias removed.}
\label{DESY1}
\end{figure*}

\begin{figure}
\centering
\includegraphics[width=0.33\columnwidth]{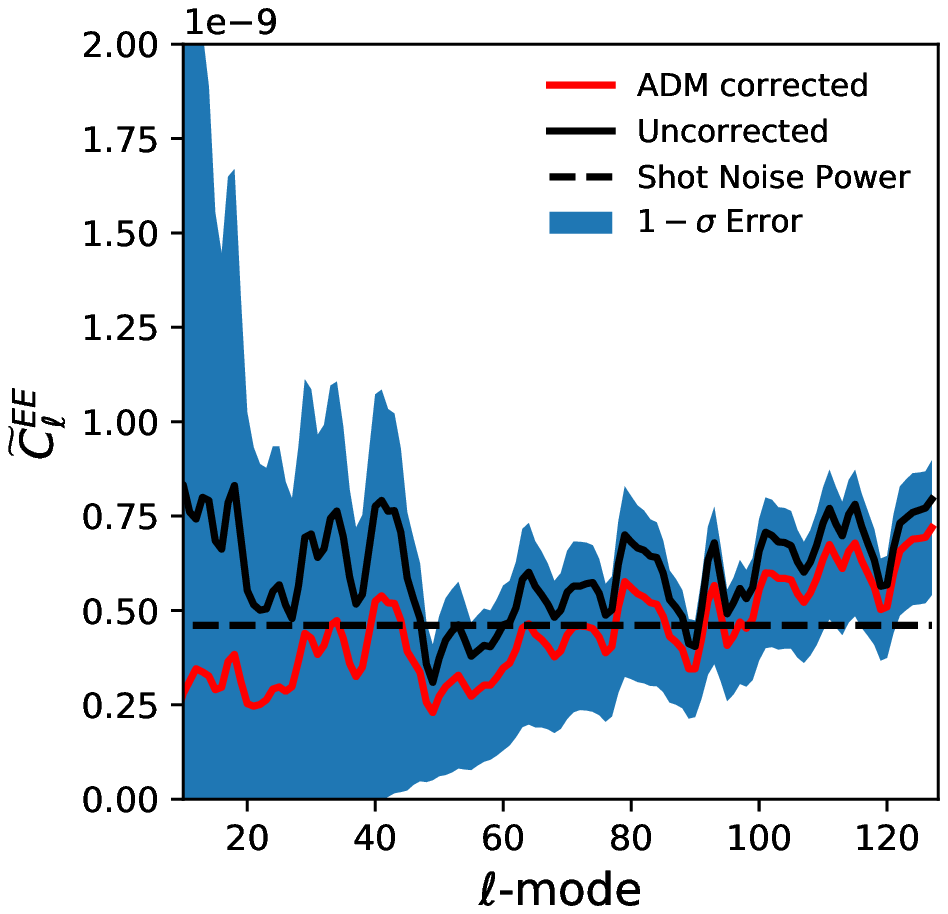}
\includegraphics[width=0.33\columnwidth]{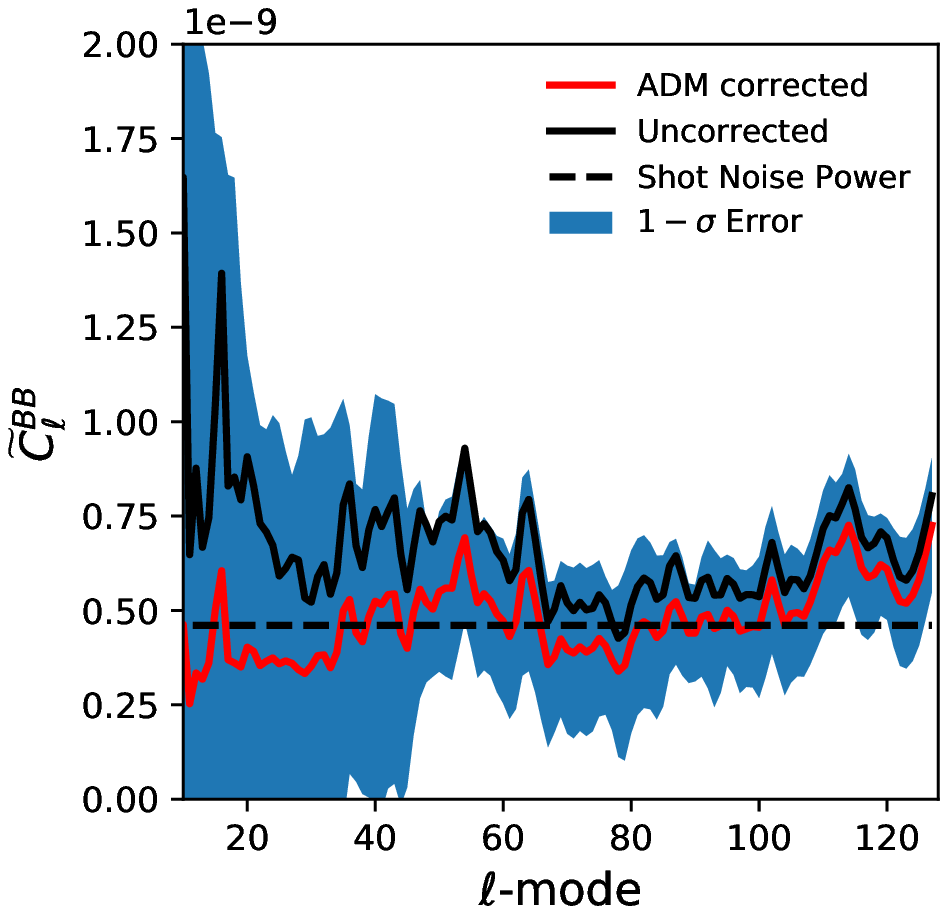}
\includegraphics[width=0.323\columnwidth]{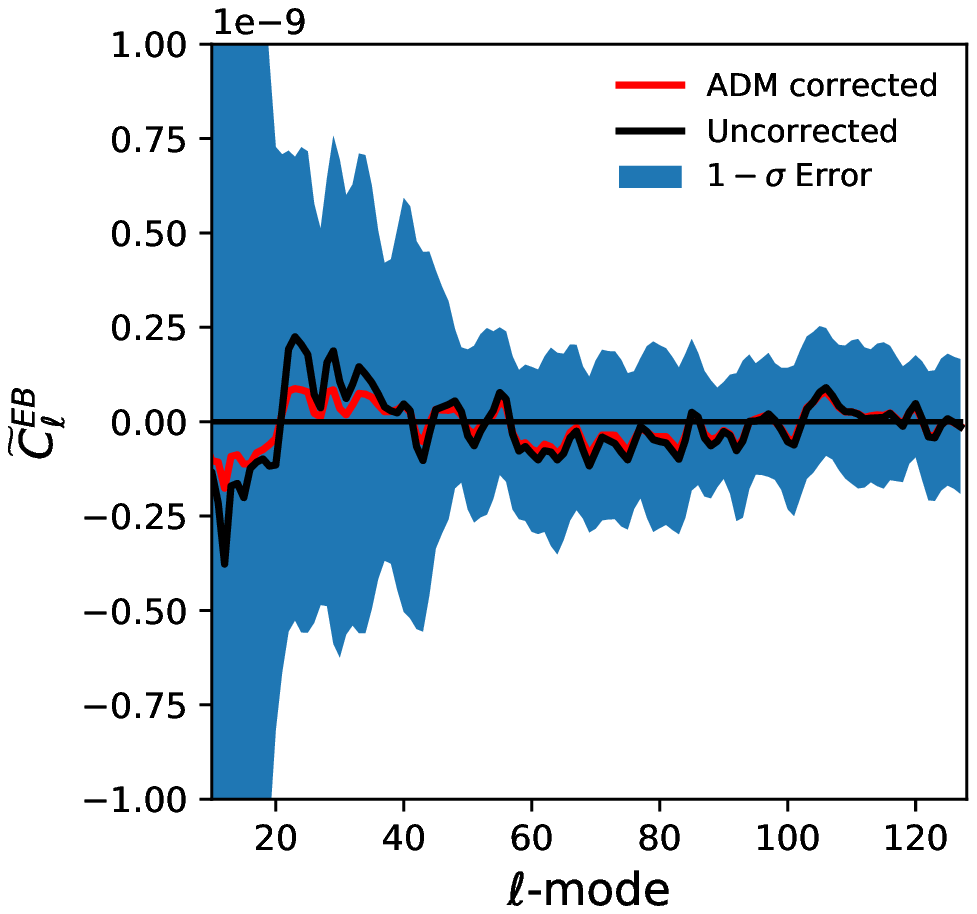}
\caption{An application to DES Year 1 data. Shown are the fractional differences between the observed power spectra and the corrected power spectrum using the additive bias field inferred from the autocorrelation discrepancy map (red) and smoothed shear field (green). Left to right are the EE power spectrum, the BB power spectrum  and the EB power spectrum. Since the EB power spectrum is consistent with zero we only show the difference. The blue filled regions show the cosmic variance and shot noise $1$-$\sigma$ confidence limits.}
\label{DESY1power}
\end{figure}

In Figure \ref{DESY1} we show the autocorrelation discrepancy map for the real and imaginary parts of the shear field, and the smoothed shear field. We find evidence for spatial variation of residual biases of order $\mathbb{R}[|c(\mathbf{\Omega})|]\simeq \mathbb{I}[|c(\mathbf{\Omega})|]\simeq 1\times 10^{-3}$ over a large proportion of the observed field, the maximum values we find are $\mathbb{R}[|c(\mathbf{\Omega})|]\simeq \mathbb{I}[|c(\mathbf{\Omega})|]\simeq 2\times 10^{-3}$. We also find some evidence for coherent spatial structure, in particular in the imaginary ($c_2$) component. In comparison with the smoothed observed shear field we find similar results, albeit at lower resolution, which is consistent with our simulation test results. In Figure \ref{DESY1power} we show the change in the observed power spectra caused by removing the estimated additive bias field, and find a $\sim 5\%$ decrease in the observed EE and BB power spectra over the scales $10<\ell<50$ (where $\ell\sim 10$ is the fundamental mode of the survey patch in this case); the change in the EB power is consistent with zero. However, we note that at these scales the observed power spectrum is essentially consistent with noise. We note that the spatial gradient of the autocorrelation discrepancy maps is in the same direction as variations in exposure time, airmass, and sky brightness, as well as $r$-band depth \citep{Drlica_Wagner_2018}, which warrants further investigation that we leave for future work.

\section{Conclusions}
\label{Conclusions}
We present a new validation test of cosmic shear data which is the autocorrelation discrepancy map, and show that, to first order, and if multiplicative biases are removed, that the autocorrelation discrepancy map corresponds to the anisotropic part of the absolute value of the residual additive bias field. We test this using simulations and find that the autocorrelation discrepancy map is indeed indicative of the additive bias field, and that such a map is complimentary to the approach of smoothing the observed shear field. We apply the method to DES Year 1 and find evidence for residual additive biases over the survey area of at  most $\pm 2\times 10^{-3}$. The method we present should be able to empirically inform the modelling of spatially additive biases which until now has relied on \emph{a priori} modelling or proxy observables with which additive biases may be correlated. We also expect the method to perform well on galaxy clustering statistics and on CMB temperature anisotropy studies because in both cases systematic effects are in general proportional to contaminating fields, rather than  the gradient of such fields as is the case in cosmic shear.
It remains to be demonstrated whether this method can achieve the required accuracy for Stage-IV dark energy surveys, and a combination of template modelling \citep[e.g.][]{2016ApJS..226...24L} (which if the templates match the true underlying additive biases should perform well) followed by removal of any residual biases via such a method is something to explore. We present the basic method of the approach, but further work could include such statistics in more sophisticated analyses such as Bayesian Hierarchical Modelling, and a generalisation to $3\times 2$-point statistics. 
\vspace{-0.3cm}
\acknowledgements
{\scriptsize \emph{Acknowledgements:} TDK acknowledges funding from the European Union’s Horizon 2020 research and innovation programme under grant agreement No 776247. ACD acknowledges funding from the Royal Society. PLT acknowledges support for this work from a NASA Postdoctoral Program Fellowship. Part of the research was carried out at the Jet Propulsion Laboratory, California Institute of Technology, under a contract with the National Aeronautics and Space Administration. We thank the developers of {\tt SSHT}, {\tt massmappy}, and {\tt CAMB} for making their code publicly available. We thank Henk Hoekstra for helpful comments.} {\scriptsize \emph{DES Acknowledgements:} This project used public archival data from the Dark Energy Survey (DES). Funding for the DES Projects has been provided by the U.S. Department of Energy, the U.S. National Science Foundation, the Ministry of Science and Education of Spain, the Science and Technology FacilitiesCouncil of the United Kingdom, the Higher Education Funding Council for England, the National Center for Supercomputing Applications at the University of Illinois at Urbana-Champaign, the Kavli Institute of Cosmological Physics at the University of Chicago, the Center for Cosmology and Astro-Particle Physics at the Ohio State University, the Mitchell Institute for Fundamental Physics and Astronomy at Texas A\&M University, Financiadora de Estudos e Projetos, Funda{\c c}{\~a}o Carlos Chagas Filho de Amparo {\`a} Pesquisa do Estado do Rio de Janeiro, Conselho Nacional de Desenvolvimento Cient{\'i}fico e Tecnol{\'o}gico and the Minist{\'e}rio da Ci{\^e}ncia, Tecnologia e Inova{\c c}{\~a}o, the Deutsche Forschungsgemeinschaft, and the Collaborating Institutions in the Dark Energy Survey.
The Collaborating Institutions are Argonne National Laboratory, the University of California at Santa Cruz, the University of Cambridge, Centro de Investigaciones Energ{\'e}ticas, Medioambientales y Tecnol{\'o}gicas-Madrid, the University of Chicago, University College London, the DES-Brazil Consortium, the University of Edinburgh, the Eidgen{\"o}ssische Technische Hochschule (ETH) Z{\"u}rich,  Fermi National Accelerator Laboratory, the University of Illinois at Urbana-Champaign, the Institut de Ci{\`e}ncies de l'Espai (IEEC/CSIC), the Institut de F{\'i}sica d'Altes Energies, Lawrence Berkeley National Laboratory, the Ludwig-Maximilians Universit{\"a}t M{\"u}nchen and the associated Excellence Cluster Universe, the University of Michigan, the National Optical Astronomy Observatory, the University of Nottingham, The Ohio State University, the OzDES Membership Consortium, the University of Pennsylvania, the University of Portsmouth, SLAC National Accelerator Laboratory, Stanford University, the University of Sussex, and Texas A\&M University.
Based in part on observations at Cerro Tololo Inter-American Observatory, National Optical Astronomy Observatory, which is operated by the Association of Universities for Research in Astronomy (AURA) under a cooperative agreement with the National Science Foundation.}

\bibliographystyle{mnras}
\bibliography{sample.bib}

\section*{Appendix A: Further Tests}
\begin{figure*}
\centering
\includegraphics[width=0.33\columnwidth]{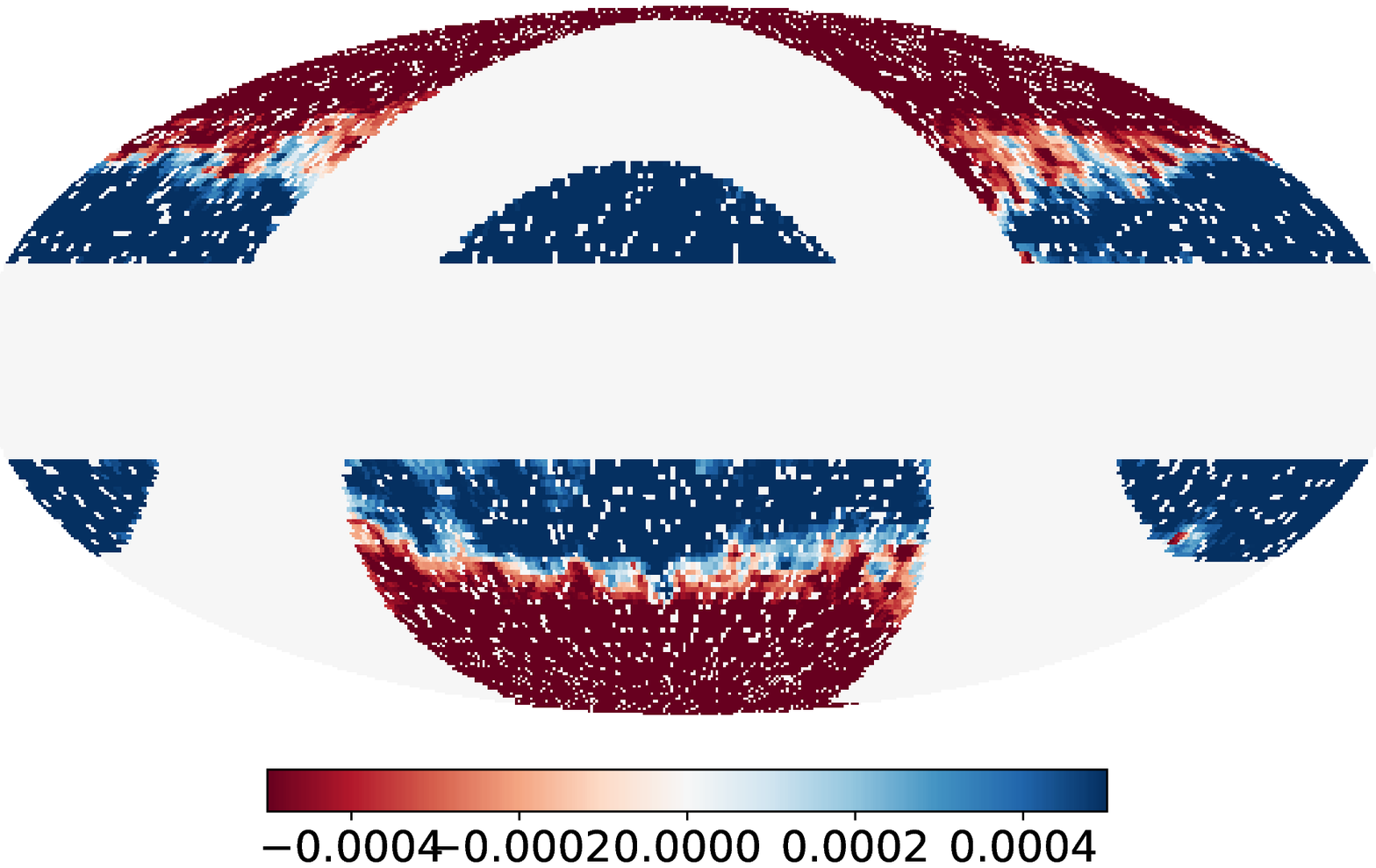}
\includegraphics[width=0.33\columnwidth]{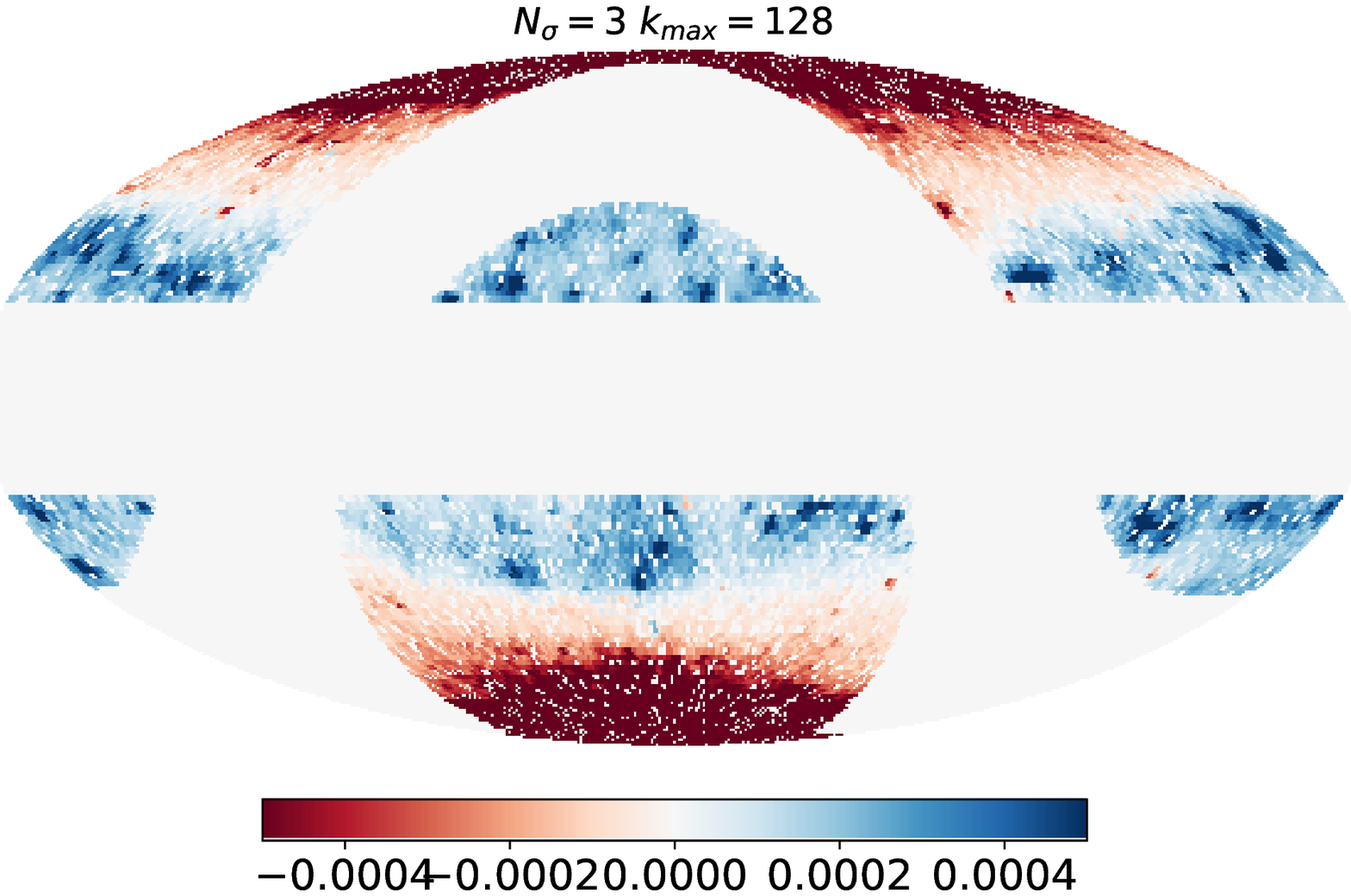}
\includegraphics[width=0.33\columnwidth]{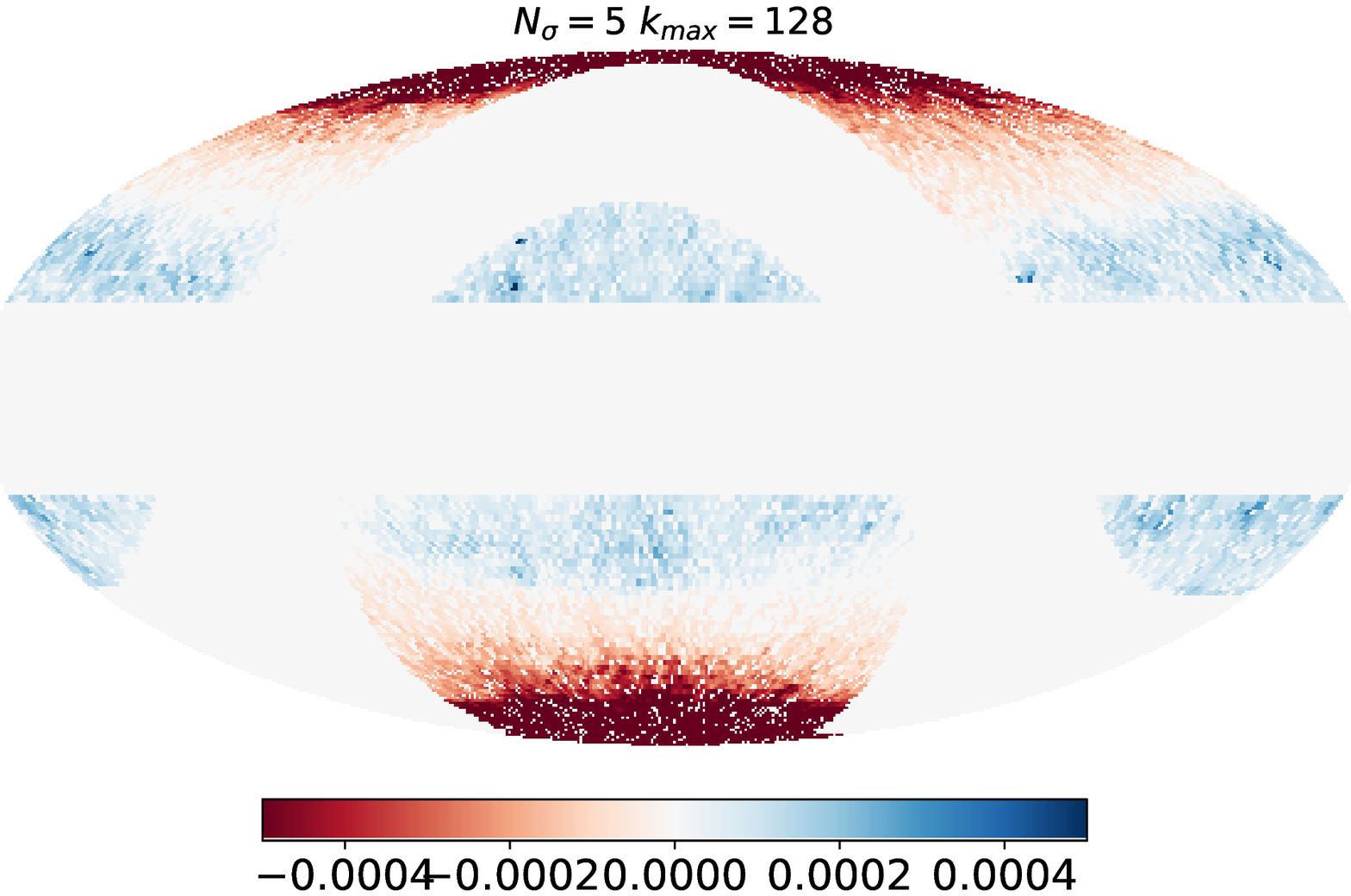}
\caption{For the simple galactic plane pattern we show the autocorrelation discrepancy map for varying $N_{\sigma}=1,3,5$ (left to right), this can be compared to Figure 2 (top left panel).}
\label{Nk}
\end{figure*}

\begin{figure*}
\centering
\includegraphics[width=0.33\columnwidth]{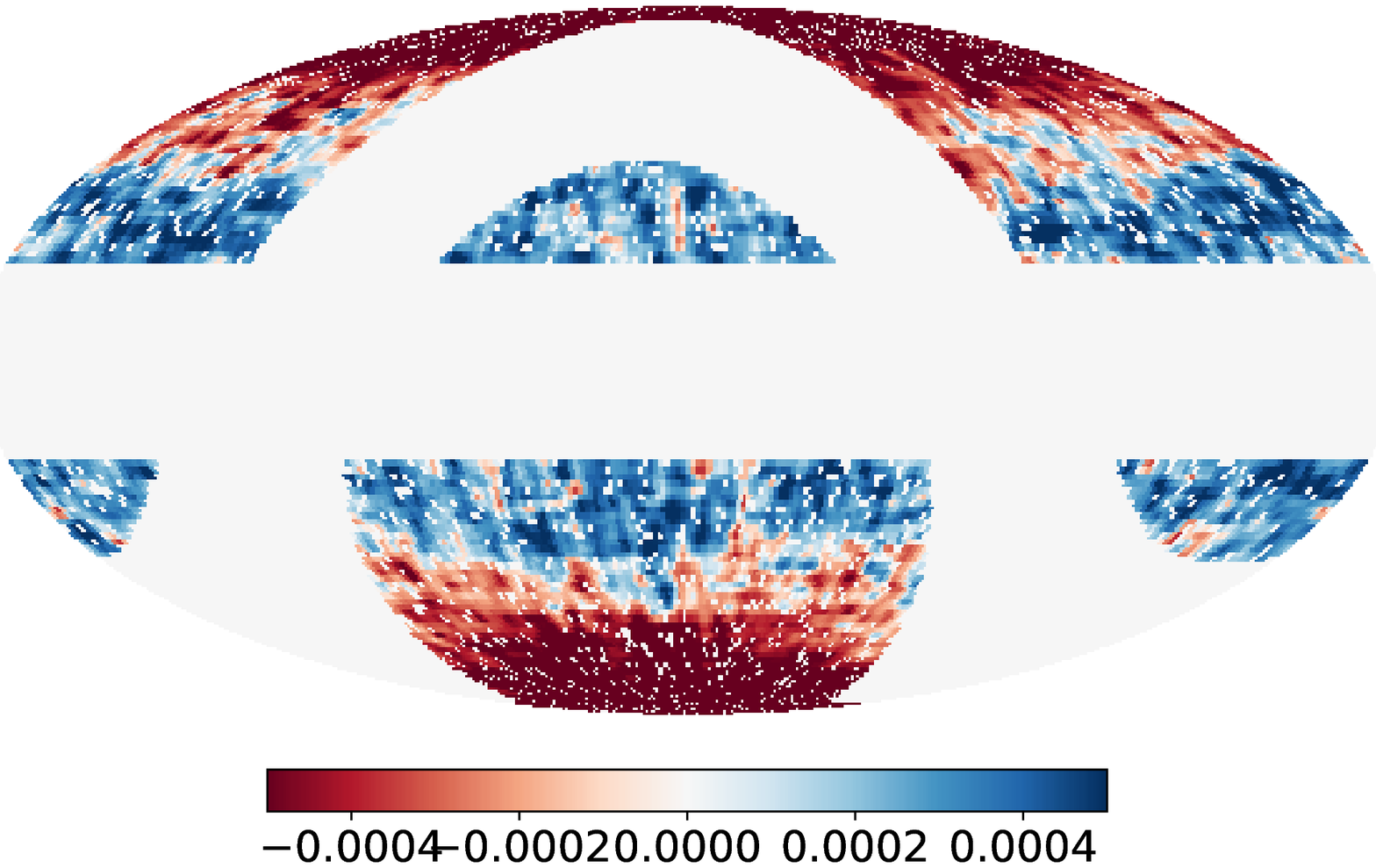}
\includegraphics[width=0.33\columnwidth]{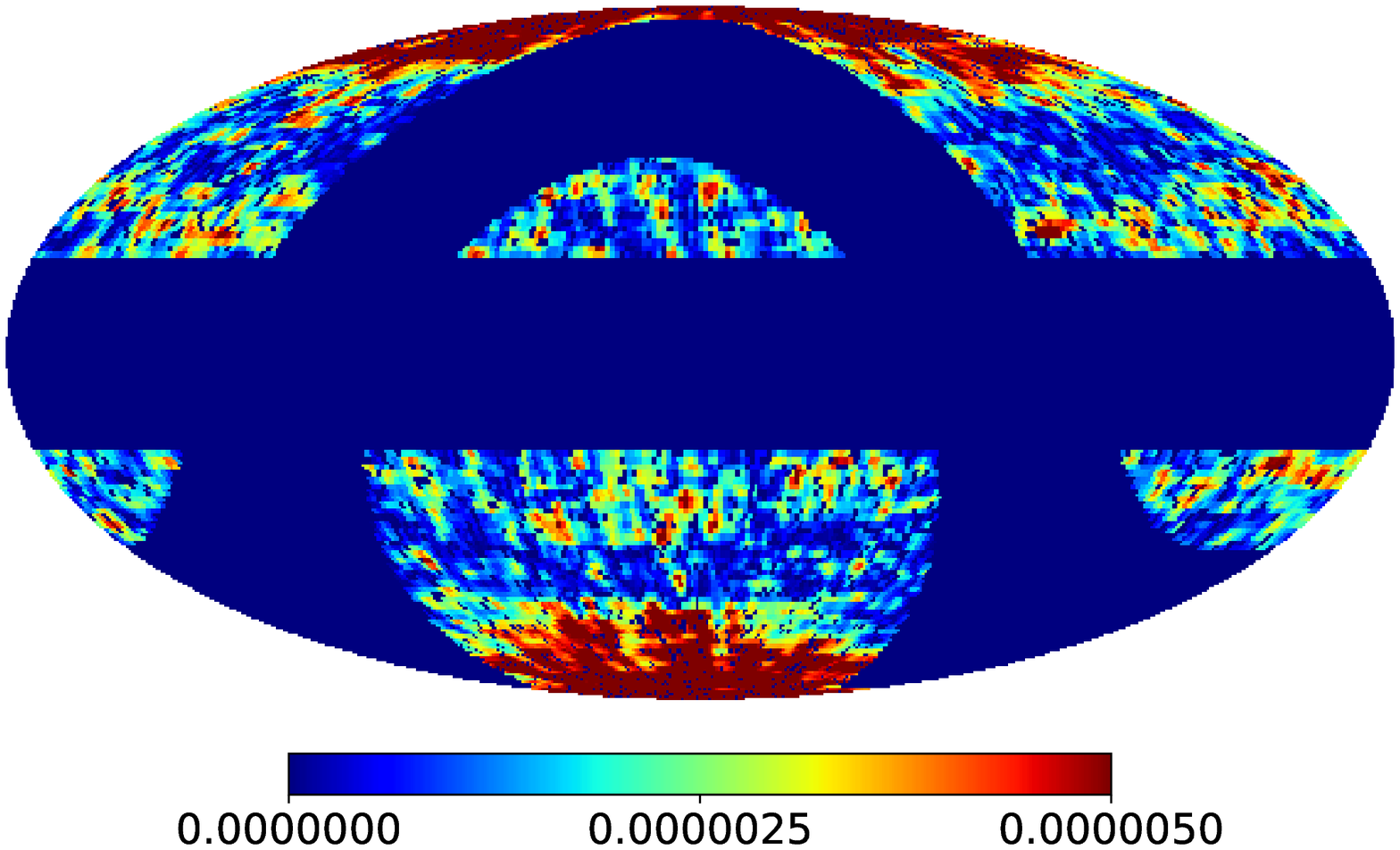}
\includegraphics[width=0.33\columnwidth]{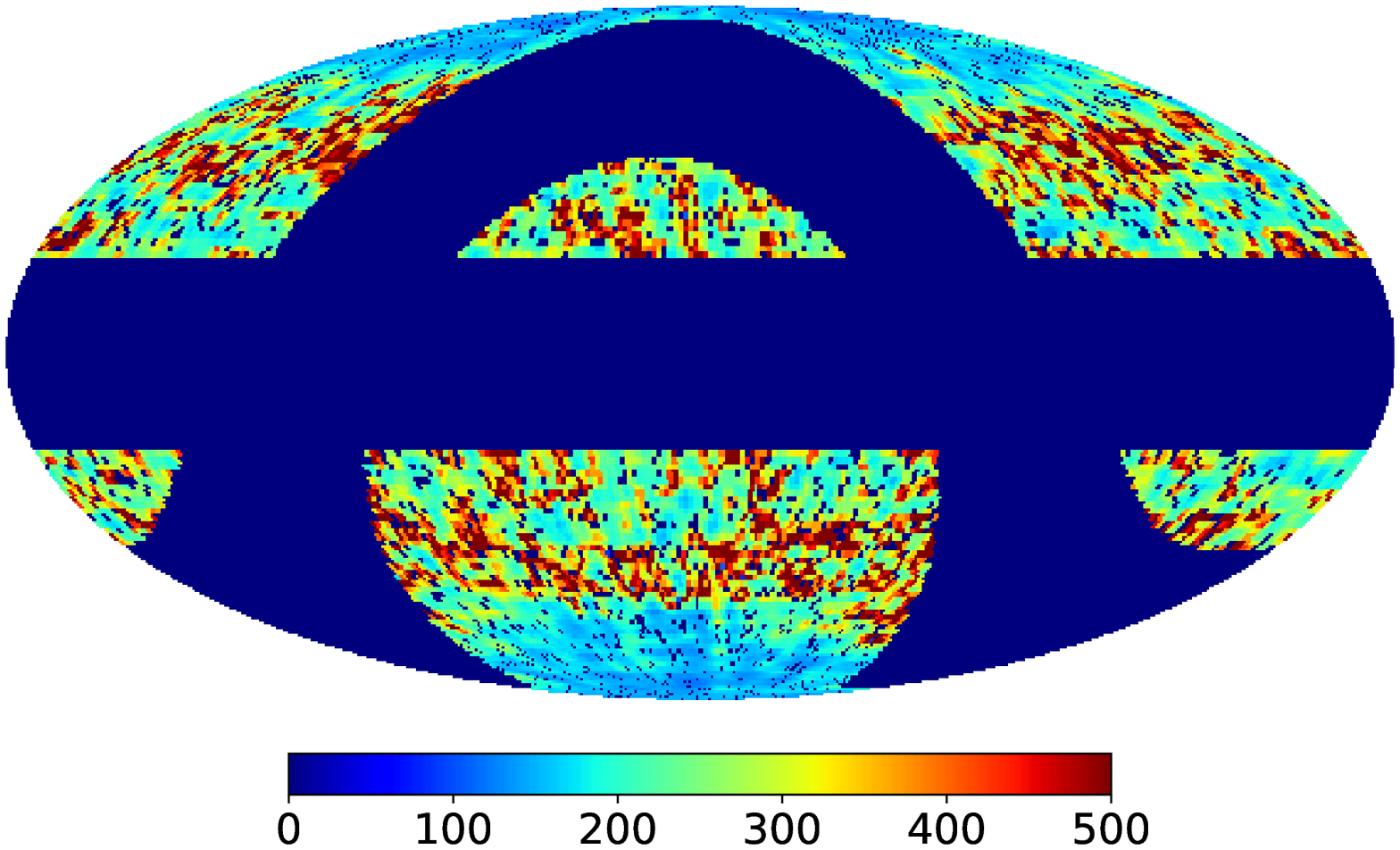}\\
\caption{A test of the error on an autocorrelation discrepancy map. The left panel shows the mean over $100$ groups of $N_{\rm sim}=100$ realisations, and the middle panel shows the standard deviation over the $100$ groups representing the estimate of the error on a single realisation, and the right panel shows the signal to noise (mean divided by standard deviation). We alter the colour in the middle and right panels relative to the other plots in this paper for clarity.}
\label{errortest}
\end{figure*}

In this Appendix we show some further tests of the autocorrelation discrepancy map statistic.
\vspace{-0.2cm}
\begin{center}
\emph{Sensitivity}    
\end{center}
\vspace{-0.3cm}
The autocorrelation discrepancy map method has $N_{\sigma}$ as a free parameter, for a given maximum multipole $L$ and $k_{\rm max}=L$. We investigate this for the simple galactic plane case in Figure \ref{Nk}. For $N_{\sigma}$ we find a too-small value leads to noisy recovery since the isotropic part of the observed field is included in the map, and with too-large a value the probability of including a fluctuation in the statistic decreases. Therefore $N_{\sigma}\approx 3$ seems approximately optimal. In practice this free parameter could be marginalised over in a likelihood analysis.
\vspace{-0.2cm}
\begin{center}
\emph{Signal-to-noise}   
\end{center}
\vspace{-0.3cm}
To test the uncertainty associated with the autocorrelation discrepancy map we generate $100$ groups of $N_{\rm sim}=100$ realisations, using the simple galactic plane case (with $k_{\rm max}=128$ and $\sigma_c=5\times 10^{-4}$), and compute the mean and the standard deviation over the groups. In Figure \ref{errortest} we show the mean over the groups and the standard deviation over the groups; noting that with only $100$ groups of simulations the error on the error is approximately $10\%$ \citep{2013MNRAS.432.1928T}. We find that the mean error on any given realisation is approximately $7\times 10^{-6}$, which is an equivalent a mean signal-to-noise of approximately $120$. A typical realisation can be found in Figure \ref{tests}. As $N_{\rm sim}$ increases this would improve still further.
\begin{figure*}
\centering
\includegraphics[width=0.33\columnwidth]{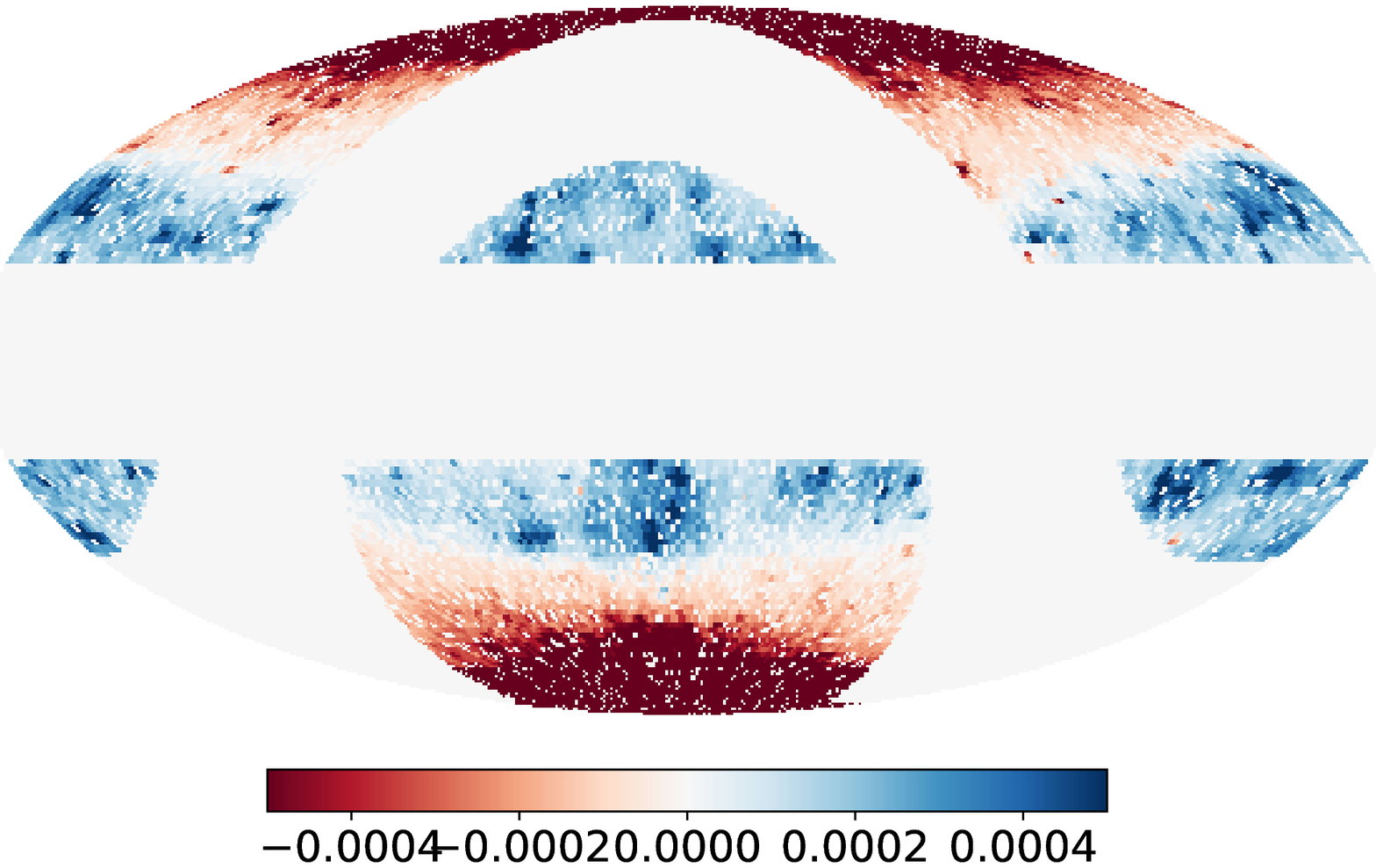}
\includegraphics[width=0.33\columnwidth]{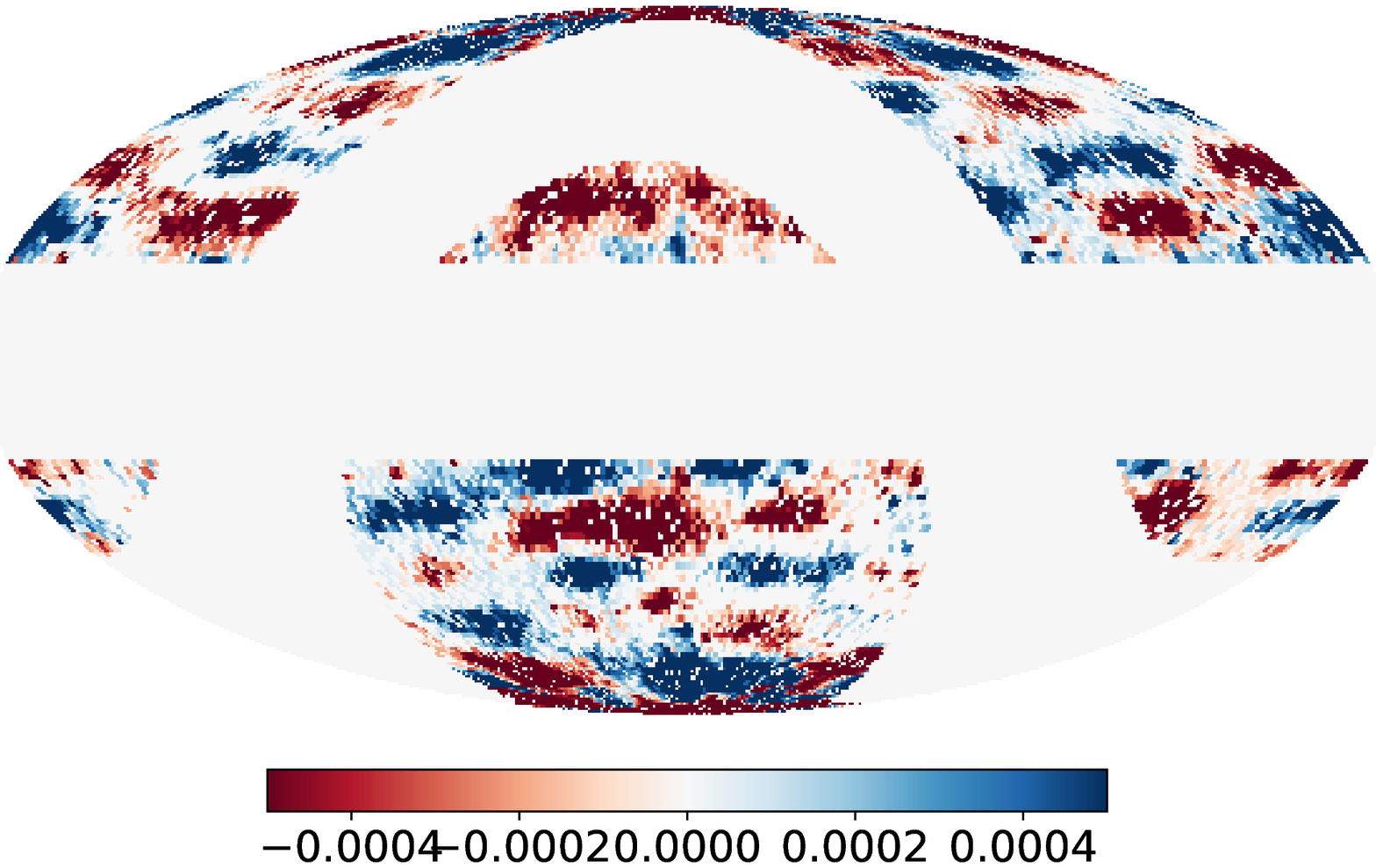}
\includegraphics[width=0.33\columnwidth]{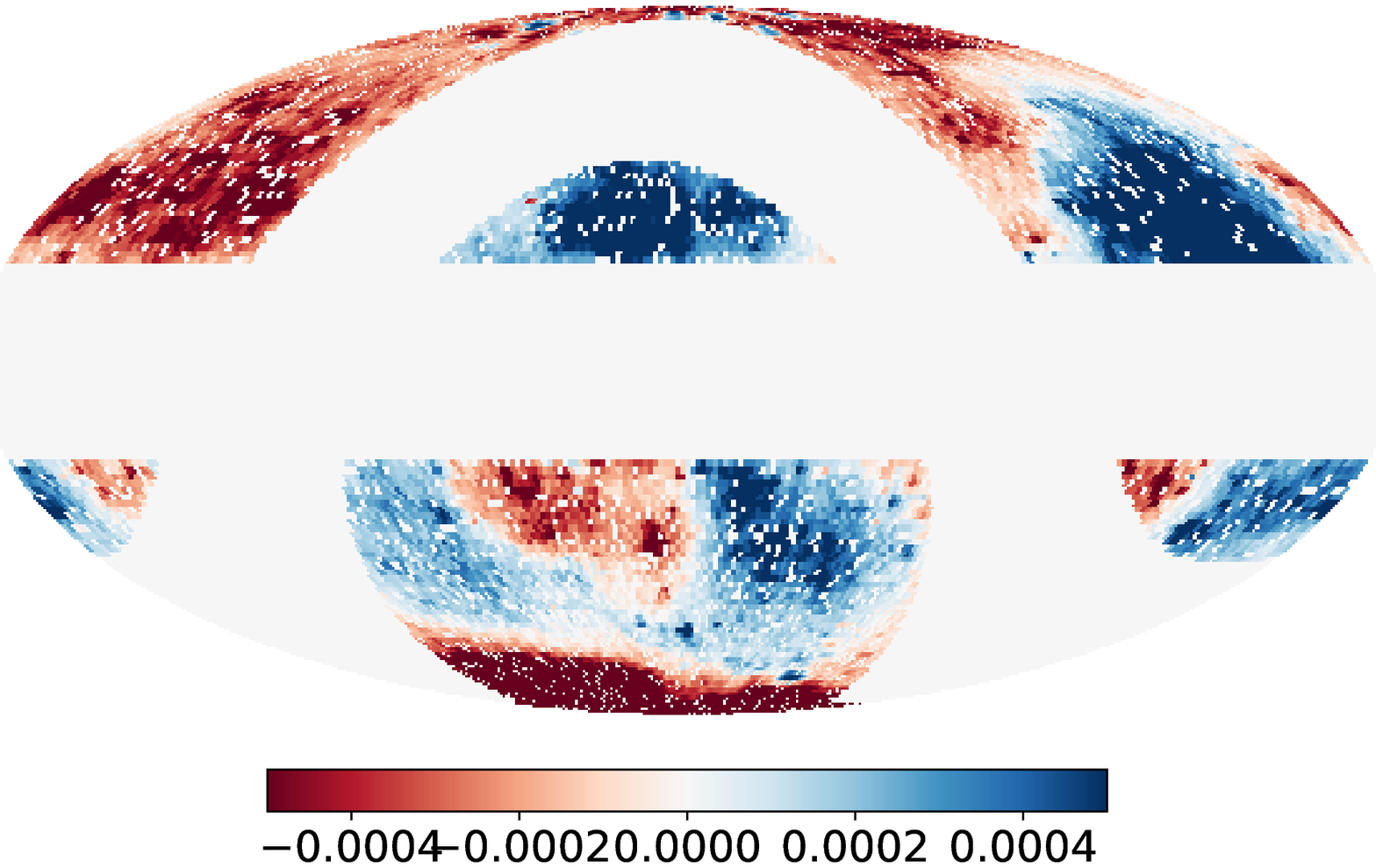}
\caption{A re-analysis of the results shown in Figure \ref{tests} except using log-normal random shear fields instead of Gaussian random shear fields.}
\label{lognormal}
\end{figure*}
\vspace{-0.2cm}
\begin{center}
\emph{Log-normal fields}
\end{center}
\vspace{-0.3cm}
The autocorrelation discrepancy map statistic presented in this paper relies on statistics that depend on deviations in variance of the spherical harmonic coefficients. For a Gaussian random field these statistics are analytic and well-defined (equations \ref{mu} and \ref{var}), and we generalise to the masked case through simulations. If the spherical harmonic coefficients were not Gaussian distributed then higher-order moments of their distribution may affect the statistic. This however should not be a problem, because the spherical harmonic coefficients are created through summing over the shear field (equation \ref{gtoe}) hence even for non-Gaussian fields the central limit theorem should result in approximately Gaussian distributed spherical harmonic coefficients \citep[and empirically this is found to be the case e.g.][]{2019PhRvD.100b3519T}. We test this by generating log-normal fields, instead of Gaussian random fields, by exponentiating the Gaussian-distributed shear field in our tests $\gamma(\mathbf{\Omega})\rightarrow {\rm exp}[-\gamma(\mathbf{\Omega})]-\lambda$ (where $\lambda$ ensures the mean of the exponentiated field is zero), and re-analysing the ability of the autocorrelation correlation discrepancy to extract the additive bias field. We show the results in Figure \ref{lognormal} and find no change in the reconstructed maps. 
\vspace{-0.2cm}
\begin{center}
\emph{Rotation of Local Coordinate System}    
\end{center}
\vspace{-0.3cm}
Additive bias fields are in general not rotationally invariant with respect to a rotation of the local coordinate system. Spin fields on the sphere have local directions defined relative to the North pole i.e. for a field $\widetilde\gamma(\mathbf{\Omega})=\widetilde\gamma_1(\mathbf{\Omega})+{\rm i}\widetilde\gamma_2(\mathbf{\Omega})$ the negative $\widetilde\gamma_1(\mathbf{\Omega})$ is orientated North-South and so forth. If one applies a local coordinate rotation $\chi \in [0,2\pi)$ then a spin-2 field is transformed via 
\begin{eqnarray}
\label{eqnrot}
[\widetilde\gamma_1(\mathbf{\Omega})\pm\widetilde\gamma_2(\mathbf{\Omega})]'={\rm e}^{\mp {\rm i}2\chi}[\widetilde\gamma_1(\mathbf{\Omega})\pm\widetilde\gamma_2(\mathbf{\Omega})],
\end{eqnarray}
where a dash refers to values in the rotated frame. This is not a global rotation on the sphere but rather a rotation by $\chi$ in the tangent plane centred on the spherical coordinates $\mathbf{\Omega}=(\theta, \phi)$, were $\theta$ is a colatitude and $phi$ is longitude (or R.A. and dec). A rotation of the global coordinate system will make no change since this is just a relabelling $\theta$ and $\phi$. When using an  autocorrelation discrepancy map to reconstruct the additive bias field we extract the real and imaginary ($c_1$ and $c_2$) separately (equation \ref{Dalpha}). When doing this this is a coordinate-dependent procedure. The $c_1$ reconstructed in this way should be the same as the \emph{true} $c_1$ component of the additive bias field, defined with respect to the coordinate system used, but since the additive field will have a different amplitude in rotated frame it may, or may not, be picked up by the filtering procedure.

To test this we create a bias field that has different $c_1$ and $c_2$ components, as defined in the fiducial coordinate system. We set $c_1={\rm SGP}$ as being the simple galactic plane field (left-hand panels of Figure 2), that we denote by ``SGP'' and $c_2=-{\rm SGP}$ as being the simple galactic plane field but with the sign flipped (i.e. positive at the poles in the fiducial coordinate frame rather than negative). With no rotation with respect to the fiducial coordinate frame the real part of the autocorrelation discrepancy map corresponds to the result in Figure 3. However if we apply a rotation then the real part of reconstructed map will now contain a mixture of the simple galactic plan and its negative (see equation \ref{eqnrot}). In this set up if $\chi=45\pi/180/2$ ($22.5$ degrees)  then $c_1'=0$ is zero and $c_2'=-\sqrt{2}[{\rm SGP}]$. If $\chi=45\pi/180/4$ ($11.25$ degrees) then $c_1'=0.54[{\rm SGP}]$ is zero and $c_2'=-1.31[{\rm SGP}]$.

\begin{figure*}
\centering
\includegraphics[width=0.33\columnwidth]{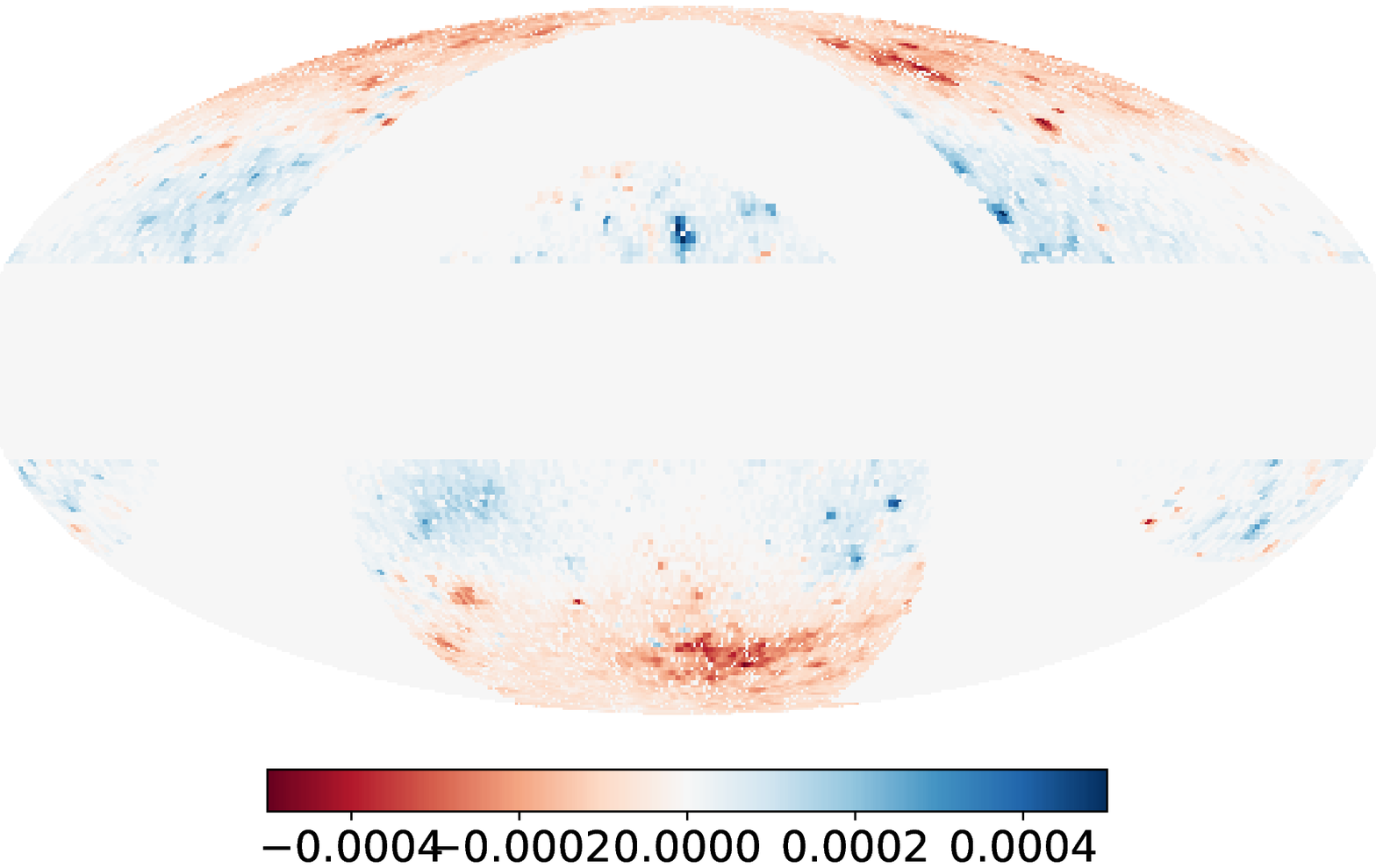}
\includegraphics[width=0.33\columnwidth]{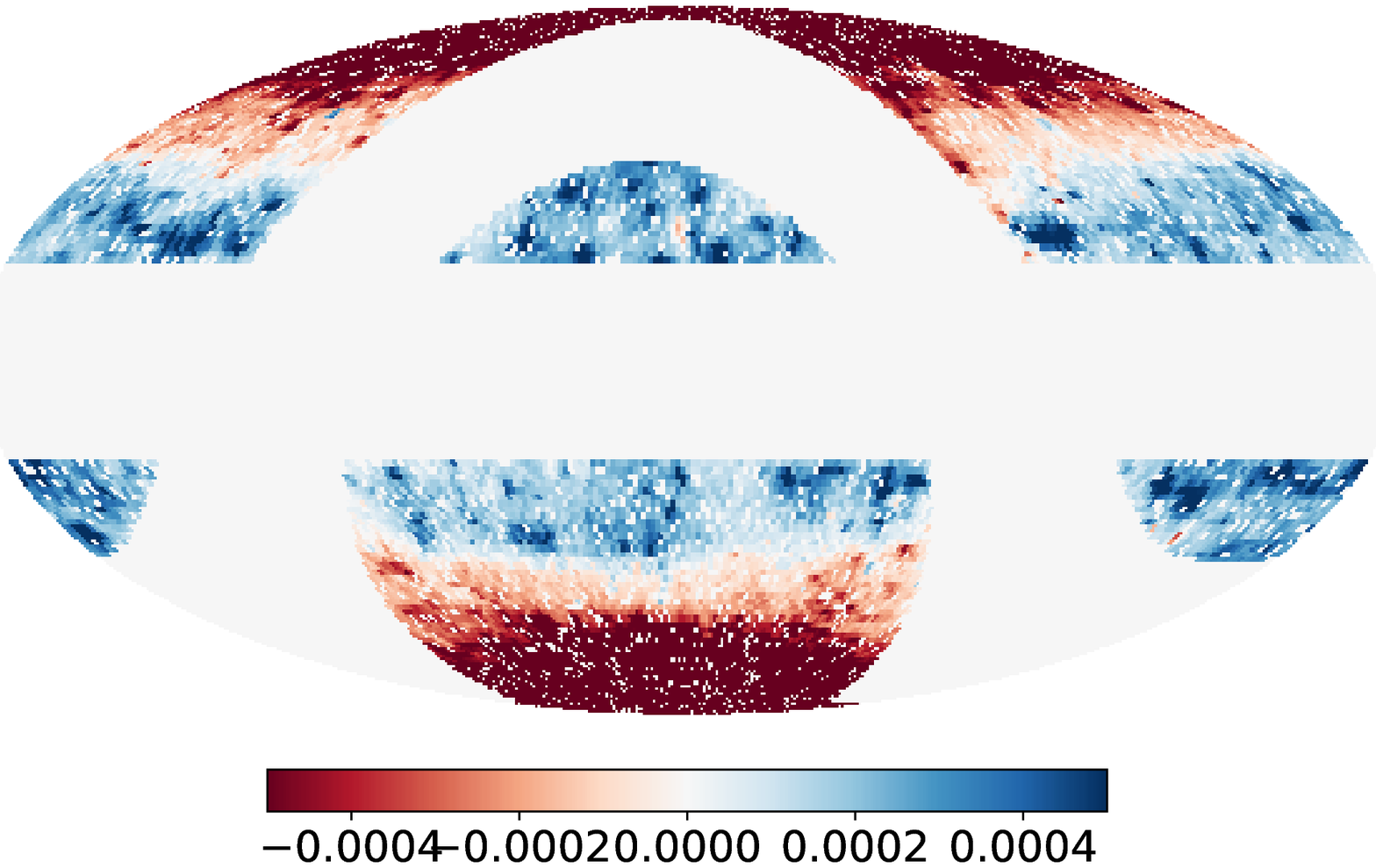}
\includegraphics[width=0.33\columnwidth]{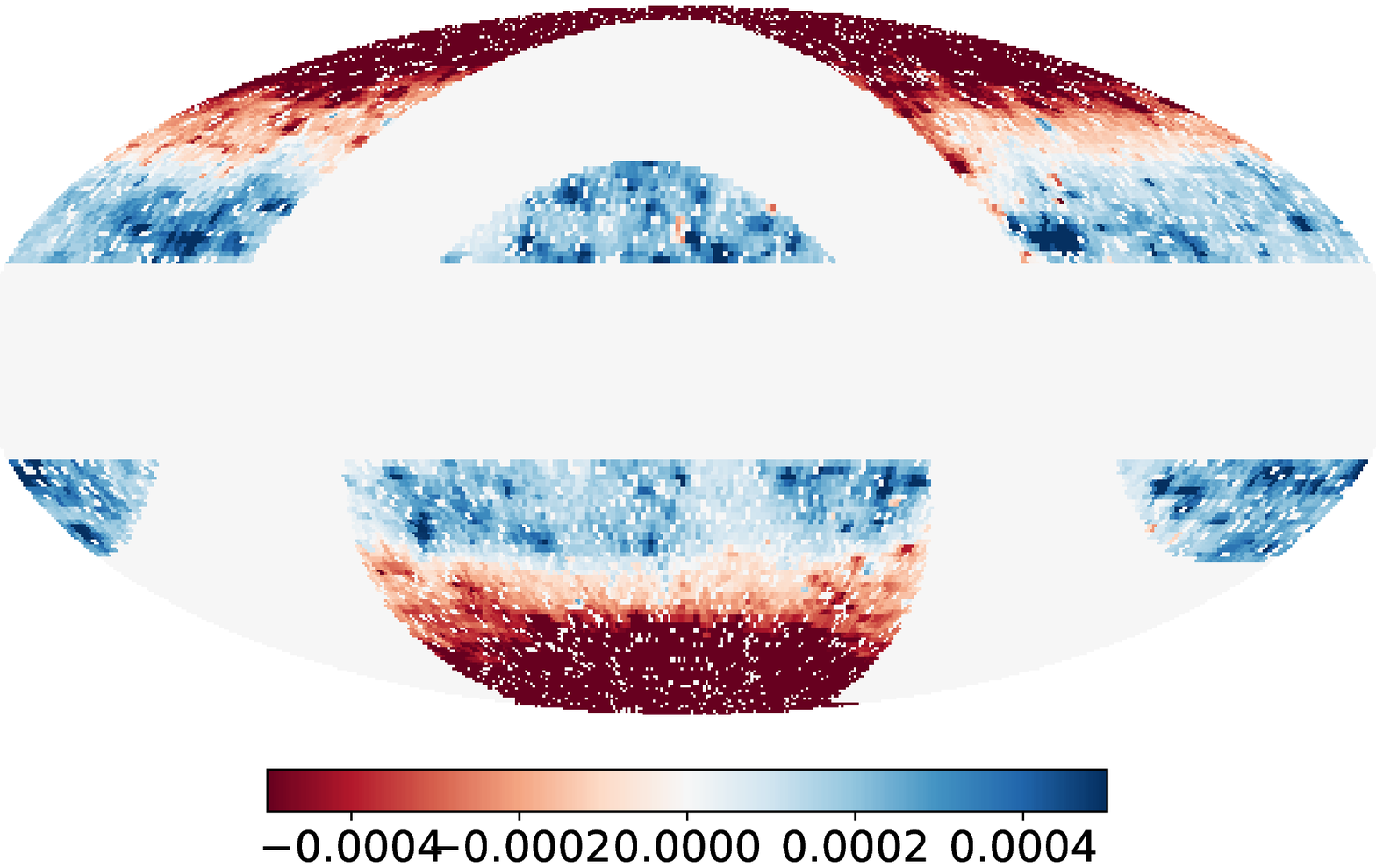}
\caption{A test of the sensitivity of the method to coordinate frame rotations. In the left hand panel we show the real part of the reconstructed additive bias field when a rotation of $11.25$ degrees is applied; this should correspond to an amplitude of $0.54$ times the simple galactic plane model. In the middle panel we show the the real part of the reconstructed additive bias field when an inverse rotation is applied back to the fiducial coordinate frame. In the right panel we show the same as the middle panel except with a rotation of $22.5$ degrees. The middle and right panels should correspond to the left hand column, third row, in Figure 3.}
\label{rotations}
\end{figure*}

In Figure \ref{rotations} we apply a $\chi=45\pi/180/4$ ($11.25$ degree) rotation to the unmasked observed shear field and show the real component of reconstructed additive bias field, which should be $c_1'=0.54[{\rm SGP}]$. We find indeed that the reconstructed real part of the field is detected at this amplitude. We then apply the inverse transform of equation (\ref{eqnrot}) and show the real part in the middle panel, which should correspond to the left hand panel in Figure 2; we find this to be the case. In the right hand panel we repeat this exercise for $\chi=45\pi/180/2$ ($22.5$ degrees) and again find that when rotated back into the original coordinate frame the reconstruction is essentially unaffected. 

\begin{figure*}
\centering
\raisebox{0.3\height}{\includegraphics[width=0.33\columnwidth]{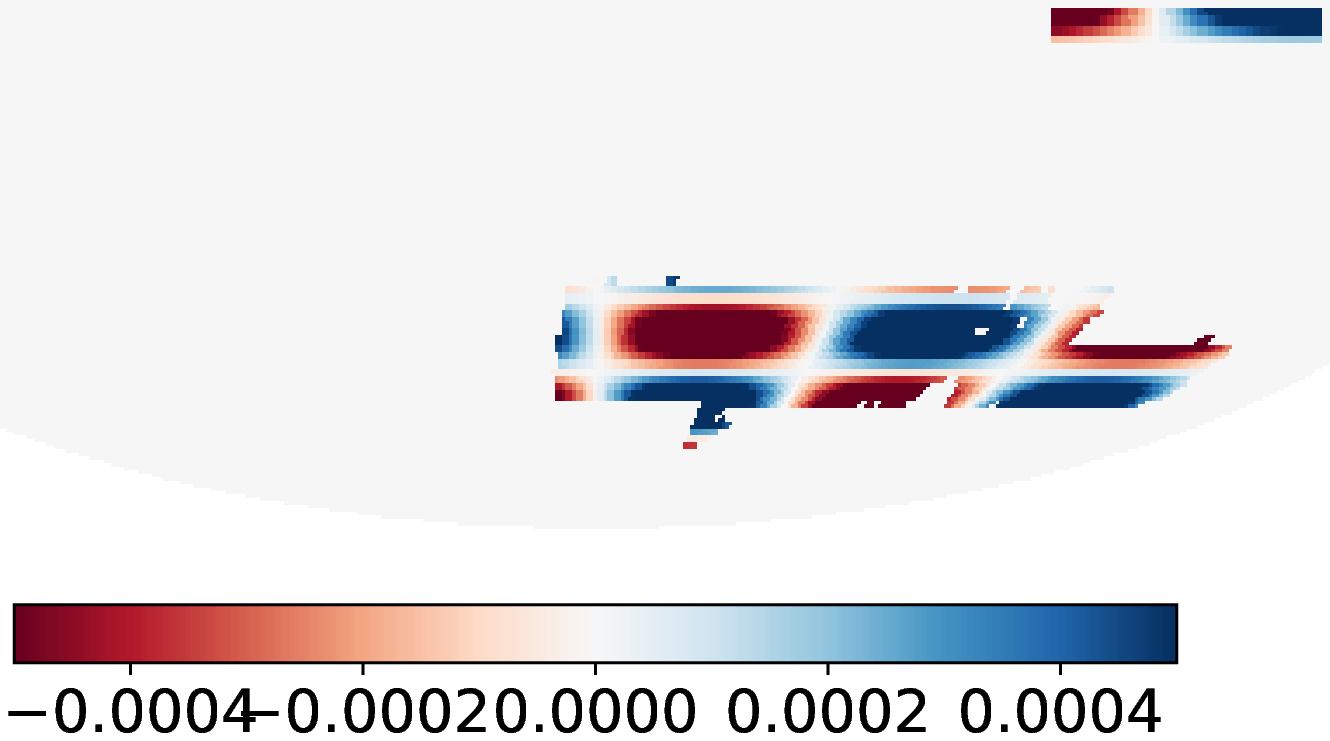}}
\raisebox{0.3\height}{\includegraphics[width=0.33\columnwidth]{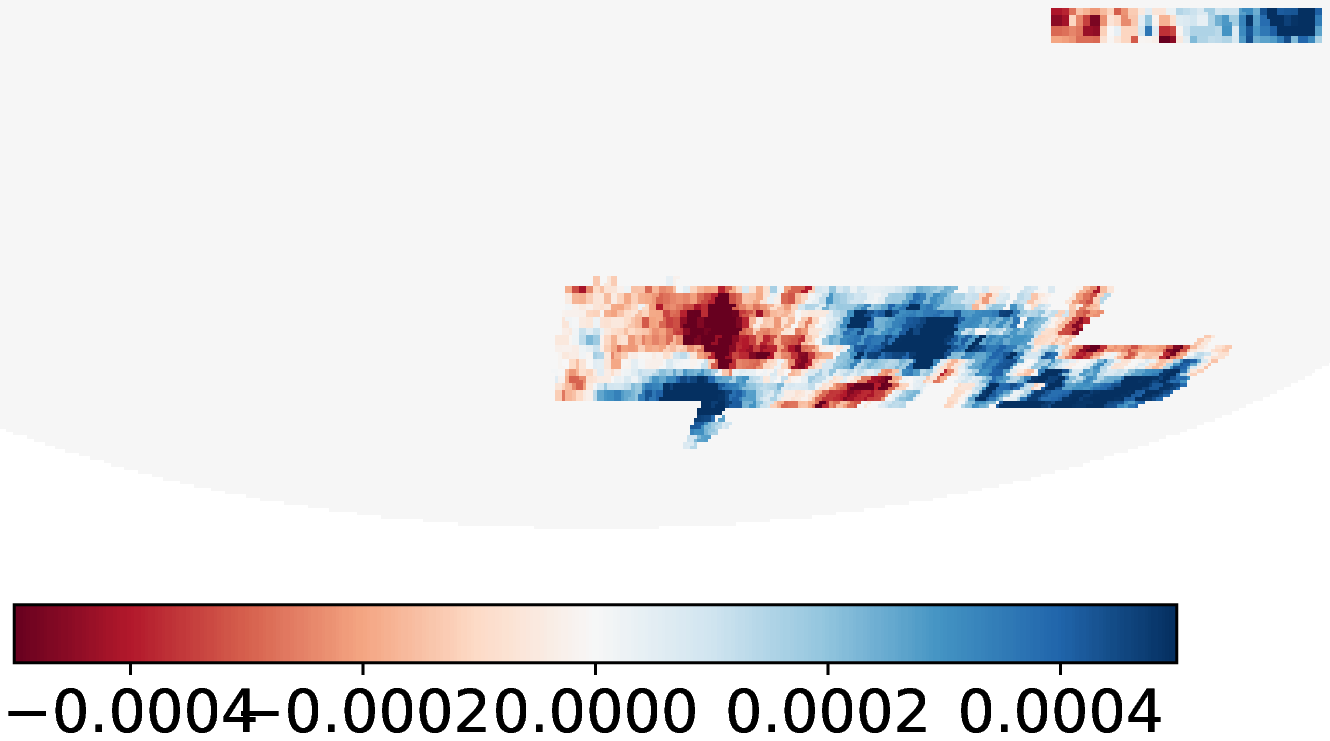}}
\includegraphics[width=0.33\columnwidth]{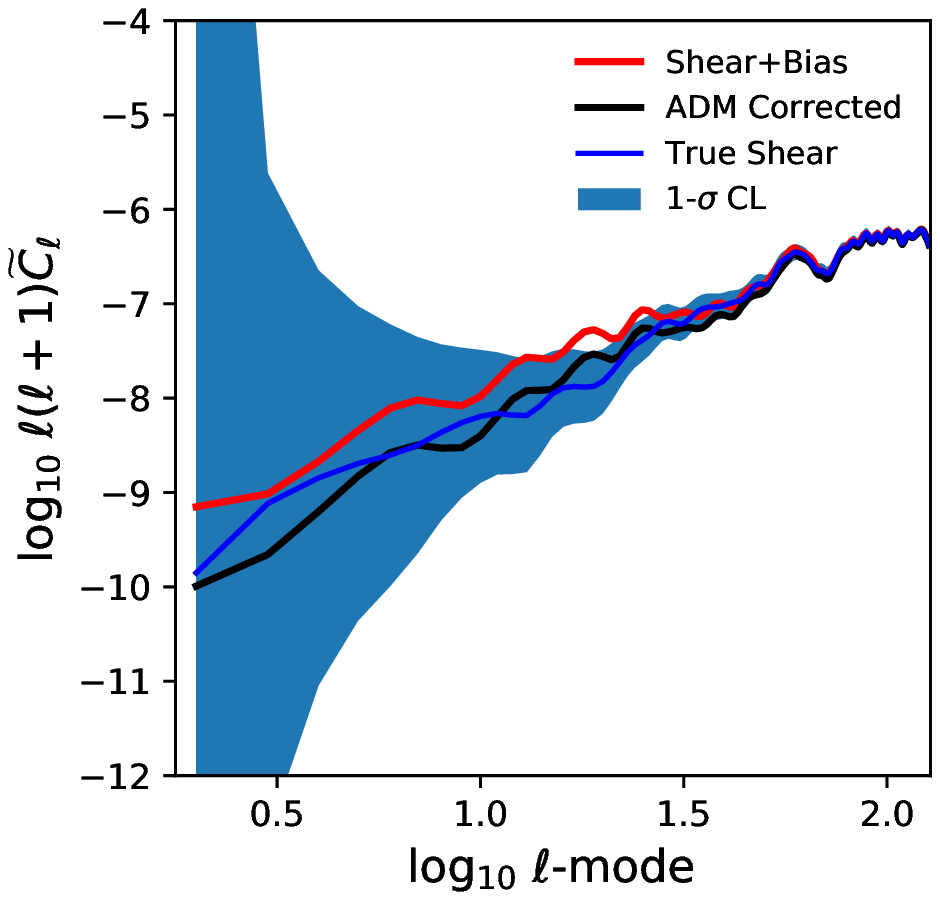}
\caption{A repeat of the test demonstrated in Figure 3, for the simple patch pattern, but using a DES Year 1 survey mask (see Section \ref{Application to Data}). The right panel shows the real part of the masked additive field, the middle panel shows the reconstructed field, and the right panel shows the correction to the power spectrum. See Figure 3 for more details. We note that the error about the true shear increases due to the smaller survey area.}
\label{despatch}
\end{figure*}

We note here however that the choice of coordinate frame $\chi$ is an additional free parameter choice for this methodology, along with the threshold $N_{\sigma}$, and as such may have to be marginalised over in a more sophisticated application. 
\vspace{-0.2cm}
\begin{center}
\emph{Small Masked Region Test}    
\end{center}
\vspace{-0.3cm}
For very small fields that are heavily masked on the sphere the angular mode mixing may be particularly strong. Whilst there is no reason that a spherical harmonic based pseudo-$C_{\ell}$ approach would not work in such a domain, we nonetheless test this by applying a DES Year 1 type mask (see \ref{Application to Data}). In Figure \ref{despatch} we show the reconstruction of the simple patch pattern masked with the DES Year 1 mask, using $N_{\sigma}=3$, and find a similar quality of reconstruction to Figure 3, which used large-scale masks.

\label{lastpage}
\end{document}